\documentclass[final,1p,times,authoryear]{elsarticle}
\usepackage{color}
\usepackage{latexsym}
\usepackage{todonotes}
\usepackage{graphicx}
\usepackage{color}
\usepackage{float}
\usepackage{caption}
\usepackage{subcaption}
\usepackage{algpseudocode}
\usepackage[linesnumbered, boxed]{algorithm2e}
\usepackage{natbib}
\usepackage{tikz}
\usepackage{amsthm}

\newcommand{\radon}{\mathfrak{R}}

\usetikzlibrary{intersections}
\usetikzlibrary{arrows, decorations.pathmorphing, backgrounds, fit , petri}
\usepackage{amsmath,mathtools,calc}

\usetikzlibrary{calc, intersections, through }
\usetikzlibrary{arrows}
\usetikzlibrary{positioning}
\usetikzlibrary{patterns,decorations.pathreplacing}

\newtheorem{thm}{Theorem}
\newtheorem{lem}[thm]{Lemma}

\newcommand{\N}{\mathbb{N}}

\newcommand{\F}{\mathfrak{F}}
\newcommand{\B}{\mathfrak{B}}

\newcommand{\x}{\mathbf{x}}

\newcommand{\R}{\mathbb{R}}

\makeatletter
\providecommand*{\diff}%
{\@ifnextchar^{\DIfF}{\DIfF^{}}}
\def\DIfF^#1{%
\mathop{\mathrm{\mathstrut d}}%
\nolimits^{#1}\gobblespace}
\def\gobblespace{%
\futurelet\diffarg\opspace}
\def\opspace{%
\let\DiffSpace\!%
\ifx\diffarg(%
\let\DiffSpace\relax
\else
\ifx\diffarg[%
\let\DiffSpace\relax
\else
\ifx\diffarg\{%
\let\DiffSpace\relax
\fi\fi\fi\DiffSpace}

\let\oldnl\nl
\newcommand{\nonl}{\renewcommand{\nl}{\let\nl\oldnl}}

\newcommand{\indi}[1]{_{\smash{\mathrlap{#1}}}}

\journal{Computational and Applied Mathematics}

\begin{document}

\begin{frontmatter}


\title{Algorithm for the reconstruction of dynamic objects in CT-scanning using optical flow \tnoteref{tnote1}}

 \author[UA]{Koen Ruymbeek\corref{cor1}\fnref{label1}}
 \ead{koen.ruymbeek@cs.kuleuven.be}
 \cortext[cor1]{Corresponding author}
 \fntext[label1]{Present adress: Department of Computer Science, KU Leuven, Celestijnenlaan 200A, 3001 Leuven, Belgium}
 \author[UA]{Wim Vanroose} 
 \tnotetext[tnote1]{This research did not receive any specific grant from funding agencies in the public, commercial, or
not-for-profit sectors.}


\address[UA]{Department of Mathematics and Computer Science, University of Antwerp, Middelheimlaan 1, 2020 Antwerp, Belgium}


\begin{abstract}
Computed Tomography is a powerful imaging technique that allows non-destructive visualization of the interior of physical objects in different scientific areas. In traditional reconstruction techniques the object of interest is mostly considered to be static, which gives artefacts if the object is moving during the data acquisition. In this paper we present a method that, given only scan results of multiple successive scans, can estimate the motion and correct the CT-images for this motion assuming that the motion field is smooth over the complete domain using optical flow. The proposed method is validated on simulated scan data. The main contribution is that we show we can use the optical flow technique from imaging to correct CT-scan images for motion.
\end{abstract}


\end{frontmatter}


\section{Introduction}

\paragraph{CT} Computed Tomography (CT) is a powerful imaging technique that allows
non-destructive visualization of the interior of physical objects in
medical applications, bio-mechanical research, material science,
geology, etc. In current applications, a certain imaging resource
and detector, e.g. an X-ray source and detector, are used to acquire
two-dimensional projection images of an object, each measured from different
directions. From these projections, a three-dimensional virtual
reconstruction can then be computed. We refer to
\cite{webb1990watching} for a review on the origin of computed
tomography.

\paragraph{Reconstruction techniques} In practice, the most commonly used analytical method for CT
reconstruction is filtered back projection \cite{pan2009commercial}. A major drawback of this method is its inflexibility to different
experimental set-ups and its inability to include reconstruction
constraints, which can be used to exploit possible prior information
to improve the reconstruction of the object. Iterative Algebraic Reconstruction Techniques
(ARTs) are a powerful alternative to the aforementioned
analytical method by describing the reconstruction
problem as a system of linear equations. Algebraic reconstruction
methods include SIRT \cite{gregor2008computational},
SART \cite{andersen1984simultaneous} and DART \cite{batenburg2011dart}
and the general class of Krylov solvers such as CG, BiCGStab, GMRES,
CGLS and LSQR (of which the latter two are applicable to non-square
systems), an overview of which can be found in
\cite{simoncini2007recent}.

\paragraph{Time-dependent CT} 
In both the analytical and algebraic methods, the object of interest
is traditionally considered to be static. However, when the object is moving or changing form during the data
acquisition process, the flow (direction of movement) also needs to be
reconstructed.  This requires the calculation of a full 3D
reconstruction of both the object and the flow field from the
projection measurements over a period of time, yielding a 4D
tomographic reconstruction problem, i.e. 3D in space plus 1D for time.

During the previous years, significant progress is made to account for
motion during the acquisition process. A first approach is to model
the motion in the reconstruction model \cite{mooser2013estimation, van2012combined, li2005motion, van2014region}.  A
second method sorts the data in subsets such that each subset
contains data acquired from a static object. This technique is used, for example when there
is periodic motion such as breathing \cite{lu2006comparison} or
heartbeats. Within each subset, where the object does not change, a
reconstruction is performed. Finally, if the motion is known upfront, it is possible to compensate for the motion of the object in the CT-scanning \cite{hahn2014efficient} by using the method of the approximate inverse \cite{sota_4}.

\paragraph{Optical flow}
Throughout this paper, we extract the motion using optical flow from the scan data. This is a widely used technique in imaging (see \cite{bardow2016simultaneous, rol_shut} and it exploits the differences between the images to identify patterns of motion. Let $f(x,y,t)$ and $f(x,y,t+\Delta t)$ be two pictures taken with a small time difference $\Delta t$ (we consider 2D images  for convenience). Assuming that for every $(x,y)$ holds that $f(x,y,t) = f(x+\Delta x, y+ \Delta y, t+ \Delta t)$ for some $\Delta x$ and $\Delta y$ (this is \emph{the brightness constancy constraint}), we can deduce the linear system $\frac{\partial f}{\partial x} v_x+\frac{\partial f}{\partial y} v_y = -\frac{\partial f}{\partial t}$ using Taylor series. The unknowns $v_x$ and $v_y$ are the $x$ and $y$ components of the optical flow or the velocity components of the object. There are many ways to use optical flow to estimate the motion from a sequence of images. The most widely used methods are the differential methods that can be classified into local methods like the Lucas-Kanade technique \cite{LK}, global methods like the Horn-Schunck approach \cite{HS} and some extensions \cite{HS_TV}. Local methods calculate the flow per point whereas global methods solve one equation for the entire domain. Recently, methods were developed that are a combination of a local and a global method \cite{Bruhn2005} and also Newton-Krylov methods with regularization have been applied to this problem \cite{mang2015inexact}. 


\paragraph{Outline} The paper is structured as follows. Section \ref{sec:notation} introduces the general notations and basic concepts used throughout this article. In section~\ref{sec:mot_est} we look at how we can retrieve the motion given just the scan results from a theoretical as well as from a practical point of view. In section~\ref{sec:cor_images} we present our method to correct CT-scan images for the motion. Numerical results on both the motion estimation and the correcting of the images are shown in section~\ref{sec:numer_results}. Finally, we draw some conclusions and we review some research possibilities in section \ref{sec:conclusion}. The main contribution is that techniques in imaging to detect motion can also be used to detect the motion in CT-scan images. The numerical results show that it gives also in practice the desired results. 

We note discretisations of analytic variables with bold small letters and matrices with bold big letters.
\section{Notation and key concepts} \label{sec:notation}
We describe everything for 2D CT-scanning, but all definitions can easily be expanded to 3D.
\subsection{Modelling scan data} 
\begin{figure}[H]
\begin{center}
\definecolor{qqttzz}{rgb}{0.,0.2,0.6}
\definecolor{ffqqqq}{rgb}{1.,0.,0.}
\definecolor{wqwqwq}{rgb}{0.3764705882352941,0.3764705882352941,0.3764705882352941}
\definecolor{qqwwtt}{rgb}{0.,0.4,0.2}
\definecolor{cqcqcq}{rgb}{0.7529411764705882,0.7529411764705882,0.7529411764705882}
\begin{tikzpicture}[line cap=round,line join=round,>=triangle 45,x=1.0cm,y=1.0cm, scale = 0.4]
\draw[->,color=black] (-6.5,0.) -- (6.5,0.);
\foreach \x in {-6.,-4.,-2.,2.,4.,6.}
\draw[shift={(\x,0)},color=black] (0pt,2pt) -- (0pt,-2pt);
\draw[->,color=black] (0.,-6.5) -- (0.,6.5);
\foreach \y in {-6.,-4.,-2.,2.,4.,6.}
\draw[shift={(0,\y)},color=black] (2pt,0pt) -- (-2pt,0pt);
\clip(-6.5,-6.5) rectangle (6.5,6.5);
\fill[color=wqwqwq,fill=wqwqwq,fill opacity=0.15] (-2.,2.) -- (2.,2.) -- (2.,-2.) -- (-2.,-2.) -- cycle;
\draw [->] (4.,-4.) -- (-4.,4.);
\draw [line width=1.6pt] (-4.,4.)-- (-1.8767425171580223,6.123257482841978);
\draw [line width=1.6pt] (-6.120212235300748,1.8797877646992514)-- (-4.,4.);
\draw [line width=1.6pt] (4.,-4.)-- (6.121320343559642,-1.8786796564403572);
\draw [line width=1.6pt] (4.,-4.)-- (1.8767425171580223,-6.123257482841978);

\draw [->] (6.121320343559642,-1.8786796564403572) -- (-1.8767425171580223,6.123257482841978);
\draw [->] (1.8786796564403572,-6.121320343559642) -- (-6.120212235300748,1.8797877646992518);
\draw (-5.4137237355989685,2.586276264401031)-- (-6.120212235300748,1.8797877646992514);
\draw (-5.4137237355989685,2.586276264401031)-- (-4.708063044045177,3.291936955954822);
\draw (-4.,4.)-- (-3.3103120588520896,4.68968794114791);
\draw (-3.3103120588520896,4.68968794114791)-- (-2.604843361244298,5.395156638755703);
\draw (1.8786796564403572,-6.121320343559642)-- (2.6134170602299553,-5.386582939770044);
\draw (2.6134170602299553,-5.386582939770044)-- (3.3201346584769653,-4.679865341523034);
\draw (4.,-4.)-- (4.708565290393692,-3.2914347096063072);
\draw (4.708565290393692,-3.2914347096063072)-- (5.416572963622752,-2.583427036377248);

\draw [->] (4.708565290393692,-3.2914347096063072) -- (-3.31031205885209,4.689687941147911);
\draw [->] (3.3201346584769653,-4.679865341523034) -- (-4.708063044045177,3.2919369559548217);
\draw [->] (2.6134170602299553,-5.386582939770044) -- (-5.4137237355989685,2.5862762644010306);
\draw (-2.,2.)-- (2.,2.);
\draw (2.,2.)-- (2.,-2.);
\draw (2.,-2.)-- (-2.,-2.);
\draw (-2.,-2.)-- (-2.,2.);
\draw [line width=1.pt,color=ffqqqq] (-4.242640687119286,-4.242640687119285)-- (0.,0.);
\draw [line width=1.6pt,color=qqttzz] (0.,0.)-- (1.39198102414633,1.3919810241463295);
\draw [shift={(0.,0.)},->,line width=1.pt,color=ffqqqq] (0.:4.5) arc (0.:45.:4.5);

\draw (3.309725262577558,-4.690274737422442)-- (4.,-4.);
\draw [->,line width=1.pt,color=ffqqqq] (1.39198102414633,1.3919810241463295) -- (4.242640687119286,4.242640687119285);
\draw [->,line width=1.pt,color=qqwwtt] (5.416572963622752,-2.583427036377248) -- (-2.6048433612442974,5.395156638755703);
\begin{small}
\draw[color=qqttzz] (1.163719105045583,0.618104745191834) node {$u$};
\draw[color=ffqqqq] (4.874487153830124,1.7526686212088711) node {$\alpha$};
\end{small}
\end{tikzpicture}
\caption{Schematic representation of the 2D Radon transformation. The grey box is the domain $\Omega$ of the object $f$} \label{intro_fig_CT_scan}
\end{center}
\end{figure}
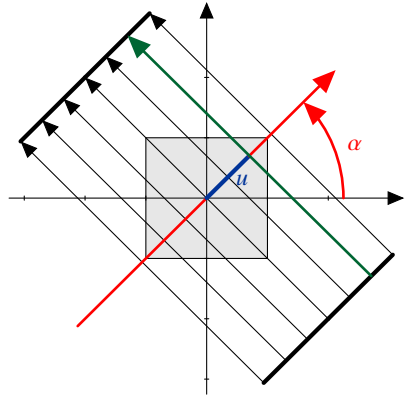

\begin{figure}[H]  
        \begin{subfigure}[b]{0.5\textwidth}
        \includegraphics[scale=0.5, clip]{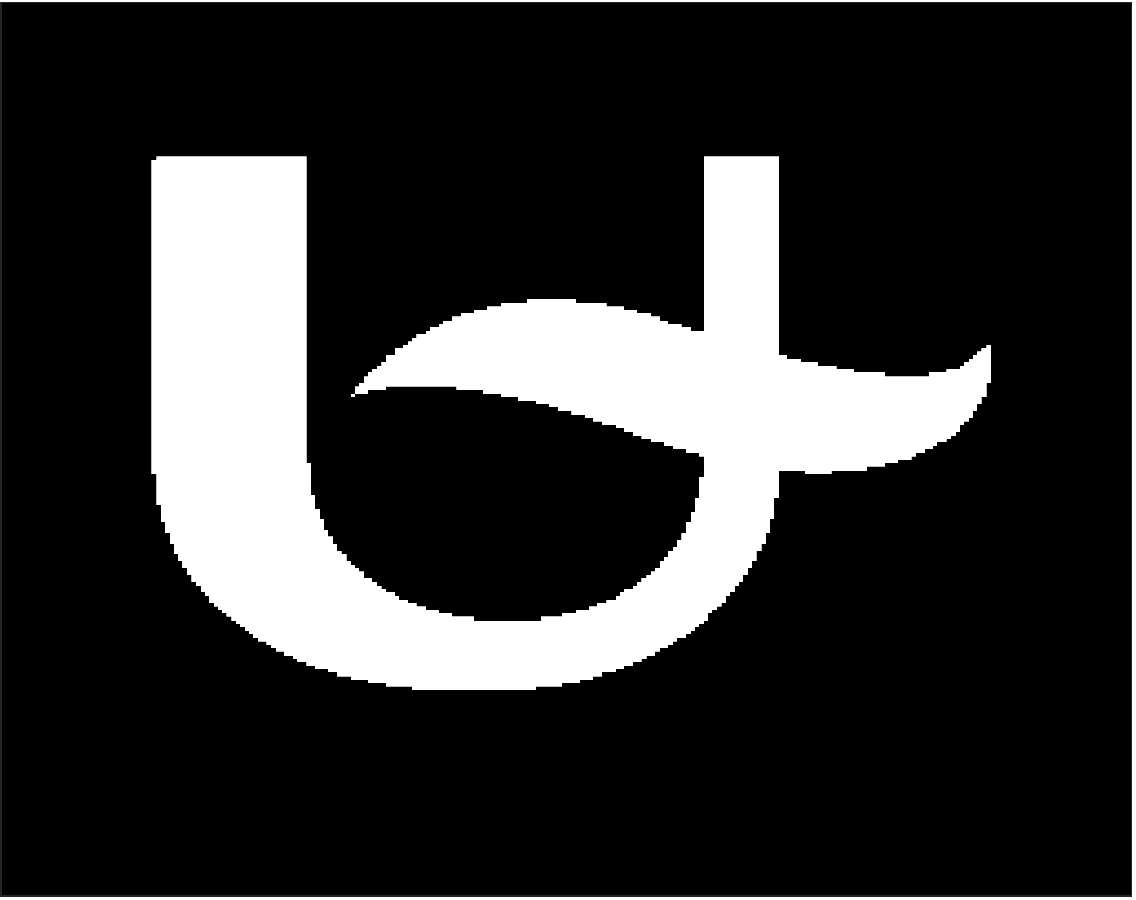}    
        \caption{}
        \end{subfigure} \hspace{0.25cm}
        \begin{subfigure}[b]{0.5\textwidth}
        \includegraphics[scale=0.47,clip]{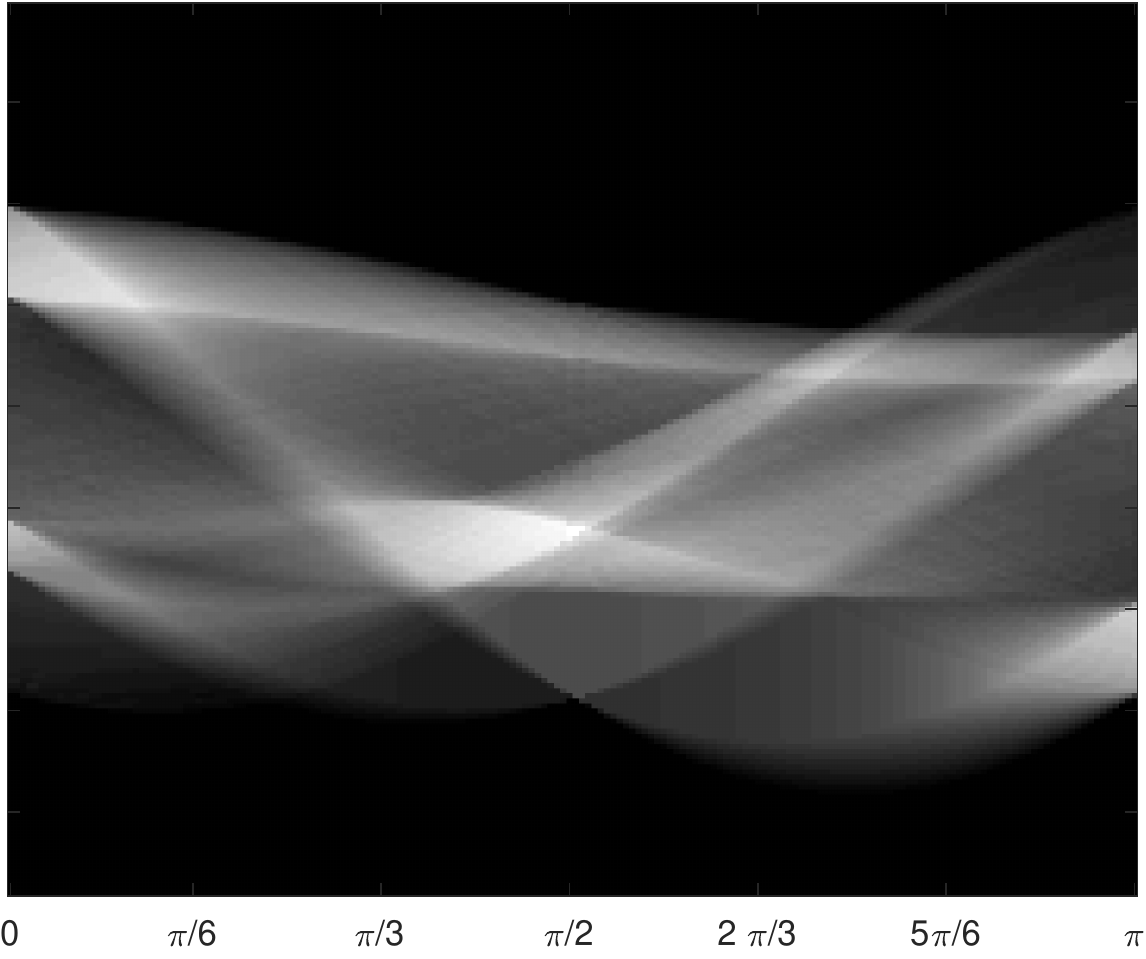}
        \caption{}
        \end{subfigure}
\caption{(a) The image $f$ on a $256 \times 256$ pixel grid at time point $0$ where black represents a zero and white represents the value 1. (b) Sinogram of the example on the left. Given this sinogram, we want to retrieve the object on the left. The x-axis respectively y-axis represents the different angles (between 0 and $\pi$) respectively the different detectors. The color shows the value of the detected X-ray.} \label{intro_vb}
\end{figure}

Let $f$ be a time-dependent object, i.e. a function $f: \R^2 \times [0,T_f] \rightarrow \R$ where $[0, T_f]$ is the time interval. For notational purposes, let $f_t: \R^2 \rightarrow \R$ be the object $f$ at time $t$. We assume, for all $t$, that $f_t$ is an element of $L^2(\R^2)$ and that it is zero outside a square domain $\Omega \subset \R^2$ around the center. 
We first define the used scan model for a stationary object (independent of $t$). In the next paragraph we extend this to time-dependent objects.
The scan data per X-ray are modelled by the so-called \emph{Radon transformation}. The Radon transform  for a specific angle $\alpha \in [0, 2 \pi]$ and shift $u \in \R$ is defined as the integral
\begin{equation}
  \radon f(\alpha,u) = \int_{L(\alpha,u)} f(x,y)\diff x \diff y \label{eq:radon}
\end{equation}
where $L(\alpha, u) = \{ (x,y) \in \R^2 | x \cos(\alpha)+y
\sin(\alpha) = u\}$. This integration area is the line
perpendicular to the direction $(\cos(\alpha), \sin(\alpha))$ at a
shift $u$ of the origin. An illustration of this definition can be
found in figure~\ref{intro_fig_CT_scan}. A full scan consists of projection data over angles in an interval of size $\pi$ and shifts $u$ such that the X-ray bunch covers the hole domain $\Omega$. The set of all projection data is called a \emph{sinogram} and is thus defined as
\begin{equation}
\radon f: [0, \pi[ \rightarrow \R: (\alpha,u) \mapsto \radon f(\alpha,u) \label{eq:sinogram}
\end{equation}
for a stationary object. An example is given in figure \ref{intro_vb}.

If the object moves or changes form we need to generalise the aforementioned definitions. We assume that all projection data for an angle $\alpha$ (so for all shifts $u$) is recorded instantaneously but that each angle has a different recording time $t$. Each complete scan has a total scanning time $\Delta t$ and we assume we scan consecutively $m$ times, so the total scan time is $m \Delta t$. The relation between the angle and the time we acquire data for this angle, is given by 
\begin{align*}
T: [0, m \pi[ \rightarrow [0, m \Delta t[: \alpha \mapsto \dfrac{\alpha}{\pi} \Delta t.
\end{align*}
This means the Radon transform \eqref{eq:radon} is generalised as
\begin{equation}
  \radon_{T(\alpha)} f(\alpha,u) = \int_{L(\alpha,u)} f(x,y; T(\alpha))\diff x \diff y.
\end{equation}
For notational purposes, we define $$t_i := T(i \pi), i = 0, \ldots, m,$$
as the start of the $i + 1 \,$-th scan and/or the end of the $i \,$-th scan. We make the sinograms continually so we define the sinogram at time $t \in [ \dfrac{\Delta t}{2}, m \Delta t - \dfrac{\Delta t}{2}[$ as $$\radon_t^{\delta t} f: [ \dfrac{t \pi}{\Delta t} -\pi/2, \dfrac{t \pi}{\Delta t} + \pi/2[ \times \R  \rightarrow \R: (\alpha, u) \mapsto \radon_{T(\alpha)+ \delta t} f( \alpha, u) $$
where the constant $\delta t$ is the time shift. Mostly $\delta t$ is equal to zero, then we just write $\radon_t f$ instead of $\radon_t^0 f$. 

For the reconstruction of the data, we make use of \emph{the filter backprojection theorem} see for example \cite{math_ct}. The \emph{backprojection} of a sinogram $\radon f$ of a time-independent function $f \in L^2(\R^2)$ is defined by 
\begin{align*}
f^\text{rec} = \B \radon g(x,y) & := \int^{\pi}_{0} \int^{\infty}_{-\infty} |\rho| \F_1\big( \radon f(\alpha,u) \big) \big( \rho \big) \exp \left( 2\pi i \rho  \left( x \cos(\alpha) + y \sin(\alpha) \right)\right) \diff \rho \diff \alpha \\
& = \int^{\pi}_{0} \F_1^{-1} \Big( |\rho| \F_1\big( \radon f(\alpha, u)\big) \big( \rho \big) \Big)\Big(x \cos(\alpha) + y \sin(\alpha) \Big)  \diff \alpha
\end{align*}
where $\F_1$ is the 1-dimensional Fourier transform.
Note that the domain of integration for $\alpha$ just needs to be an interval of length $\pi$ and that the interval itself depends on the angles for which we have projection data.

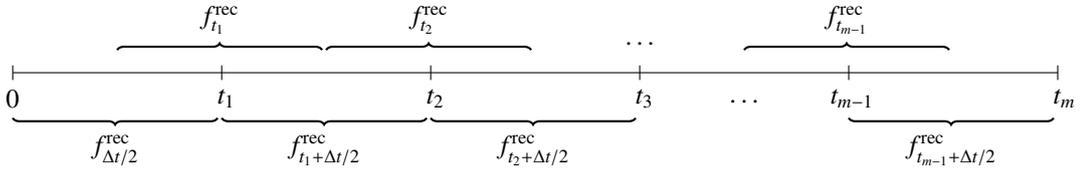
\begin{figure}[H]
\begin{tikzpicture}[scale = 0.55]
	\draw (0,0) -- (25,0);
	\draw (4*5,5pt) -- (4*5,-5pt) node[anchor = north east]  {$t\indi{m-1}$};
	\draw (5*5,5pt) -- (5*5,-5pt) node[anchor = north] {  $t\indi{m}$};
	\draw (0,5pt) -- (0,-5pt) node[anchor = north] {  $0$};
	\foreach \x in {1,2,3}
		\draw (5*\x,5pt ) -- (5*\x ,-5pt ) node[anchor = north] {$t\indi{\x}$};
	\node at (3*5+2.5,-0.7cm) {$\hdots$};
	\node at (3*5,0.7cm) {$\hdots$};
	\draw [thick,decoration={brace,mirror, raise=0.6cm},decorate] (5*4,0) -- (5*5-0.1,0) 
	node [midway,anchor = north, yshift=-0.7cm] {$f^{\text{rec}}_{t_{m-1} + \Delta t/2}$};
		\draw [thick,decoration={brace,mirror, raise=0.6cm},decorate] (0,0) -- (5-0.1,0) 
	node [midway,anchor = north, yshift=-0.7cm] {$f^{\text{rec}}_{\Delta t/2}$};
	\foreach \x in {1,2}
		\draw [thick,decoration={brace,mirror,raise=0.6cm},decorate] (\x*5,0) -- (\x*5+5-0.1,0) 
	node [pos=0.5,anchor=north,yshift=-0.7cm] {$f^{\text{rec}}_{t_{\x}+ \Delta t/2}$};
	
		\draw [thick,decoration={brace,raise=0.3cm},decorate] (5*4-2.5,0) -- (5*4+5-2.6,0) 
	node [pos=0.5,yshift=0.7cm] {$f^{\text{rec}}_{t_{m-1}}$};
	\foreach \x in {1,2}
		\draw [thick,decoration={brace,raise=0.3cm},decorate] (\x*5 - 2.5,0) -- (\x*5+5-2.6,0) 
	node [pos=0.5,yshift=0.7cm] {$f^{\text{rec}}_{t_{\x}}$};
\end{tikzpicture}
	\caption{Schematic representation of the notation for some time points. The range of the brace represents the time period the data is measured for the reconstruction of $f^{\text{rec}}_t$. }\label{fig:sinogram}
\end{figure}

For a time-dependent object $f$, we define the reconstructed object $f^{\text{rec}}_t$ at time $t$ as 
\begin{align}
f_t^{\text{rec}}(x,y) & = \B \radon_t f(x,y) = \int_{\frac{t \pi}{\Delta t} - \frac{\pi}{2}}^{\frac{t \pi}{\Delta t} + \frac{\pi}{2}} \F_1^{-1} \Big( |\rho| \F_1\big( \radon_t f(\alpha, u)\big) \big( \rho \big) \Big)\Big(x \cos(\alpha) + y \sin(\alpha) \Big) \diff \rho \diff \alpha. \label{eqn:f_rec}
\end{align} 
This is in fact the filtered backprojection theorem applied on the time-dependent sinogram $\radon_t f$.  Note that the operator $\radon_t$ has a non-trivial null space $\aleph \radon_t$ which means that  $\B \radon_t (.)$ has a non-empty null space too. This means that in theory, it is possible that a reconstruction is not unique. In practice, the motion is small enough to assume that this does not causes any problem. An illustration of these definitions can be found in figure \ref{fig:sinogram}. 

\subsection{Discretization of Radon transformations} \label{sec:discr_radon}
A discrete sinogram for stationary objects \eqref{eq:sinogram} can be constructed by dividing the volume of the unknown object in pixels.  With each pixel center $(x_i,y_j)$, we associate an unknown value $f_{ij}$ and we assume that every pixel is a square. We take in each scan projection data for 180 angles uniformly distributed over $[0,\pi[$. The line integral of a single Radon transform (see \eqref{eq:radon}) can now be approximated by a weighted sum over these pixels. These weights are the length of the line segment of the projection direction through the pixel (as it needs to approximate an integral) and are entered in a large sparse matrix to form a linear algebra problem $\mathbf{A} \mathbf{x} = \mathbf{b}$. In this matrix $\mathbf{A}$ each row contains the weights of one X-ray and the vectors $\mathbf{x}$ and $\mathbf{b}$ are the unknown pixel values and the discrete measurements respectively. The solution of this linear system is in fact the discretized version of the reconstructed image $f^\text{rec}$. In this context, the matrix $\mathbf{A}$ is called the \emph{projection operator}. There are multiple ways to discretize Radon transformations and to construct this matrix $\mathbf{A}$. We use in this paper the projection scheme described by Joseph in \cite{joseph1982improved}. In practice we generate this matrix using the Astra Toolbox, see \cite{palenstijn2011performance}. In section \ref{sec:cor_images} we explain how we can adapt this matrix $\mathbf{A}$ for the motion.

\subsection{Evolution} \label{sec:evolution}
We assume that each pixel moves in a 2D flow field
$v(x,y)$ that is yet unknown and that needs to be extracted from
the measurements. 
In this paper we assume that the evolution of the object $f$ itself can be modelled by the \emph{optical flow PDE}
\begin{equation}\label{eq:evolution}
   \dfrac{\partial f(x,y;t)}{\partial t}  + v(x,y) \cdot \nabla f(x,y;t) = 0    \quad \text{for all} \quad x,y \in \Omega, 
\end{equation}
where the flow field $v(x,y)$ represents the motion of $f$ on a time interval of size $\Delta t$. The operator $\nabla$ represents the gradient. If we model the flow of $f$ by this equation, we accept that the \emph{brightness constancy constraint} 
\begin{equation} \forall x,y,t \quad \exists v_x, v_y: f(x,y,t) = f(x + \Delta t v_x, y+\Delta t v_y, t + \Delta t) \label{eq:const_constr} \end{equation} holds for a given time period $\Delta t$. In our case this $\Delta t$ is the scan time of one complete scan.
For a specific $x$ and $y$, \eqref{eq:evolution} now follows from the Taylor expansion 
$$f(x+ \Delta t v_x, y + \Delta t v_y, t + \Delta t) \approx f(x,y,t) + \dfrac{\partial f(x,y;t)}{\partial t} + \Delta t v_x \dfrac{\partial f(x,y;t)}{\partial x} + \Delta t v_y \dfrac{\partial f(x,y;t)}{\partial y}$$
and \eqref{eq:const_constr}, if we define $v(x,y) = ( v_x, v_y)$.

This flow field is unknown and needs to be extracted from the measurements. We assume that this flow field is, for short time horizons, independent of the time $t$. 
This assumption can be made because a CT-scan is applied fast and changes in the flow happens smoothly at a slower time scale (so changes are small in short time periods). 

We describe the method to estimate the motion for objects at different time points, in section~\ref{sec:mot_est} we describe how we use this method if only scan results are available.

To estimate the flow field $v(x,y)$, we use the Horn-Schunck method \cite{HS}. 
This method is a global method (one equation for the entire domain) that prefers flow fields that are smooth.

Here we minimize the following expression with respect to the flow field $v(x,y)$, 
\begin{equation} E(v(x,y)) = \int \int_{\Omega} \left( \Delta t \dfrac{\partial f(x,y;t)}{\partial t}  + v(x,y) \cdot \nabla f(x,y;t) \right)^2 + \lambda \left \| \nabla  v(x,y) \right \|_2^2 \diff x \diff y \label{optflow_HS} \end{equation}
with $\lambda > 0$ a regularisation parameter.
The first part of the integral is the optical flow expression (see~\eqref{eq:evolution}) and the other part is a regularisation term. This means we are searching for a solution $v(x,y)$ such that it satisfies the optical flow equation and that the solution itself is smooth. The extent to which the solution needs to be smooth, is represented by the parameter $\lambda$.
Denote with $v_x$ and $v_y$ the restriction of $v(x,y)$ to the first respectively second variable.
If we minimize \eqref{optflow_HS} analytically by the Euler-Lagrange equations, we get 
\begin{equation}
\left\{ \begin{aligned}
\frac{\partial f}{\partial x}\left(\frac{\partial f}{\partial x} v_x +\frac{\partial f}{\partial y} v_y+\Delta t\frac{\partial f}{\partial t}\right) - \lambda \left( \frac{\partial^2 v_x}{\partial x^2} + \frac{\partial^2 v_x}{\partial y^2} \right)= 0 \\
\frac{\partial f}{\partial y}\left(\frac{\partial f}{\partial x} v_x +\frac{\partial f}{\partial y} v_y+ \Delta t \frac{\partial f}{\partial t}\right) - \lambda \left(\frac{\partial^2 v_y}{\partial x^2} + \frac{\partial^2 v_y}{\partial y^2}\right) = 0. \\
\end{aligned}\right. \label{eqn:solution_HS}
\end{equation}
A proof for \eqref{eqn:solution_HS} can be found in \cite{HS_conv}.
We want to discretize \eqref{eqn:solution_HS} in order to make it useful for doing calculations. We use second order central formulas for all derivatives and we set Neumann boundary conditions on our domain. This means we need three images to make an estimation of the applied motion. 
We denote by $\mathbf{D}_i$ and $\mathbf{D}^2_i$ the matrix we use to estimate the first respectively the second derivative in the $i$th direction. Further we define $\mathbf{z}_1 \odot \mathbf{z}_2$ as the pointwise multiplication and  $\mathbf{z}^2_\odot$ as the pointwise multiplication with itself. Similar to the discretisation of an object, we associate with the centre the horizontal and vertical component $\mathbf{v}_x(x_i,y_j)$ and $\mathbf{v}_y(x_i,y_j)$ of the discretised flow field $\mathbf{v}(x_i, y_j)$. If we denote with $\mathbf{D}_t \mathbf{f}$ the approximated time derivative with time step $\Delta t$, it is verifiable that the discretisation of \eqref{eqn:solution_HS} leads to a system  
\begin{equation} \mathbf{A}_{HS}  \begin{bmatrix}
\text{vec}( \mathbf{v}_x)\\ 
\text{vec}( \mathbf{v}_y)
\end{bmatrix} = \mathbf{b}_{HS} \label{eqn::HS} \end{equation}
with 
$$\mathbf{A}_{HS} = \begin{bmatrix}
\text{diag}\left(\left(\mathbf{D}_x \mathbf{f}_t \right)_\odot^2\right) -\lambda \left( \mathbf{D}_x^2 + \mathbf{D}_y^2 \right)  & \text{diag}\left(\mathbf{D}_x \mathbf{f}_t \odot \mathbf{D}_y \mathbf{f}_t\right) \\
\text{diag}\left(\mathbf{D}_x \mathbf{f}_t \odot \mathbf{D}_y \mathbf{f}_t \right) & \text{diag}\left(\left(\mathbf{D}_y \mathbf{f}_t \right)_\odot^2\right) -\lambda \left( \mathbf{D}_x^2 + \mathbf{D}_y^2 \right) 
\end{bmatrix} \in \R^{2 n^2  \times 2 n^2} \quad \text{and}$$ 

$$\mathbf{b}_{HS} = \begin{bmatrix}
-\Delta t \mathbf{D}_x \mathbf{f}_t \odot \mathbf{D}_t \mathbf{f}\\ 
-\Delta t \mathbf{D}_y \mathbf{f}_t \odot \mathbf{D}_t \mathbf{f}
\end{bmatrix} \in \R^{2 n^2  \times 1}.$$ 

%
We denote with $\mathbf{v}$ the exact motion and with $\hat{\mathbf{v}}$ the estimated motion.
In this work, we set the parameter $\lambda$ equal to 1. This choice gives us in practice good reconstructions. In future work this can be set automatically. 
  
\section{Motion estimation from scan results} \label{sec:mot_est}
\subsection{Theoretical derivation}
In section \ref{sec:evolution} we have seen how we can estimate the motion in three successive images. In practice, we do not have images but only scan results, so we need to estimate the motion starting from the reconstructed images $f_t^\text{rec}$ (see \eqref{eqn:f_rec}). Before we can prove the optical flow equation for the reconstructions (see theorem  \ref{the:rec_optic_flow}) we first need some lemmas.

\begin{lem} \label{lem:dRdrho}
For $f \in L^2(\R^2)$, it holds, for all angles $\alpha$ and $u$, that
$$ \dfrac{\partial \radon f}{\partial u}(\alpha, u) = \cos(\alpha) \radon \frac{\partial f}{\partial x}(\alpha, u)  + \sin(\alpha) \radon \frac{\partial f}{\partial y}(\alpha, u).$$
\begin{proof}
We make use of the following transformation and its inverse in the next calculations.  
\begin{equation} \begin{bmatrix}
u \\ 
w
\end{bmatrix} = \begin{bmatrix}
\cos(\alpha) & \sin(\alpha) \\
-\sin(\alpha) & \cos(\alpha)
 \end{bmatrix}
\begin{bmatrix}
x \\ 
y
\end{bmatrix} \quad \text{ and } \quad \begin{bmatrix}
x \\ 
y
\end{bmatrix} = \begin{bmatrix}
\cos(\alpha) & -\sin(\alpha) \\
\sin(\alpha) & \cos(\alpha)
 \end{bmatrix}
\begin{bmatrix}
u \\ 
w
\end{bmatrix} \label{anrec_eqn7}. \end{equation}
We obtain that
\begin{align*}
\dfrac{\partial \radon f}{\partial u}(\alpha, u) &= \dfrac{\partial }{\partial u} \int_{L(\alpha, u)} f(x,y) \diff x \diff y \\
& = \dfrac{\partial }{\partial u} \int_{-\infty}^\infty f\big( \cos(\alpha) u - \sin(\alpha)w, \sin(\alpha) u + \cos(\alpha) w\big) \diff w \\
& = \int_{-\infty}^\infty \dfrac{\partial }{\partial u} f\big( \cos(\alpha) u - \sin(\alpha)w, \sin(\alpha) u + \cos(\alpha) w\big) \diff w \\
& = \int_{-\infty}^\infty \cos(\alpha) \dfrac{\partial f}{\partial x}\big( \cos(\alpha)u - \sin(\alpha)w, \sin(\alpha)u + \cos(\alpha)w \big) + \\ 
& \qquad  \qquad \sin(\alpha) \dfrac{\partial f}{\partial y}\big( \cos(\alpha)u- \sin(\alpha)w, \sin(\alpha)u + \cos(\alpha)w \big)  \diff w \\
& =  \cos(\alpha) \int_{L(\alpha, u)}  \dfrac{\partial f}{\partial x} \diff x \diff y + \sin(\alpha) \int_{L(\alpha, u)}  \dfrac{\partial f}{\partial y}  \diff x \diff y.
\end{align*}
\end{proof}
\end{lem}

We can use this result in the next lemma which gives us a link between the derivatives of the reconstructions $f_t^\text{rec}$ and the derivatives of the object $f_t$.

\begin{lem} \label{lem:dfrecdx}
For the reconstruction of the object $f$ at time $t \in [\Delta t/2, m \Delta t - \Delta t/2]$, it holds that
$$\frac{\partial f^{\text{rec}}_t}{\partial x} = \B \radon_t \dfrac{\partial f}{\partial x} + \B \sin(\alpha) \left( \cos(\alpha) \radon_t \dfrac{\partial f}{\partial y} - \sin(\alpha) \radon_t \dfrac{\partial f}{\partial x} \right) $$ and
$$\frac{\partial f^{\text{rec}}_t}{\partial y} = \B \radon_t \dfrac{\partial f}{\partial y} - \B \cos(\alpha) \left( \cos(\alpha) \radon_t \dfrac{\partial f}{\partial y} - \sin(\alpha) \radon_t \dfrac{\partial f}{\partial x} \right).$$
\begin{proof}
For every $(x,y) \in \Omega$ holds that
\begin{align}
 \frac{\partial f^{\text{rec}}_t}{\partial x}(x,y) & = \int_{\frac{t \pi}{\Delta t} - \frac{\pi}{2}}^{\frac{t \pi}{\Delta t} + \frac{\pi}{2}} \int_{-\infty}^{\infty} |\rho|\F_1 \big( \radon_t f(\alpha, u)   \big)\big(\rho \big) \frac{\partial }{\partial x} \exp \big( 2 \pi i \rho \left(x \cos(\alpha) + y \sin(\alpha) \right)\big) \diff \rho \diff \alpha \nonumber \\
& = \int_{\frac{t \pi}{\Delta t} - \frac{\pi}{2}}^{\frac{t \pi}{\Delta t} + \frac{\pi}{2}} \int_{-\infty}^{\infty} \cos(\alpha) |\rho|  2 \pi i \rho \F_1 \big( \radon_t f(\alpha, u) \big)\big(\rho \big) \exp \left( 2 \pi i \rho \big(x \cos(\alpha) + y \sin(\alpha) \right)\big) \diff \rho \diff \alpha \nonumber \\
& = \int_{\frac{t \pi}{\Delta t} - \frac{\pi}{2}}^{\frac{t \pi}{\Delta t} + \frac{\pi}{2}} \int_{-\infty}^{\infty} \cos(\alpha) |\rho| \F_1 \Big( \dfrac{\partial \radon_t f(\alpha,u)}{\partial u}\Big)\big(\rho \big) \exp \left( 2 \pi i \rho \big(x \cos(\alpha) + y \sin(\alpha) \right)\big) \diff \rho \diff \alpha \nonumber \\
& =  \B \bigg( \cos(\alpha) \dfrac{ \partial \radon_t f(\alpha,u)}{\partial u} \bigg)(x,y) \nonumber \\
& =  \B  \bigg( \radon_t \frac{\partial f}{\partial x}(\alpha, u) \cos^2(\alpha) \bigg)(x,y) + \B \bigg( \radon_t \frac{\partial f}{\partial y}(\alpha, u) \sin(\alpha) \cos(\alpha)\bigg)(x,y) \label{eqn:lem31} \\
& = \B  \bigg( \radon_t \frac{\partial f}{\partial x}(\alpha, u)\bigg)(x,y) - \B \bigg( \sin^2(\alpha) \radon \frac{\partial f}{\partial x}(\alpha, u)\bigg)(x,y) + \nonumber \\
& \quad \quad \B \bigg(\sin(\alpha) \cos(\alpha) \radon_t \frac{\partial f}{\partial y}(\alpha, u) \bigg)(x,y) \nonumber
 \end{align} 
 where we have used previous lemma in \eqref{eqn:lem31}.
Similarly we can prove that 
\begin{align*}
\frac{\partial f^{\text{rec}}_t}{\partial y} & =  \B  \sin(\alpha)\dfrac{\partial \radon_t f(\alpha,u)} {\partial u} \\
& = \B \radon_t \dfrac{\partial f}{\partial y} - \B \cos(\alpha) \left( \cos(\alpha) \radon_t \dfrac{\partial f}{\partial y} - \sin(\alpha) \radon_t \dfrac{\partial f}{\partial x} \right).
\end{align*}
\end{proof}
\end{lem}

In our main theorem \ref{the:rec_optic_flow} we encounter the term $ \cos(\alpha) \radon_t \dfrac{\partial f}{\partial y}(\alpha,u)  -  \sin(\alpha)  \radon_t\dfrac{\partial f}{\partial x}(\alpha,u)$. It is a consequence of following lemma that this term is equal to zero.

\begin{lem} \label{lem:extra_term}
For every angle $\alpha$ it holds, for all $u$, that
$$ \cos(\alpha) \radon \dfrac{\partial f}{\partial y}(\alpha,u)  -  \sin(\alpha) \radon  \dfrac{\partial f}{\partial x} (\alpha, u) = 0.$$
\begin{proof}
We use the same transformation \eqref{anrec_eqn7} as in lemma \ref{lem:dRdrho}.
For a particular angle $\alpha$ and $u \in \R$, we obtain
\begin{align}
& \cos(\alpha) \radon \dfrac{\partial f}{\partial y}(\alpha,u)  -  \sin(\alpha) \radon  \dfrac{\partial f}{\partial x} (\alpha, u) \nonumber \\
= & \int_{L(\alpha,u)} \dfrac{\partial f}{\partial y} \cos(\alpha) - \dfrac{\partial f}{\partial x} \sin(\alpha) \diff x \diff y \nonumber \\
= &  \int_{-\infty}^{\infty} \dfrac{\partial f}{\partial y} \big( \cos(\alpha) u - \sin(\alpha) w, \sin(\alpha) u + \cos(\alpha) w \big) \cos(\alpha) - \nonumber \\ 
&  \quad \dfrac{\partial f}{\partial x}\big(\cos(\alpha) u - \sin(\alpha) w, \sin(\alpha) u + \cos(\alpha)w \big) \sin(\alpha) \diff w \nonumber \\
= & \int_{-\infty}^{\infty} \dfrac{\partial f}{\partial w}\big(\cos(\alpha) u - \sin(\alpha) w, \sin(\alpha) u + \cos(\alpha)w\big) \diff w  \nonumber \\
= & [f(\cos(\alpha) u - \sin(\alpha) w, \sin(\alpha) u + \cos(\alpha) w )]^{\infty}_{-\infty} \nonumber \\
= & 0 \label{eqn:iszero}
\end{align}
where \eqref{eqn:iszero} follows from the fact that $f$ is zero outside a closed area $\Omega$.
\end{proof}
\end{lem}

The following lemma give us the necessary relation between the object and its reconstruction.

\begin{lem} \label{lem:verband_frec_f}
It holds for every $t$ and $z \in \mathbb{Z}$ that
$$ \B \radon_t^{z \Delta t} f = f_{t+z \Delta t}^{\text{rec}}.$$
\begin{proof}
For every $z \in \mathbb{Z}$ holds that
\begin{align*}
&  \B \radon_t^{z \Delta t} f \\
= & \int_{\frac{\pi t}{\Delta t} - \frac{\pi}{2}}^{\frac{\pi t}{\Delta t} + \frac{\pi}{2}} \int_{-\infty}^{\infty} |\rho| \F_1 \Big( \radon_t^{z \Delta t} f(\alpha,u) \Big) \Big( \rho \Big)  \exp \big( 2 \pi i \rho \left( x \cos(\alpha) + y \sin(\alpha) \right) \big)  \diff \rho \diff \alpha  \\
= & \int_{\frac{\pi t}{\Delta t} + z \pi - \frac{\pi}{2}}^{\frac{\pi t}{\Delta t} + z \pi + \frac{\pi}{2}} \int_{-\infty}^{\infty} |\rho| \F_1 \Big( \radon_t^{z \Delta t} f(\alpha - z \pi,u) \Big) \Big( \rho \Big) \exp \big( 2 \pi i \rho \left( x \cos(\alpha-z\pi) + y \sin(\alpha-z\pi) \right) \big) \diff \rho \diff \alpha  \\
= & \int_{\frac{\pi t}{\Delta t} + z\pi- \frac{\pi}{2}}^{\frac{\pi t}{\Delta t} + z \pi + \frac{\pi}{2}} \int_{-\infty}^{\infty} |\rho| \F_1 \Big( \radon_{t+z\Delta t} f(\alpha,(-1)^z u) \Big) \Big(\rho \Big) \exp \big( 2 \pi i \left( (-1)^z \rho \right) \left( x \cos(\alpha) + y \sin(\alpha) \right) \big)   \diff \rho \diff \alpha \\
= & \int_{\frac{\pi t}{\Delta t} + z \pi - \frac{\pi}{2}}^{\frac{\pi t}{\Delta t} + z \pi + \frac{\pi}{2}} \int_{-\infty}^{\infty} |(-1)^z \rho| \F_1 \Big( \radon_{t+z \Delta t} f(\alpha,u) \Big) \Big((-1)^z \rho \Big) \exp \big( 2 \pi i \left( (-1)^z \rho \right) \left( x \cos(\alpha) + y \sin(\alpha) \right) \big)   \diff \rho \diff \alpha \\
= & f_{t+z \Delta t}^{\text{rec}}
\end{align*}
\end{proof}
\end{lem}

We prove an alternative on the optical flow differential equation \eqref{eqn::HS}.  Instead of $\frac{\partial f}{\partial t}$ we take the second-order Taylor estimation $\dfrac{ f_{t+ \Delta t} - f_{t-\Delta t} }{2 \Delta t}$.
\begin{thm} (Optical flow equation for the reconstructions)\label{the:rec_optic_flow}

\begin{enumerate}
\item If for all $t \in [\Delta t, (m-1) \Delta t]$  
\begin{align}
v_x \frac{\partial f_t}{\partial x} + v_y \frac{\partial f_t}{\partial y} +  \dfrac{f_{t+\Delta t} - f_{t-\Delta t}}{2 } & = 0 \label{eqn:optflow_1}
\end{align}
then it holds for all $t \in [3 \frac{\Delta t}{2}, (m - \frac{3}{2})\Delta t]$ 
\begin{align*}
v_x \frac{\partial f^{\text{rec}}_t}{\partial x} + v_y \frac{\partial f^{\text{rec}}_t}{\partial y} + \dfrac{f^{\text{rec}}_{t+\Delta t} - f^{\text{rec}}_{t-\Delta t}}{2 } & = 0.
\end{align*}
\item If for all  $t \in [3 \frac{\Delta t}{2},(m - \frac{3}{2}) \Delta t]$ 
\begin{align*}
v_x \frac{\partial f^{\text{rec}}_t}{\partial x} + v_y \frac{\partial f^{\text{rec}}_t}{\partial y} + \dfrac{f^{\text{rec}}_{t+\Delta t} - f^{\text{rec}}_{t-\Delta t}}{2 } & =0 \nonumber
\end{align*}
then it holds for $t \in [\Delta t, (m-1) \Delta t]$ that
$$v_x \frac{\partial f_t}{\partial x} + v_y \frac{\partial f_t}{\partial y} +  \dfrac{f_{t+\Delta t} - f_{t-\Delta t}}{2 } = g $$
for a certain $g \in \aleph \B \radon_t$ (= null space of operator $\B \radon_t$) .
\end{enumerate}

\begin{proof}
\begin{enumerate}
\item 
If we apply the operator $\B \radon_t$ on \eqref{eqn:optflow_1} we obtain
\begin{align}
\B \radon_t \left( v_x \dfrac{\partial f_t}{\partial x} + v_y \dfrac{\partial f_t}{\partial y} +  \dfrac{f_{t+\Delta t} - f_{t-\Delta t}}{2 } \right) & = 0 \nonumber \\ 
& \Updownarrow \nonumber \\
v_x \B \radon_t \dfrac{\partial f_t}{\partial x} + v_y \B \radon_t \dfrac{\partial f_t}{\partial y} + \dfrac{f^\text{rec}_{t+\Delta t} - f^\text{rec}_{t-\Delta t}}{2 } & = 0 \label{eqn:verband_frec_f} \\
& \Updownarrow \nonumber \\
v_x \dfrac{\partial f^\text{rec}_t}{\partial x} + v_y \dfrac{\partial f^\text{rec}_t}{\partial y} + \dfrac{f^\text{rec}_{t+\Delta t} - f^\text{rec}_{t-\Delta t}}{2 } & = v_x \dfrac{\partial f^\text{rec}_t}{\partial x} - v_x \B \radon_t \dfrac{ \partial f_t}{\partial x} + v_y \dfrac{\partial f^\text{rec}_t}{\partial y} - v_y \B \radon_t \dfrac{ \partial f_t}{\partial y} \nonumber \\
& \Updownarrow \nonumber \\
v_x \dfrac{\partial f^\text{rec}_t}{\partial x} + v_y \dfrac{\partial f^\text{rec}_t}{\partial y} + \dfrac{f^\text{rec}_{t+\Delta t} - f^\text{rec}_{t-\Delta t}}{2 } &  = v_x \B \sin(\alpha) \left( \cos(\alpha) \radon_t \dfrac{\partial f_t}{\partial y} - \sin(\alpha) \radon_t  \dfrac{\partial f_t}{\partial x} \right)  -  \label{eqn:dfrecdx}\\ 
& \quad \quad v_y \B \cos(\alpha) \left( \cos(\alpha)  \radon_t \dfrac{\partial f}{\partial y} - \sin(\alpha)  \radon_t \dfrac{\partial f}{\partial x} \right) \nonumber \\
& \Updownarrow \nonumber \\
v_x \dfrac{\partial f^\text{rec}_t}{\partial x} + v_y \dfrac{\partial f^\text{rec}_t}{\partial y} + \dfrac{f^\text{rec}_{t+\Delta t} - f^\text{rec}_{t-\Delta t}}{2} & = 0 \label{eqn:extra_term}
\end{align}
where \eqref{eqn:verband_frec_f}, \eqref{eqn:dfrecdx} and \eqref{eqn:extra_term} follows from respectively lemma \ref{lem:verband_frec_f}, \ref{lem:dfrecdx} and \ref{lem:extra_term}.
\item 
This follows from
\begin{align}
& \B \radon_t \left( v_x \dfrac{\partial f}{\partial x} + v_y \dfrac{\partial f}{\partial y} \right) \nonumber \\
= & v_x \frac{\partial f_t^{\text{rec}}}{\partial x} + v_y \frac{\partial f_t^{\text{rec}}}{\partial y} \nonumber \\
= & \dfrac{ f_{t+ \Delta t}^{\text{rec}} - f_{t-\Delta}^{\text{rec}}}{2} \nonumber \\
= & \B \radon_t \left( \dfrac{ f_{t+\Delta t} - f_{t-\Delta t}}{2} \right)  \nonumber.
\end{align}
\end{enumerate}
\end{proof}
\end{thm}

It is possible to achieve a similar result by approximating the time derivative by for example first order approximation formulas as long as the time mesh width is $\Delta t$ (duration of one scan) but tests showed us that the reconstruction had a significant lower quality in that case.
The function $g \in \aleph \B \R_t$ in previous theorem is considered to be negligible in practice. 
\subsection{Implementation} \label{sec:implem}

Because the optical flow equation is still valid for the reconstructions (see previous theorem), we can use the method of Horn-Schunck (see section \ref{sec:evolution}) to estimate the flow from the reconstructions $f^{\text{rec}}_{t_i-\Delta t/2}, i = 1, \hdots, m$. This means we calculate a reconstruction per scan. In fact, we are not limited to the boundaries of a scan, we can use data which is coming partly from the $i\,$-th scan and partly from the $i+1 \,$-th scan. We can reconstruct for example new images $g_i, i = 1, \hdots, m-1$ for which the data for angles between $[\pi/2 , \pi[$ is coming from the $i+1 \,$-th scan and the data for angles between $[0 , \pi/2[$ is coming from the $i \,$-th scan.
However, tests showed us that the use of more images does not add significant value and furthermore it is computationally more expensive.
From  section~\ref{sec:evolution}, we can retrieve systems $$\mathbf{A}_{HS}^i \begin{bmatrix} \text{vec}( \mathbf{v}_x)\\ 
\text{vec}( \mathbf{v}_y)
\end{bmatrix} = \mathbf{b}_{HS}^i, \quad i = 2, 3,\hdots, m-1$$ between three successive reconstructions. 
Because we assume the motion is constant over time, we know the motion between successive reconstructions is equal. This means we can see the unknown motion $\mathbf{v}$ as the solution of the big system
\begin{equation} \begin{bmatrix}
\mathbf{A}_{HS}^2\\ 
\mathbf{A}_{HS}^3\\ 
\vdots\\ 
\mathbf{A}_{HS}^{m-1}
\end{bmatrix}                                                                                                                      \begin{bmatrix}
\text{vec}( \mathbf{v}_x)\\ 
\text{vec}( \mathbf{v}_y)
\end{bmatrix} = \begin{bmatrix}
\mathbf{b}_{HS}^2\\ 
\mathbf{b}_{HS}^3\\ 
\vdots\\ 
\mathbf{b}_{HS}^{m-1}
\end{bmatrix}. \label{ctr_eqn1} \end{equation}

By using a single system, we make use of all information available in the scans. It is known in the literature that the Horn-Schunck method can estimate well small motions, but that large displacements can not be retrieved. We can deal with this by applying a coarse-to-fine scheme where we calculate the motion on a lower resolution and use this as an initial solution when calculating the motion on a higher resolution, see for example \cite{anandan1989computational}. This results in following algorithm where the upper index represents the resolution of the used image.
We assume the reconstructed images are objects on a $2^p \times 2^p$ grid for a certain $p \in \N$. The variable $d$ determines the factor we reduce the resolution in the first estimation of the motion. For the examples in section \ref{sec:numer_results} we take always $d$ equal to $3$. This means we first estimate the motion on a $\frac{2^p}{2^3} \times \frac{2^p}{2^3}$ grid.

\begin{algorithm}[H]
	\SetKwInOut{Input}{Input}
	\SetKwInOut{Output}{Output}

	\Input{\begin{enumerate}
	\item $d = $ depth: defines the resolution of the image in the first iteration 
	\item $\mathbf{f}^{\text{rec}}_{t_j-\Delta t/2}$ = reconstruction of the $i$th scan on a $2^p \times 2^p$ pixel grid, $j = 1,\hdots, m$
	\end{enumerate}}
	\Output{$\mathbf{v} = \mathbf{v}^{2^p} $: calculated motion on a $2^p \times 2^p$ pixel grid}
	
	$\mathbf{v}_{\text{old}}^{2^p/2^d}$ = 0 \\
	\For{$i = d:-1:0$}{
		$\left( \mathbf{f}^{\text{rec}}_{t_j-\Delta t/2}\right)^{2^p/2^i}, j = 1, \hdots, m$ = reduce the resolution of the image $\mathbf{f}^{\text{rec}}_{t_j+\Delta t}$ with \\
		\qquad a factor $2^i$ \\ 
		$\mathbf{v}^{2^p/2^i}$= calculated motion using \eqref{ctr_eqn1}
		with initial solution $\mathbf{v}_{\text{old}}^{2^p/2^i}$ \\
		\If{$i > 0$}{
			$\mathbf{v}_{\text{old}}^{2^p/2^{i-1}}$ = Interpolate $\mathbf{v}^{2^p/2^i}$ to a motion on a $2^p/2^{i-1} \times 2^p/2^{i-1}$ grid
		}
	}
\caption{Coarse-to-fine algorithm to estimate the motion $\mathbf{v}$ in scan reconstructions $\mathbf{f}^{\text{rec}}_{t_j-\Delta t/2}$. For simplicity we only consider images on a $2^p \times 2^p$ grid. Uper indices on the motion $\mathbf{v}$ and reconstructed objects $ \mathbf{f}^{\text{rec}}_{t_j-\Delta t/2}$ represent the resolution. If the resolution $s$ of the image cannot be written as $2^{n_1}, n_1 \in \mathbb{N}$ then we can just adapt this to the nearest resolution that can be written as $2^{n_1}$. Practically, adapting the resolution is done via interpolation. } \label{algo:motion}
\end{algorithm}

\section{Correcting images for motion} \label{sec:cor_images}
In this section we present a method to correct the reconstruction of a CT-scan for motion when we know the deformation $v(x,y)$ (or an estimate of it) that is performed during one scan. For this we only make use of one scan. The strategy is to move the data recorded at different time steps to a single reference time point and execute the reconstruction there. In fact, we look at how the X-rays are deformed by the motion over time.  A consequence is that we also need to have an estimate of the motion on shorter time intervals. We choose to interpolate linearly since no other information is available about the motion on shorter time intervals. So the quality of the reconstruction depends on the extent the linear approximation corresponds with the real deformation.
In the next calculations we choose to take the middle of the scan time interval (= $\Delta t/2$) as reference time point, but alternative choices are also possible. It will be clear that the proposed method can easily be adapted such that it reconstructs at another time point. 
To move the data to the middle of scan, we do following calculations for all angles. For the projection under an angle $\alpha$ performed at time $T(\alpha)$ now holds, for all $u$, that
\begin{align}
 \radon f_{T(\alpha)} (\alpha,u) \nonumber & = \int_{L(\alpha,u)} f_{T(\alpha)}(x,y)\diff x \diff y \nonumber \\
  & \approx \int_{L(\alpha,u)} f_{\Delta t/2} \left(x - \dfrac{-\Delta t/2+T(\alpha)}{\Delta t}v_x(x,y),y- \dfrac{-\Delta t/2+T(\alpha)}{\Delta t}v_y(x,y)\right)\diff x \diff y. \label{eqn:deform_int}
\end{align}
So instead of integrating over the path $L(\alpha, u)$, we integrate over the path 
$$L_{\text{moved}}(\alpha,u) := \left \{  \left(x - \dfrac{-\Delta t/2+T(\alpha)}{\Delta t}v_x(x,y), y- \dfrac{-\Delta t/2+T(\alpha)}{\Delta t}v_y(x,y) \right) \mid (x, y) \in L(\alpha,u) \right \}.$$
and we obtain that the integral in \eqref{eqn:deform_int} is equal to
\begin{equation} \int_{L_{\text{moved}}(\alpha,u)} f_{\Delta t/2} \left(x,y\right)\diff x \diff y. \label{eqn:new_int}\end{equation}
The objective is now to approximate the integral in \eqref{eqn:new_int} by a matrix-vector product $\mathbf{A}_{\text{moved}} \mathbf{f}$ where $\mathbf{f}$ is the discretised and vectorised image at time $\Delta t/2$. 
Just as the weights on each line in the projection operator $\mathbf{A}$ (see section \ref{sec:discr_radon}) are determined by the length of the segment of the projection line $L(\alpha,u)$ through the pixels, the weights on each row $\mathbf{A}_{\text{moved}}$ are the length of the adapted projection line $L_{\text{moved}}(\alpha, u)$ through the pixels. In practice we determine the values in the matrix $\mathbf{A}_{\text{moved}}$ with following algorithm for a projection angle $\alpha$ and $u \in \R$. Let $\Omega_\delta$ be the domain $\Omega$ with an extra border such that we are ensured that all adapted projection lines start and finish outside $\Omega$. 

\begin{algorithm}[H]
\SetKwInOut{Input}{Input}
\SetKwInOut{Output}{Output}
\Input{\begin{enumerate}
\item $\Omega_\delta$: Domain $\Omega$ with an extra border 
\item $L(\alpha, u)$: projection line for angle $\alpha$ and shift $u$
\item $T(\alpha$): Determines the time the projection under angle $\alpha$ is performed
\item $\Delta t$: Duration time of one scan
\item $\mathbf{v}$: Estimated or exact motion
\end{enumerate}}
\Output{$\mathbf{A}_{\text{moved}}$: projection matrix corrected for the motion $\mathbf{v}$}
\nonl We describe the method to determine a row in the matrix $\mathbf{A}_{\text{moved}}$ corresponding with a projection with angle $\alpha$ and shift $u$. This algorithm is repeated for all angles $\alpha$ and shifts $u$. \\

Find the intersection points $s$ of the projection lines $L(\alpha,u)$ with the pixels of the domain $\Omega_\delta$.

Calculate the mid points $m$ between every 2 successive intersection points.

As we have only values for the motion in the pixel centre, we estimate the motion $v(m)$ for the points $m$ from $\mathbf{v}$ using interpolation. Apply the motion $- \dfrac{-\Delta t/2+T(\alpha)}{\Delta t} v(m)$ on $m$ and call these points $m_{\text{moved}}$. 

Define the path $L_{\text{moved}}(\alpha,u)$ determined by the points $m_{\text{moved}}$ where we interpolate linearly between the points.

Find the intersection points $s_{\text{moved}}$ of the adapted projection line $L_{\text{moved}}(\alpha, u)$ with the pixels of the domain $\Omega$.

Determine the values in the matrix $\mathbf{A}_{\text{moved}}$ by calculating the distance between 2 successive points $s_{\text{moved}}$.
\caption{Method to correct CT-scan images for the motion. Note that we need to perform this algorithm on every projection line but it is possible to calculate this in parallel as each projection is independent. Furthermore the first two steps do not depend on the motion $\mathbf{v}$ so they can be calculated in advance. } \label{algo:correcting}
\end{algorithm}

After performing this algorithm we get a new linear system $$\mathbf{A}_{\text{moved}} \mathbf{x} = \mathbf{b}$$ with $\mathbf{x}$ the vectorized version of $f_{\Delta t/2}$. We use $f_{\Delta t/2}^{\text{corr}}$ to refer to the solution of this system using the LSQR algorithm. 

\begin{figure}[H]
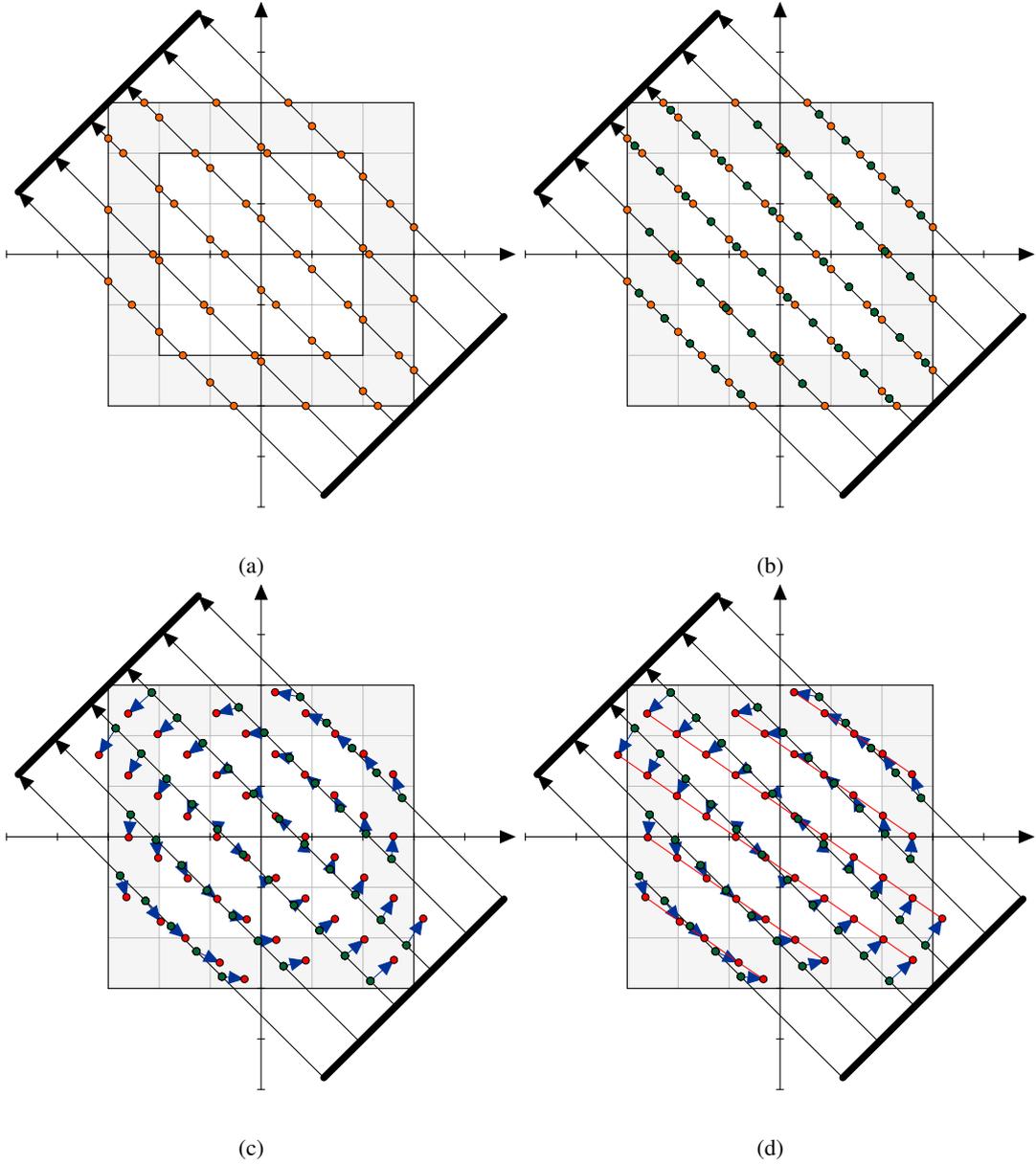
  
        \begin{subfigure}[b]{0.5\textwidth}
             \definecolor{aqaqaq}{rgb}{0.6274509803921569,0.6274509803921569,0.6274509803921569}
\definecolor{ffwwqq}{rgb}{1.,0.4,0.}
\definecolor{cqcqcq}{rgb}{0.7529411764705882,0.7529411764705882,0.7529411764705882}
\begin{tikzpicture}[line cap=round,line join=round,>=triangle 45,x=1.0cm,y=1.0cm,scale = 0.7]
\draw [color=cqcqcq, xstep=1.0cm,ystep=1.0cm] (-3.,-3.) grid (3.,3.);
\draw[->,color=black] (-5.,0.) -- (5.,0.);
\foreach \x in {-5.,-4.,-3.,-2.,-1.,1.,2.,3.,4.}
\draw[shift={(\x,0)},color=black] (0pt,2pt) -- (0pt,-2pt);
\draw[->,color=black] (0.,-5.) -- (0.,5.);
\foreach \y in {-5.,-4.,-3.,-2.,-1.,1.,2.,3.,4.}
\draw[shift={(0,\y)},color=black] (2pt,0pt) -- (-2pt,0pt);
\clip(-5.,-5.) rectangle (5.,5.);
\fill[color=aqaqaq,fill=aqaqaq,fill opacity=0.1] (-3.,3.) -- (-2.,2.) -- (-2.,-2.) -- (-3.,-3.) -- cycle;
\fill[color=aqaqaq,fill=aqaqaq,fill opacity=0.1] (-3.,3.) -- (3.,3.) -- (2.,2.) -- (-2.,2.) -- cycle;
\fill[color=aqaqaq,fill=aqaqaq,fill opacity=0.1] (3.,3.) -- (2.,2.) -- (2.,-2.) -- (3.,-3.) -- cycle;
\fill[color=aqaqaq,fill=aqaqaq,fill opacity=0.1] (-3.,-3.) -- (-2.,-2.) -- (2.,-2.) -- (3.,-3.) -- cycle;
\draw (2.,-2.)-- (2.,2.);
\draw (-2.,2.)-- (2.,2.);
\draw (-2.,-2.)-- (2.,-2.);
\draw (-2.,-2.)-- (-2.,2.);

\draw (-3.,3.)-- (-3.,-3.);
\draw (-3.,-3.)-- (3.,-3.);
\draw (3.,-3.)-- (3.,3.);
\draw (-3.,3.)-- (3.,3.);
\draw [line width=2.8pt] (-4.767766952966369,1.2322330470336311)-- (-1.2322330470336311,4.767766952966369);
\draw [line width=2.8pt] (1.2322330470336311,-4.767766952966369)-- (4.767766952966369,-1.2322330470336311);
\draw [->] (4.767766952966369,-1.2322330470336311) -- (-1.2322330470336311,4.767766952966369);
\draw [->] (1.2322330470336311,-4.767766952966369) -- (-4.767766952966369,1.2322330470336311);
\draw [->] (4.060660171779822,-1.9393398282201788) -- (-1.939339828220179,4.060660171779822);
\draw [->] (3.3492748675080954,-2.6507251324919046) -- (-2.6464466094067265,3.3535533905932735);
\draw [->] (2.646446609406726,-3.353553390593274) -- (-3.353553390593274,2.646446609406726);
\draw [->] (1.9393398282201786,-4.060660171779822) -- (-4.060660171779821,1.9393398282201781);
\begin{scriptsize}
\draw [fill=ffwwqq] (0.5355339059327378,3.) circle (2.0pt);
\draw [fill=ffwwqq] (1.5735923630161022,1.9720410744812524) circle (2.0pt);
\draw [fill=ffwwqq] (-0.8786796564403567,3.) circle (2.0pt);
\draw [fill=ffwwqq] (0.1213203435596431,2.) circle (2.0pt);
\draw [fill=ffwwqq] (2.5355339059327378,1.) circle (2.0pt);
\draw [fill=ffwwqq] (1.121320343559643,1.) circle (2.0pt);
\draw [fill=ffwwqq] (2.1213203435596433,0.) circle (2.0pt);
\draw [fill=ffwwqq] (-2.292893218813452,3.) circle (2.0pt);
\draw [fill=ffwwqq] (-1.292893218813452,2.) circle (2.0pt);
\draw [fill=ffwwqq] (-0.2928932188134521,1.) circle (2.0pt);
\draw [fill=ffwwqq] (0.7071067811865479,0.) circle (2.0pt);
\draw [fill=ffwwqq] (1.707106781186548,-1.) circle (2.0pt);
\draw [fill=ffwwqq] (2.707106781186548,-2.) circle (2.0pt);
\draw [fill=ffwwqq] (-2.707106781186548,2.) circle (2.0pt);
\draw [fill=ffwwqq] (-1.707106781186548,1.) circle (2.0pt);
\draw [fill=ffwwqq] (-0.7071067811865479,0.) circle (2.0pt);
\draw [fill=ffwwqq] (0.2928932188134521,-1.) circle (2.0pt);
\draw [fill=ffwwqq] (1.292893218813452,-2.) circle (2.0pt);
\draw [fill=ffwwqq] (2.292893218813452,-3.) circle (2.0pt);
\draw [fill=ffwwqq] (-2.1213203435596437,0.) circle (2.0pt);
\draw [fill=ffwwqq] (-1.1213203435596435,-1.) circle (2.0pt);
\draw [fill=ffwwqq] (-0.12132034355964352,-2.) circle (2.0pt);
\draw [fill=ffwwqq] (0.8786796564403565,-3.) circle (2.0pt);
\draw [fill=ffwwqq] (-2.5355339059327378,-1.) circle (2.0pt);
\draw [fill=ffwwqq] (-1.5355339059327378,-2.) circle (2.0pt);
\draw [fill=ffwwqq] (-0.5355339059327378,-3.) circle (2.0pt);
\draw [fill=ffwwqq] (1.,2.5355339059327378) circle (2.0pt);
\draw [fill=ffwwqq] (2.,1.5355339059327378) circle (2.0pt);
\draw [fill=ffwwqq] (3.,0.5355339059327378) circle (2.0pt);
\draw [fill=ffwwqq] (0.,2.1213203435596433) circle (2.0pt);
\draw [fill=ffwwqq] (3.,-0.8786796564403567) circle (2.0pt);
\draw [fill=ffwwqq] (1.,1.121320343559643) circle (2.0pt);
\draw [fill=ffwwqq] (2.,0.1213203435596431) circle (2.0pt);
\draw [fill=ffwwqq] (-2.,2.707106781186548) circle (2.0pt);
\draw [fill=ffwwqq] (-1.,1.707106781186548) circle (2.0pt);
\draw [fill=ffwwqq] (0.,0.7071067811865479) circle (2.0pt);
\draw [fill=ffwwqq] (1.,-0.2928932188134521) circle (2.0pt);
\draw [fill=ffwwqq] (-3.,2.292893218813452) circle (2.0pt);
\draw [fill=ffwwqq] (-2.,1.292893218813452) circle (2.0pt);
\draw [fill=ffwwqq] (-1.,0.2928932188134521) circle (2.0pt);
\draw [fill=ffwwqq] (0.,-0.7071067811865479) circle (2.0pt);
\draw [fill=ffwwqq] (1.,-1.707106781186548) circle (2.0pt);
\draw [fill=ffwwqq] (2.,-2.707106781186548) circle (2.0pt);
\draw [fill=ffwwqq] (-3.,0.8786796564403565) circle (2.0pt);
\draw [fill=ffwwqq] (-2.,-0.12132034355964352) circle (2.0pt);
\draw [fill=ffwwqq] (-1.,-1.1213203435596435) circle (2.0pt);
\draw [fill=ffwwqq] (0.,-2.1213203435596437) circle (2.0pt);
\draw [fill=ffwwqq] (-3.,-0.5355339059327378) circle (2.0pt);
\draw [fill=ffwwqq] (-2.,-1.5355339059327378) circle (2.0pt);
\draw [fill=ffwwqq] (-1.,-2.5355339059327378) circle (2.0pt);
\draw [fill=ffwwqq] (2.,-1.292893218813452) circle (2.0pt);
\draw [fill=ffwwqq] (3.,-2.292893218813452) circle (2.0pt);
\end{scriptsize}
\end{tikzpicture}
             \caption{}
        \end{subfigure} 
 \hspace{0.25cm}
        \begin{subfigure}[b]{0.5\textwidth}
             \input{geogebra/ctr_2_2}     
             \caption{}       
        \end{subfigure} \\
		\begin{subfigure}[b]{0.5\textwidth}
			\input{geogebra/ctr_2_3} 
			\caption{}                 
        \end{subfigure} \hspace{0.25cm}
        \begin{subfigure}[b]{0.5\textwidth}
            \input{geogebra/ctr_2_4}  
            \caption{}              
        \end{subfigure} 
\caption{An illustration of the previous algorithm where the grey zone is the border and each square corresponds with a pixel. We use a clockwise rotation around the center point as motion. For this particular example, the adapted projection lines $L_{\text{moved}}(\alpha, u)$ are straight lines but this is usually not the case. (a) step 1: The orange balls are the intersection points $s$ of some projection lines $L(\alpha,u)$ with the pixel edges (b) step 2: The green balls are the mid points $m$ of 2 successive intersection points $s$ (c) step 3: the motion $- \dfrac{-\Delta t/2+T(\alpha)}{\Delta t} v(m)$ (the blue arrows) applied on the mid points $m$. The obtained points (the red balls) are called $m_{\text{moved}}$ (d) step 4: Define the path $L_{\text{moved}}(\alpha,u)$ determined by the points $m_{\text{moved}}$ where we interpolate linearly between the points. }\label{ctr_fig}
\end{figure}

The accuracy of the algorithm can be further improved by taking more intermediate points between successive intersection points. When doing this, take in mind that the computational time is much higher.

\section{Numerical results} \label{sec:numer_results}
\begin{figure}[H]
  \begin{subfigure}[b]{0.33\textwidth}
        \includegraphics[scale=0.366, clip]{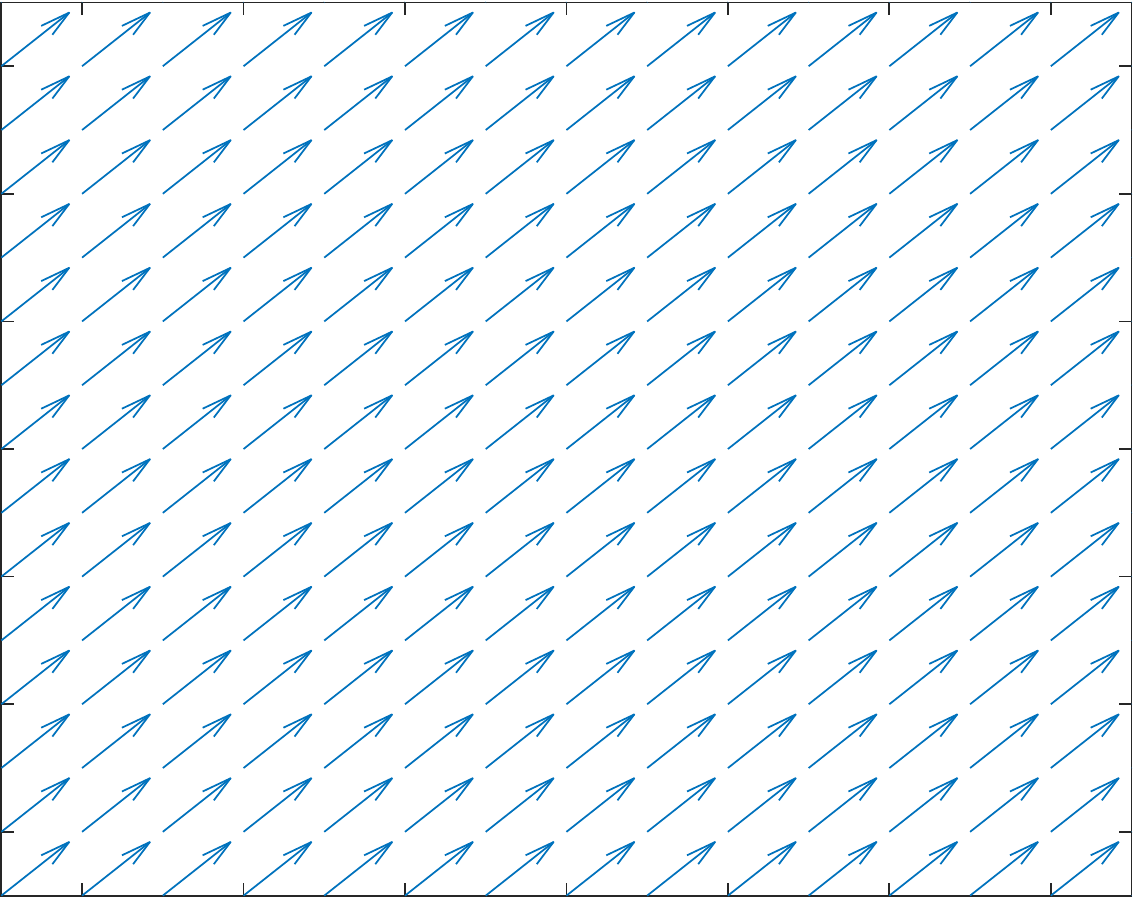}    
        \caption{}
  \end{subfigure} 
  \begin{subfigure}[b]{0.33\textwidth}
        \includegraphics[scale=0.366, clip]{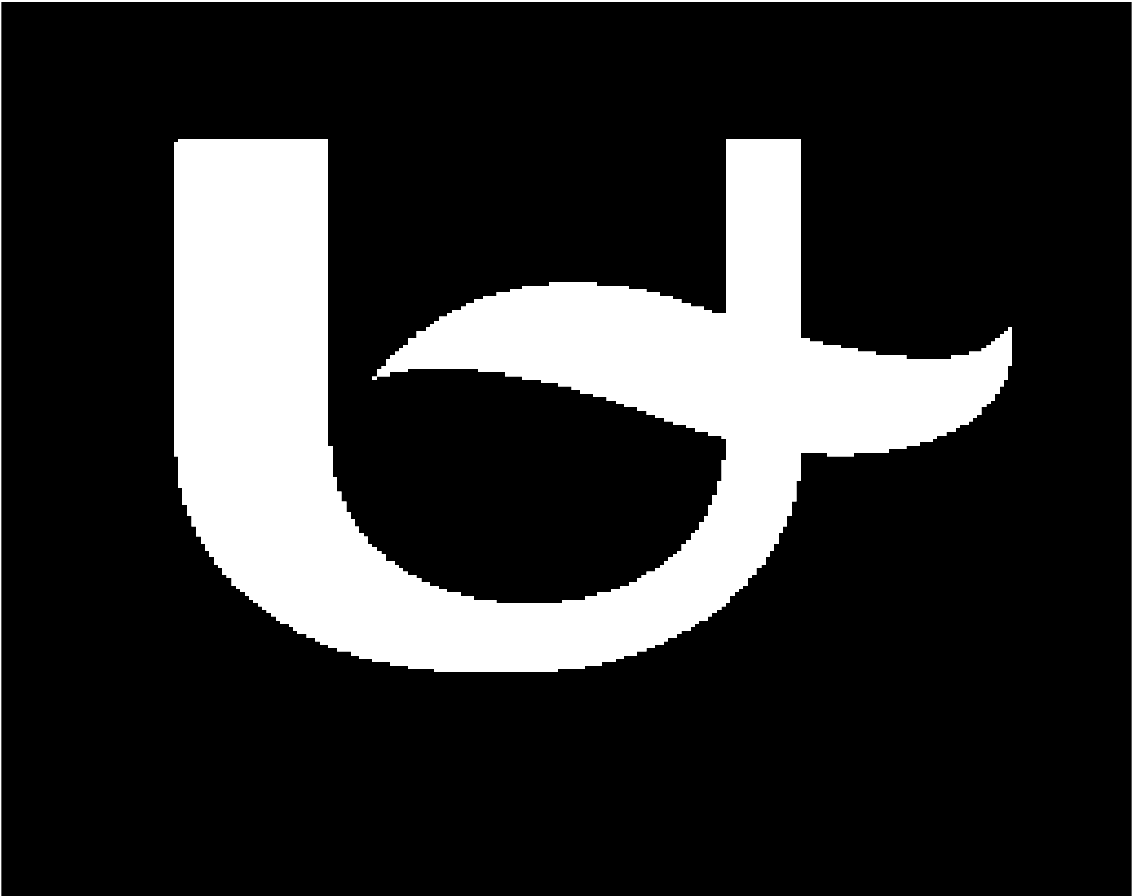}    
        \caption{}       
  \end{subfigure} 
  \begin{subfigure}[b]{0.33\textwidth}
        \includegraphics[scale=0.366, clip]{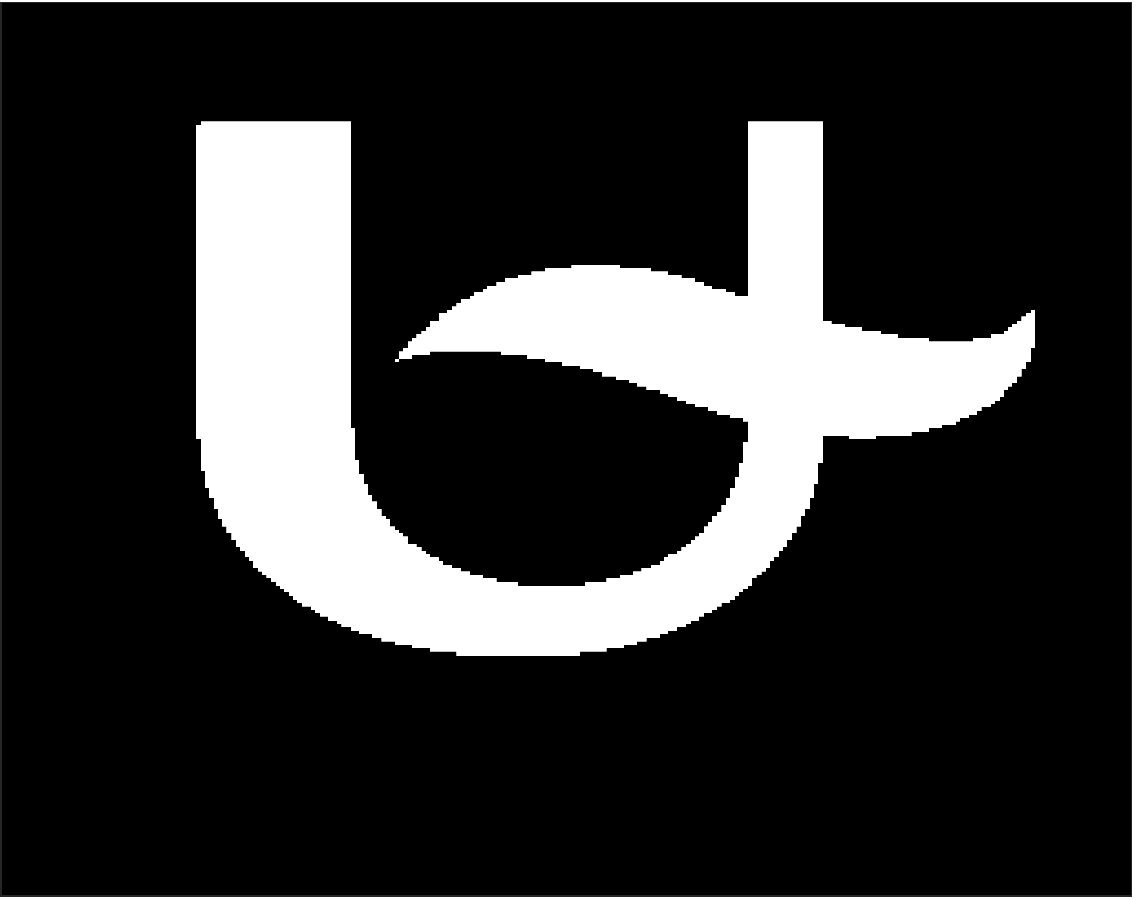}    
        \caption{}
  \end{subfigure} \\
  \begin{subfigure}[b]{0.33\textwidth}
        \includegraphics[scale=0.366,  clip]{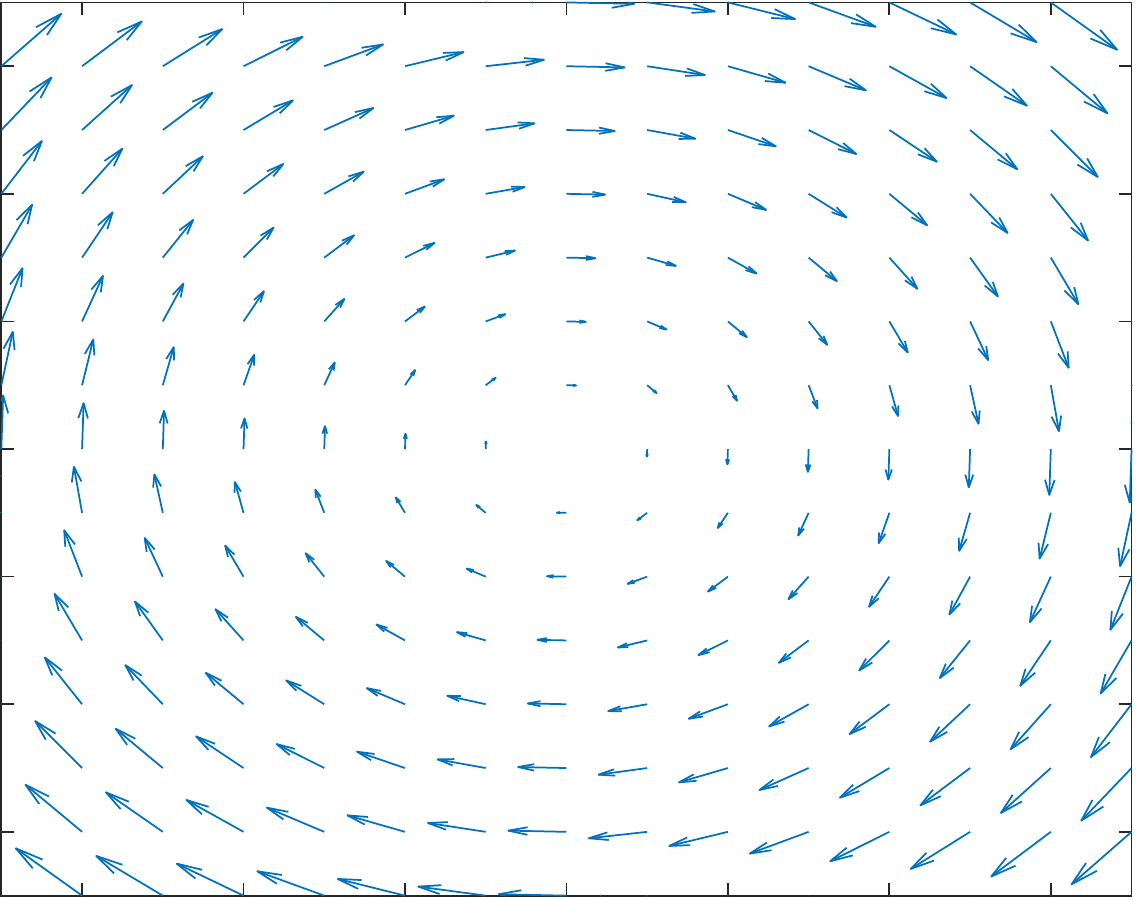}    
        \caption{}       
  \end{subfigure}
    \begin{subfigure}[b]{0.33\textwidth}
        \includegraphics[scale=0.366,  clip]{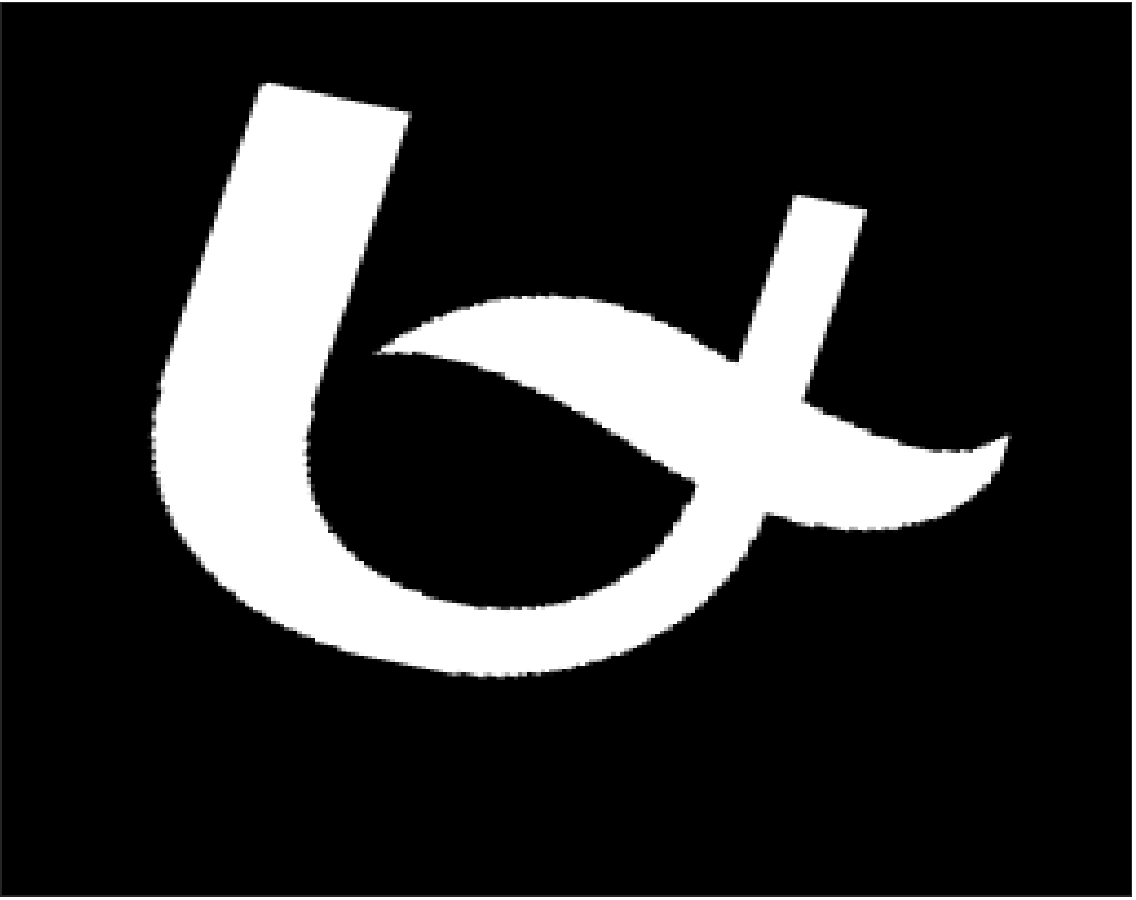}    
        \caption{}       
  \end{subfigure}
    \begin{subfigure}[b]{0.33\textwidth}
        \includegraphics[scale=0.366, clip]{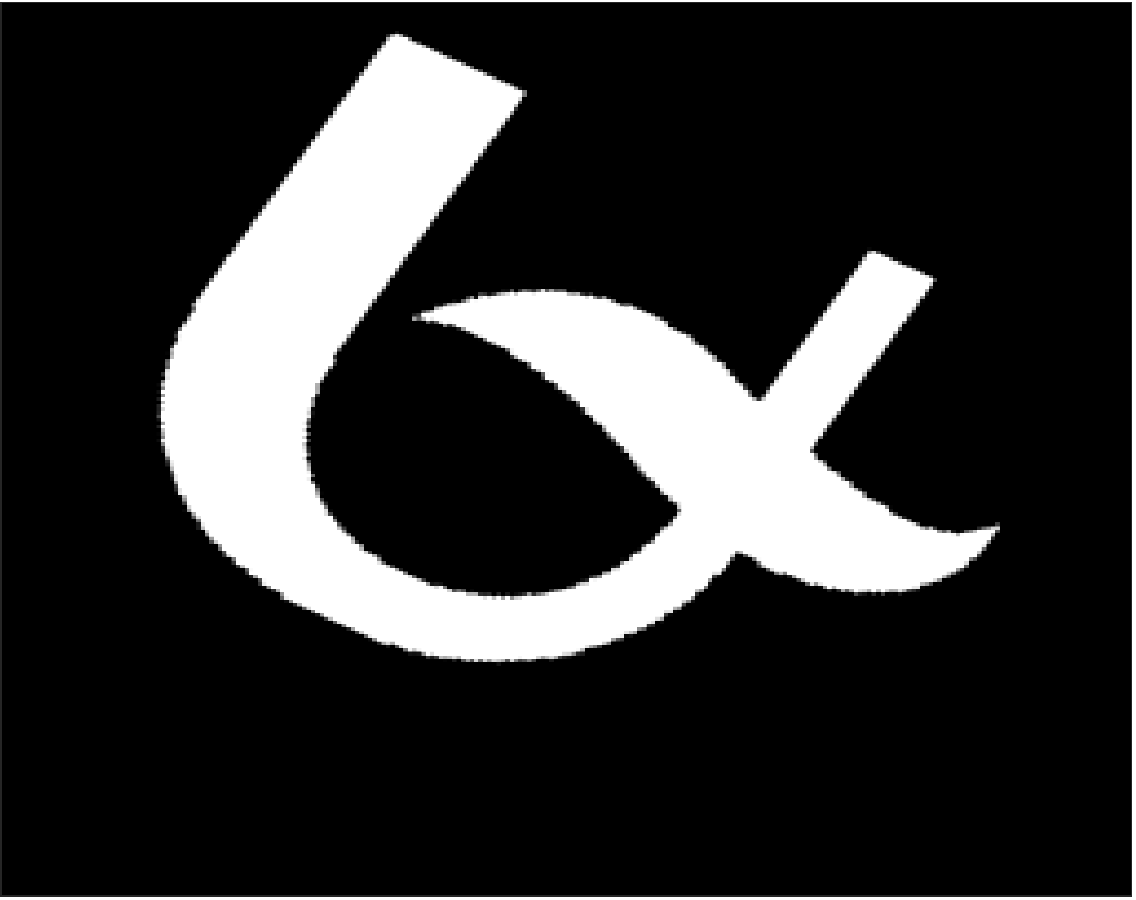}    
        \caption{}       
  \end{subfigure}
    \begin{subfigure}[b]{0.33\textwidth}
        \includegraphics[scale=0.366, clip]{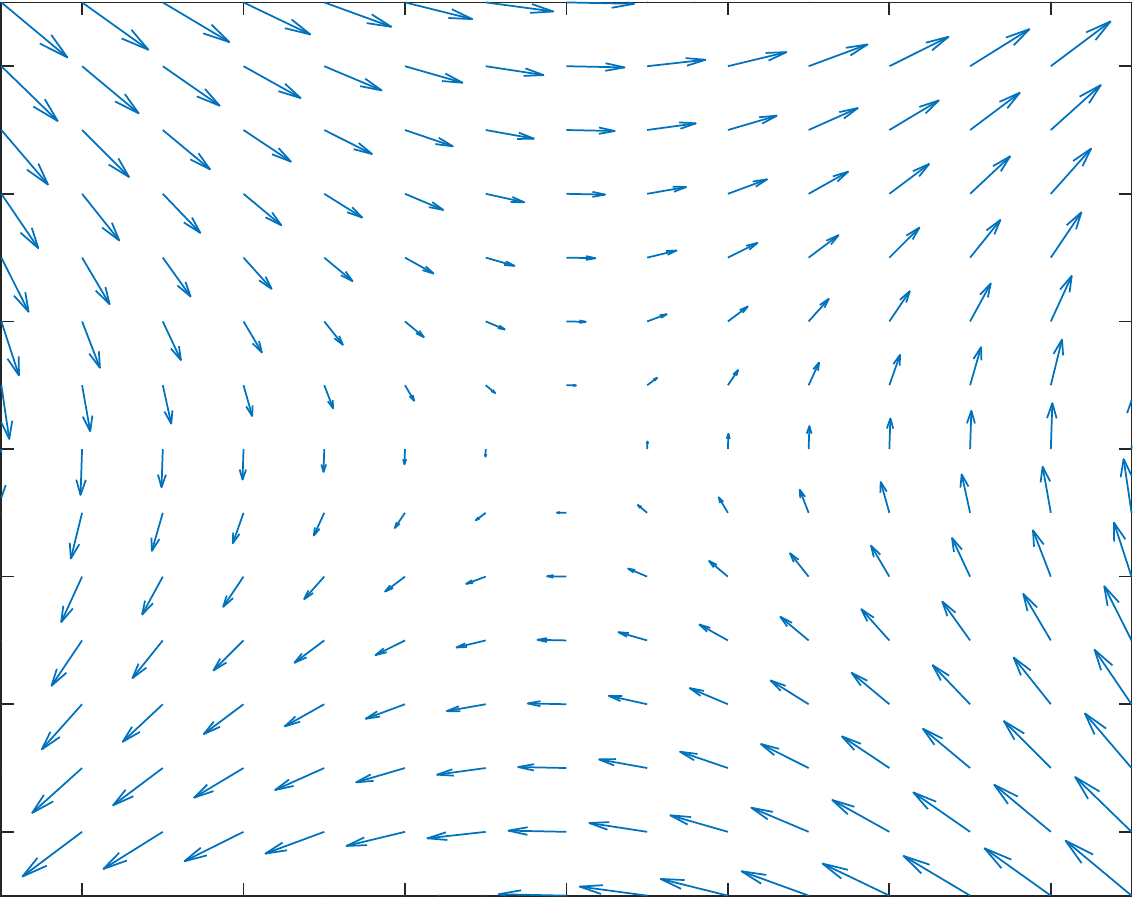}    
        \caption{}       
  \end{subfigure}
    \begin{subfigure}[b]{0.33\textwidth}
        \includegraphics[scale=0.366, clip]{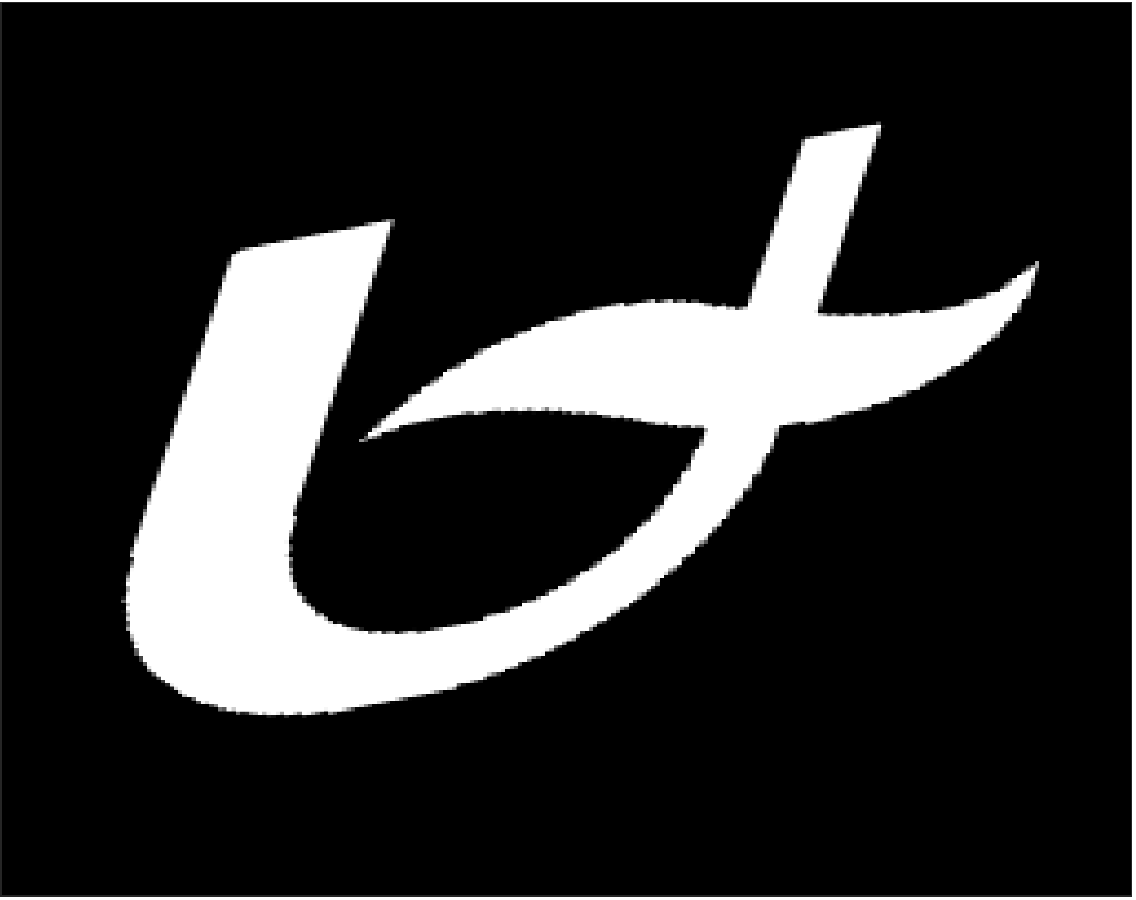}    
        \caption{}       
  \end{subfigure}
    \begin{subfigure}[b]{0.33\textwidth}
        \includegraphics[scale=0.366, clip]{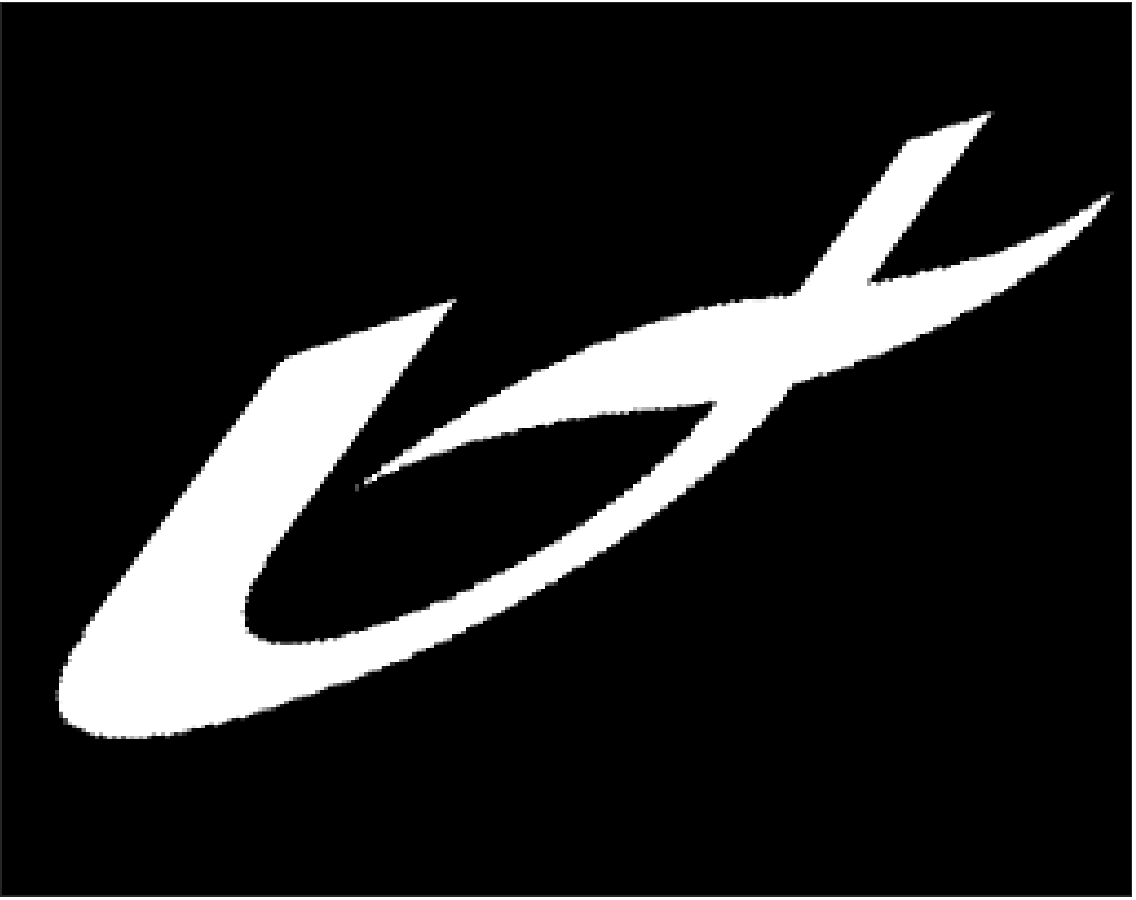}   
        \caption{}       
  \end{subfigure}
\caption{A representation of all the applied motions. Each row represents a different motion. The first column is the applied motion field. The motion is presented on the scale of one complete scan. The second column is the object $f$ at time $5 \Delta t$, so after 5 successive scans. The last column is the object $f$ at time $10 \Delta t$, so after 10 successive scans (a) - (c): The motion is a shift with 1 pixel per scan in both the horizontal and the vertical direction (d)-(f) The motion is a clockwise rotation around the origin with 3 degrees per scan (g)-(i) The applied motion is $\mathbf{v}_x(x,y) = -\left( (\cos(3)-1)x - \sin(3)y \right), \mathbf{v}_y(x,y) = \sin(3) x + (\cos(3)-1)y$.}\label{fig:examples}
\end{figure}

In this section, the algorithms on motion estimation and the correcting of CT-scan images are tested and validated on the examples in figure \ref{fig:examples}.
The object in all our simulated examples is the logo of University of Antwerp, see figure \ref{intro_vb} (a) on a $256 \times 256$ pixel grid. We choose a shift (see figure \ref{fig:examples} (a)-(c)), a rotation (see figure \ref{fig:examples} (d)-(f)) and a motion which is not an affine transformation (see figure \ref{fig:examples} (g)-(i)) as the applied motions. 
We simulated for every motion field 10 successive scans while the object was moving where per scan we acquire projection data for 180 angles uniformly distributed over $[0,\pi[$.

In the first section we discuss the results for the motion estimation based on algorithm \ref{algo:motion} described in section \ref{sec:implem}. To validate the proposed algorithm \ref{algo:correcting} in section \ref{sec:cor_images} we correct the images for the exact motion and the estimated motion. 


\subsection{Estimation of motion} \label{sec:motion_est_num}
\begin{figure}[H]
  \begin{subfigure}[b]{0.33\textwidth}
        \includegraphics[scale=0.366, clip]{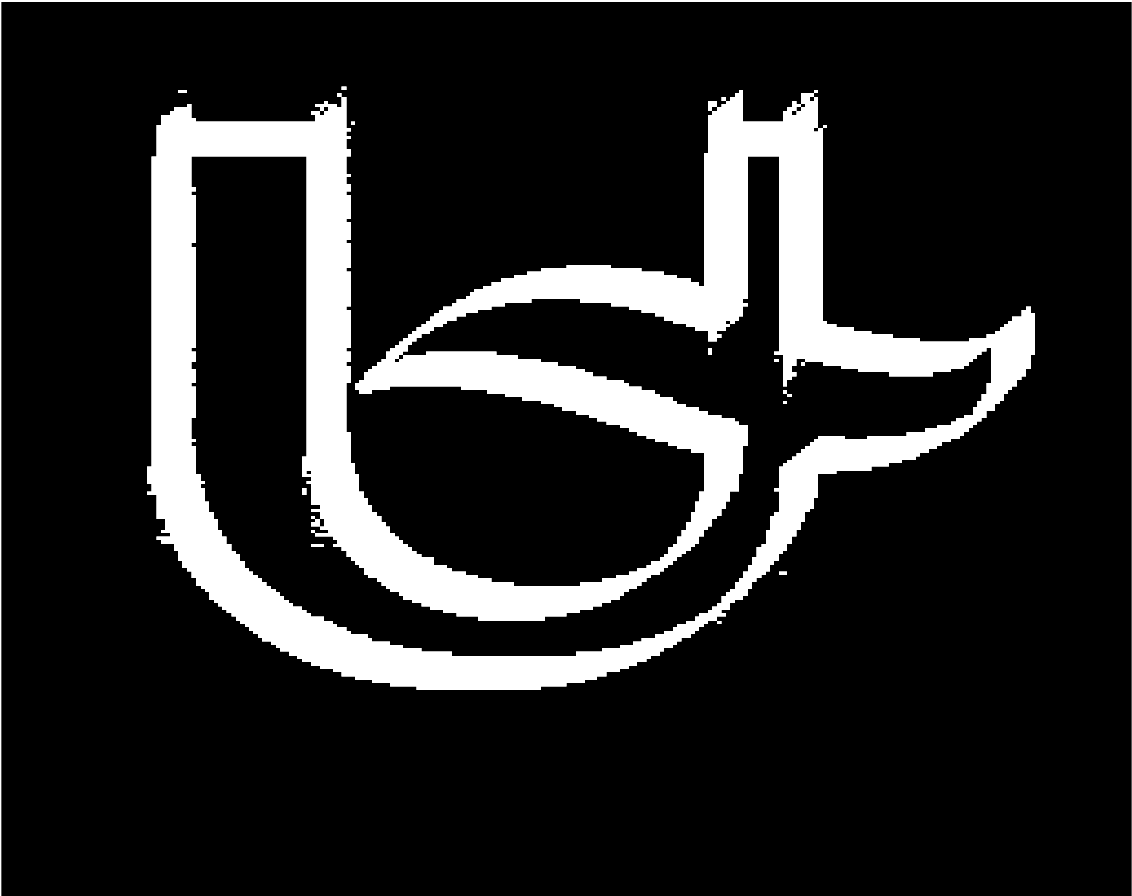}    
        \caption{}
  \end{subfigure} 
  \begin{subfigure}[b]{0.33\textwidth}
        \includegraphics[scale=0.366, clip]{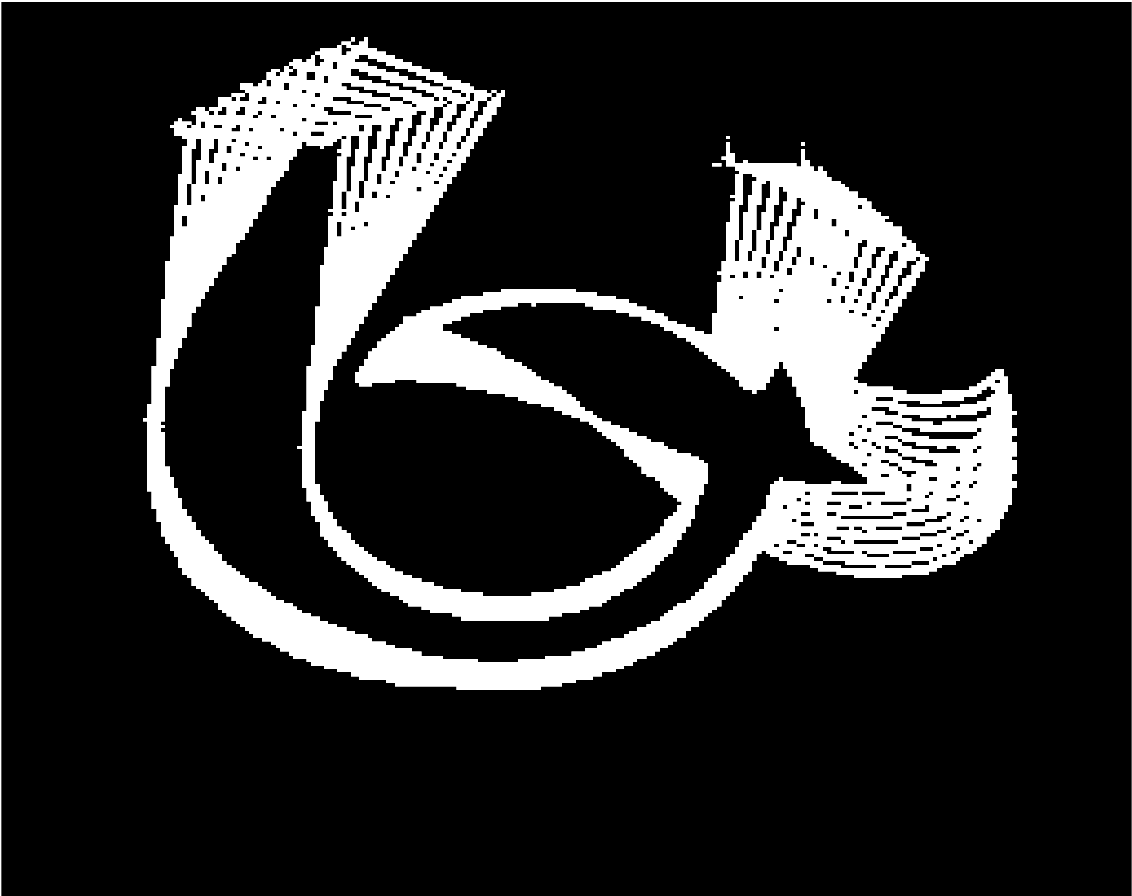}    
        \caption{}       
  \end{subfigure} 
  \begin{subfigure}[b]{0.33\textwidth}
        \includegraphics[scale=0.366, clip]{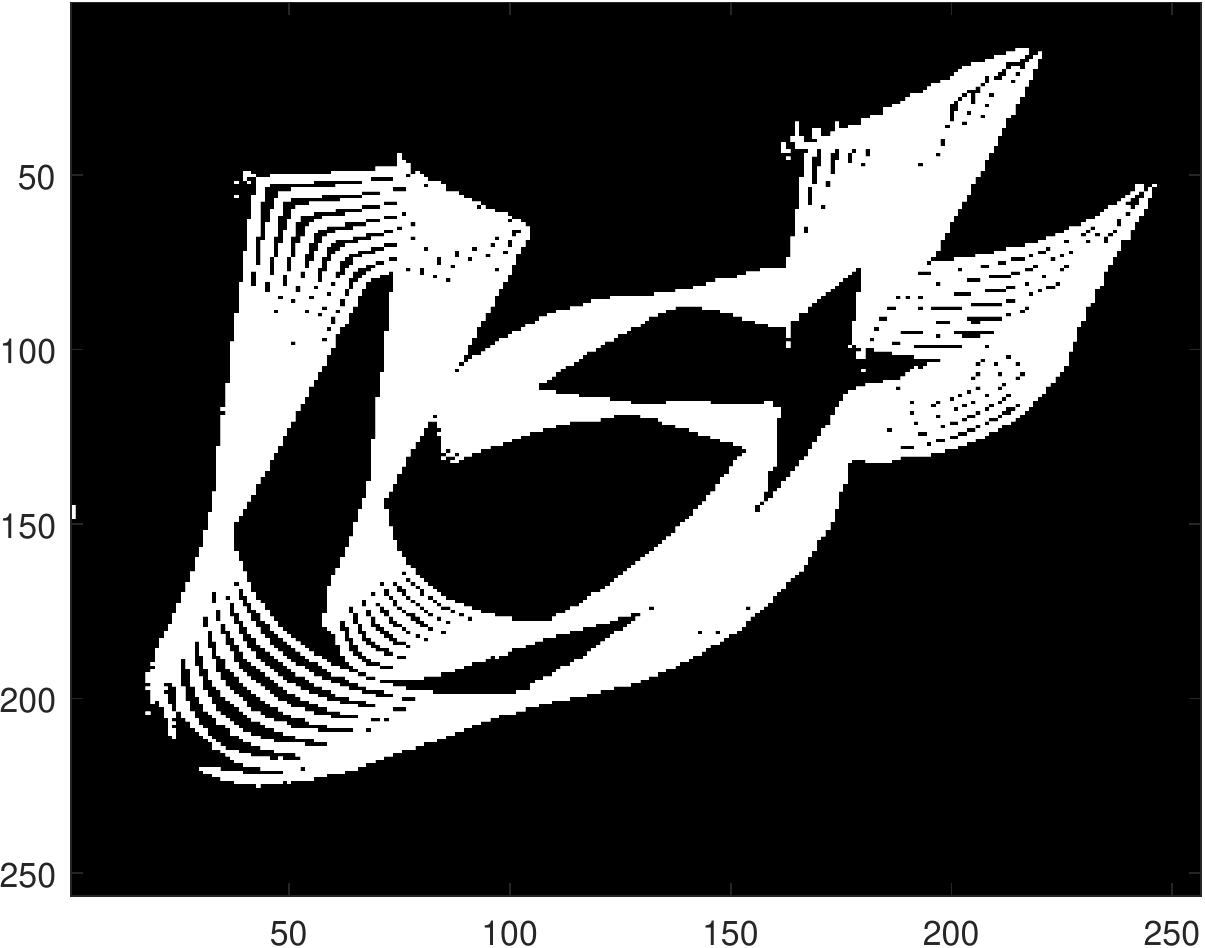}    
        \caption{}
  \end{subfigure} 
  \caption{For every example in figure \ref{fig:examples} the area $\mathfrak{A}$ (see \eqref{eqn:set_A}) is denoted in white. These are the points for which we have information about the motion. If we calculate the error $\text{RMSE}_\mathfrak{A}$, we only take these points into account. We have used $\beta$ equal to $0.15$ (a) motion 1: shift (b) motion 2: rotation (c) Motion 3. } \label{fig:RMSE_A}
\end{figure}

In this section, we check the quality of the motion estimate. Because the error of the motion estimate is less relevant in the regions where the information comes only from the regularisation, we calculate the error only for points in the set
\begin{equation}
\mathfrak{A} := \Big \{(x,y) \in \Omega | \exists t \in [0,m \Delta t[: \left| \frac{\partial f^{\text{rec}}_t}{\partial x} \right|  >  \beta  \text{ or } \left| \frac{\partial f^{\text{rec}}_t}{\partial y} \right| >  \beta  \Big \}. \label{eqn:set_A}
\end{equation}
This is the set of points for which the derivative is in absolute value, for at least one moment in the time interval, greater than a certain small default value $\beta > 0$. This corresponds with the points for which we have information about the motion. This set is represented in figure \ref{fig:RMSE_A} for each example. We define the error $\text{RMSE}_\mathfrak{A}$ as 
\begin{equation} \text{RMSE}_\mathfrak{A} := \sqrt{\sum_{(x,y) \in \mathfrak{A}} \frac{ \left( v_x(x,y) - \hat{v}_x(x,y) \right)^2 + \left( v_y(x,y) - \hat{v}_y(x,y)\right)^2}{n} } \label{eqn:RMSE}
\end{equation} with $n$ the number of elements in the set $\mathfrak{A}$ and $\hat{v}(x,y)$ is the estimated motion from algorithm \ref{algo:motion}. 

\begin{figure}[H]
  \begin{subfigure}[b]{0.33\textwidth}
        \includegraphics[scale=0.366, clip]{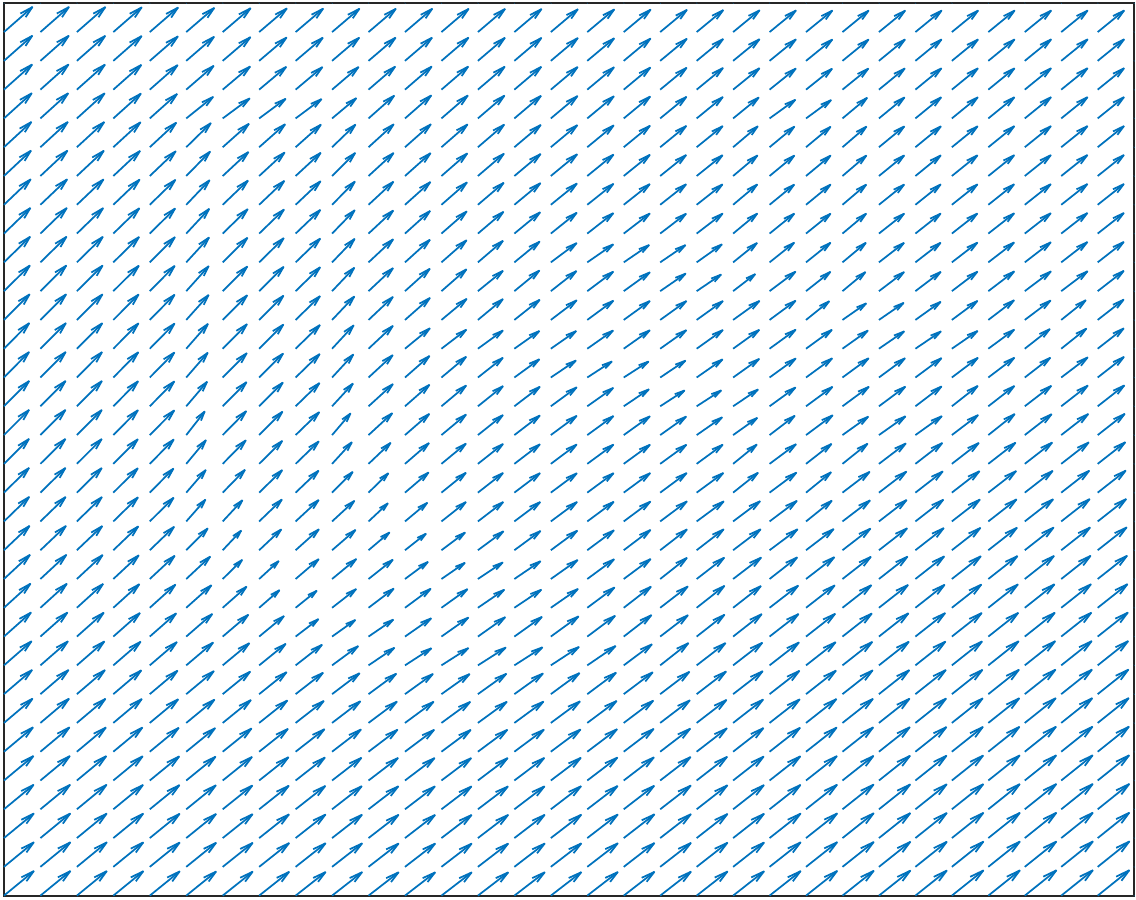}    
        \caption{}
  \end{subfigure} 
  \begin{subfigure}[b]{0.33\textwidth}
        \includegraphics[scale=0.366, clip]{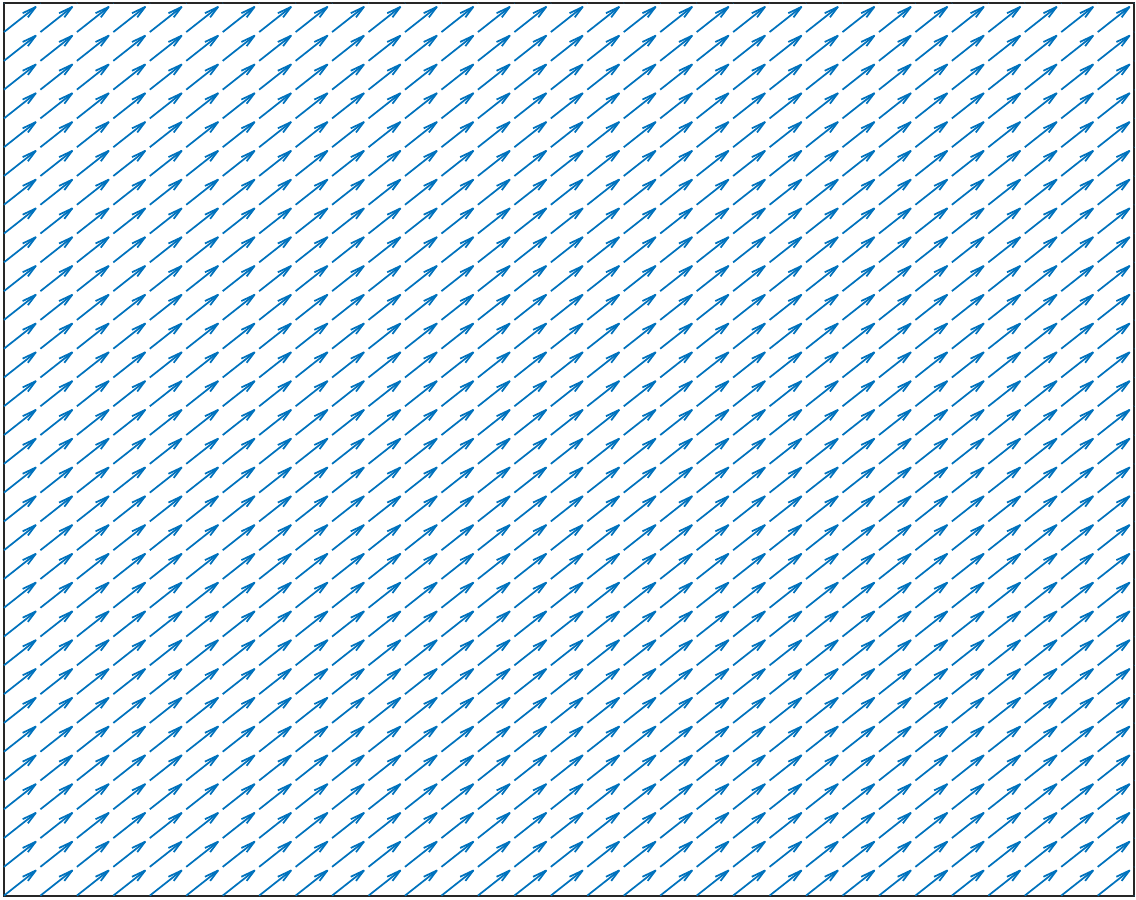}    
        \caption{}       
  \end{subfigure} 
  \begin{subfigure}[b]{0.33\textwidth}
        \includegraphics[scale=0.366,  clip]{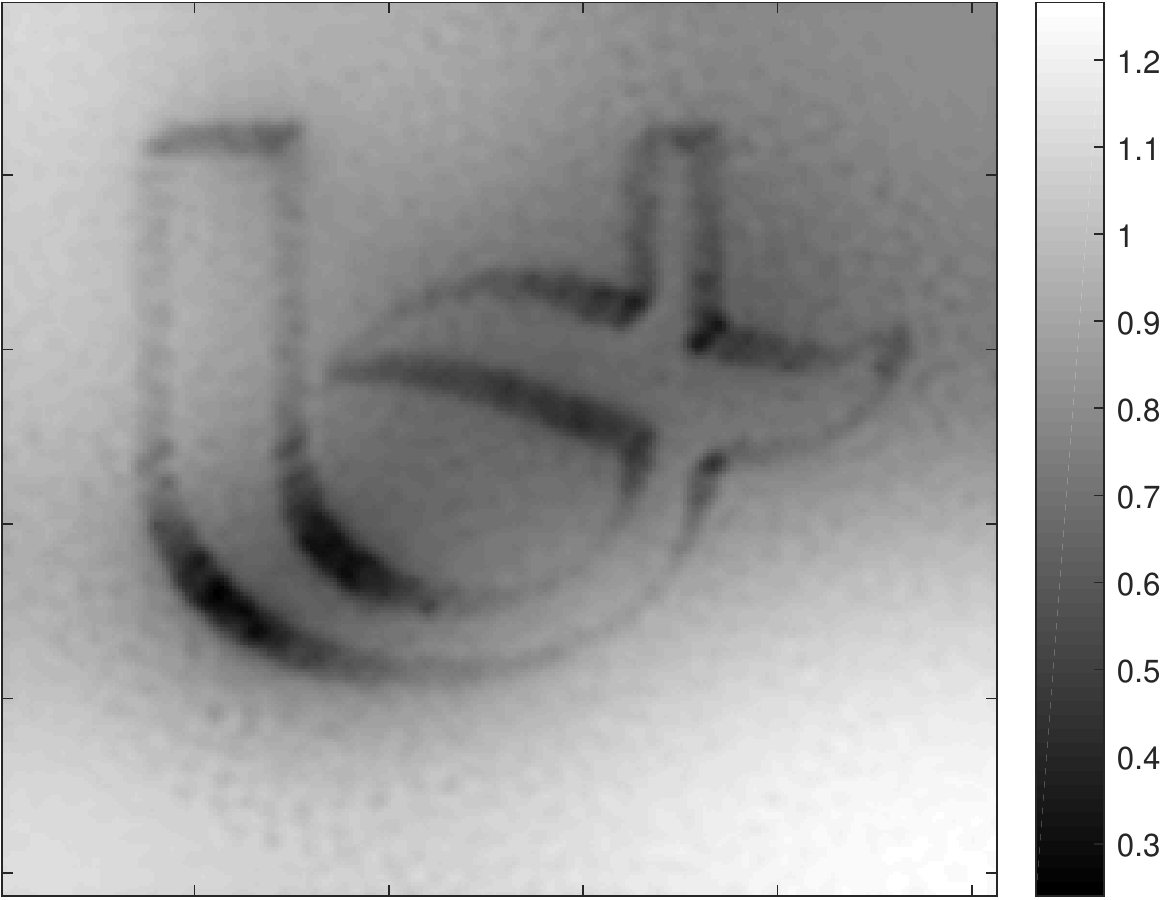}    
        \caption{}
  \end{subfigure} 
\caption{In the first column we plot the motion field of the estimated motion $\hat{\mathbf{v}}$ for the first example from figure \ref{fig:examples} where we have used the depth $d$ equal to 3 in algorithm \ref{algo:motion} and we did not add noise to the data. In the second column we plot the exact motion field of the same example. In the last column we plot the error $\sqrt{ \left( v_x(x,y) - \hat{v}_x(x,y)\right)^2 + \left( v_y(x,y) - \hat{v}_y(x,y)\right)^2}$. We can derive from the last picture that at the places where we have information about the motion (see the areas $\mathfrak{A}$ in figure \ref{fig:RMSE_A}), the motion is far better estimated than at the places where the only information comes from the regularisation.}\label{fig:motion_est}
\end{figure}

In figure \ref{fig:motion_est} the estimated motion is presented together with a plot of the error with the exact motion. In table \ref{table:motion_est} we calculate the error $\text{RMSE}_\mathfrak{A}$ for all three examples for the exact data and for data where we add some normally distributed noise with mean $0$ and standard deviation equal to $2$. We choose to calculate it using depth $d = 0$ (so without changing the resolution) and depth $d= 3$ (we start algorithm \ref{algo:motion} by estimating the motion on a resolution which is $8 (= 2^3)$ times smaller).

\begin{table}[H]
\begin{center}
\begin{tabular}{ccccc}
\hline
& depth $d = 0$ & depth $d = 3$ & depth $d = 0$ & depth $d = 3$ \\
& no noise added & no noise added & with noise & with noise \\
\hline
Motion 1: Shift & $0.3994$ & $0.6664$ &  $0.7873$ & $0.6640$ \\
Motion 2: Rotation & $3.1863$& $1.2257$ & $3.4893$ & $1.2315$ \\
Motion 3& $2.6076$ & $0.4679$ & $3.5171$ & $0.4658$ \\ 
\hline 
\end{tabular} 
\caption{The $\text{RMSE}_\mathfrak{A}$ \eqref{eqn:RMSE} when estimating the motion using the algorithm \ref{algo:motion} with depth $d$ equal to $0$ (first and third column) and depth equal to $3$ (second and last column). The noise is normally distributed with mean zero and standard deviation $2$. We can derive from this table that doing the coarse-to-fine algorithm \ref{algo:motion} improves the performance for big motions but that for smaller motion it is better to use depth $d = 0$. We see that when adding noise to the data, it is preferable to choose the depth $d$ equal to $3$.}  \label{table:motion_est}
\end{center}
\end{table}

\subsection{Correcting CT-images}
\begin{figure}[H]
   \begin{subfigure}[b]{0.33\textwidth}
        \includegraphics[scale=0.366, clip]{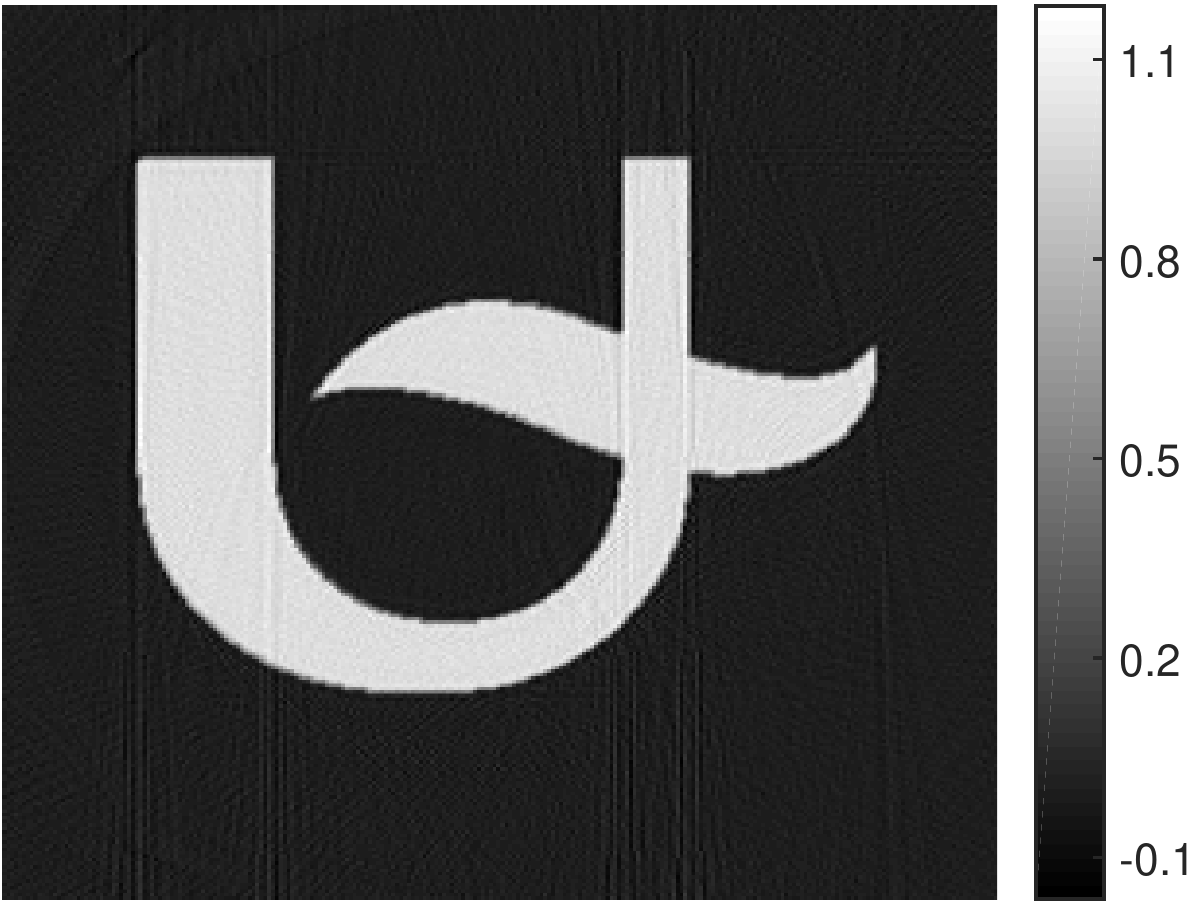}    
        \caption{}       
  \end{subfigure} 
  \begin{subfigure}[b]{0.33\textwidth}
        \includegraphics[scale=0.366, clip]{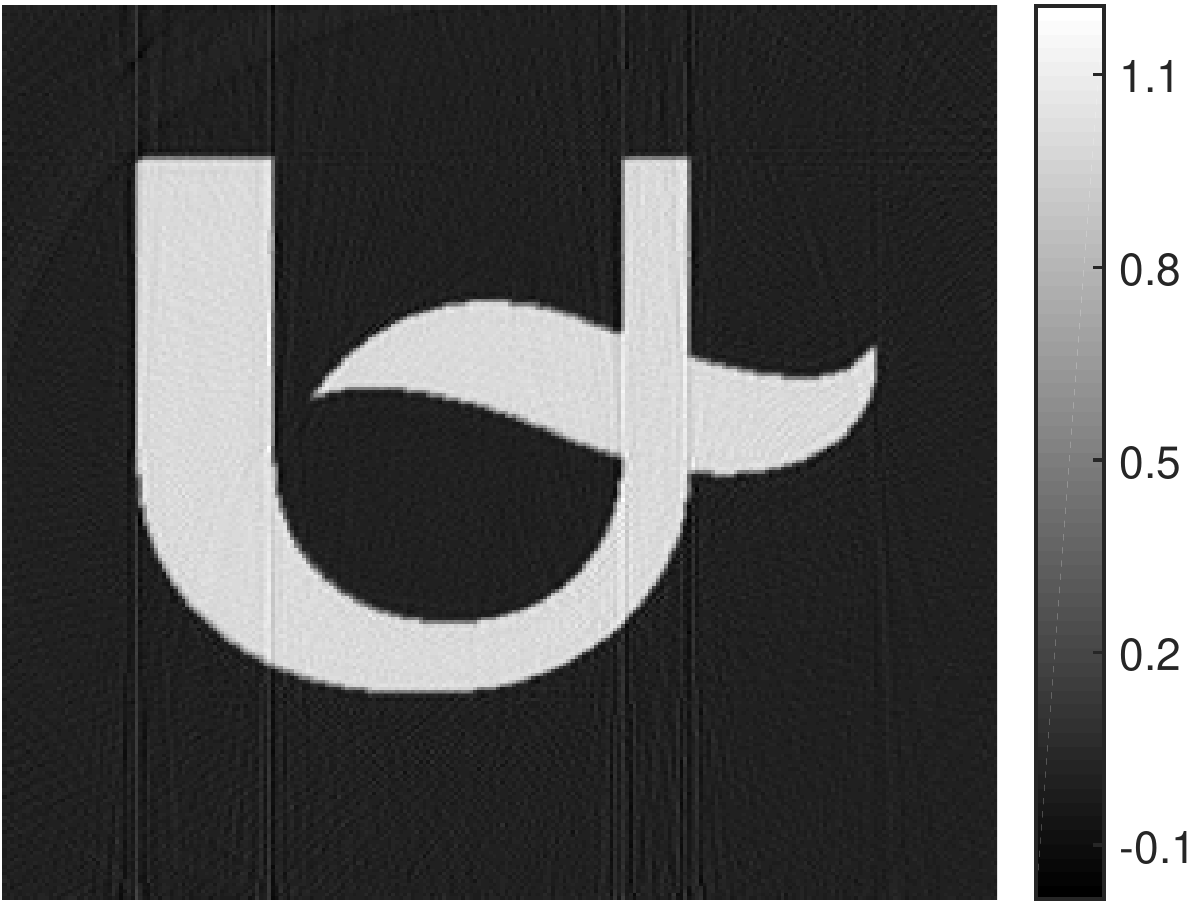}    
        \caption{}
  \end{subfigure}
  \begin{subfigure}[b]{0.33\textwidth}
        \includegraphics[scale=0.366, clip]{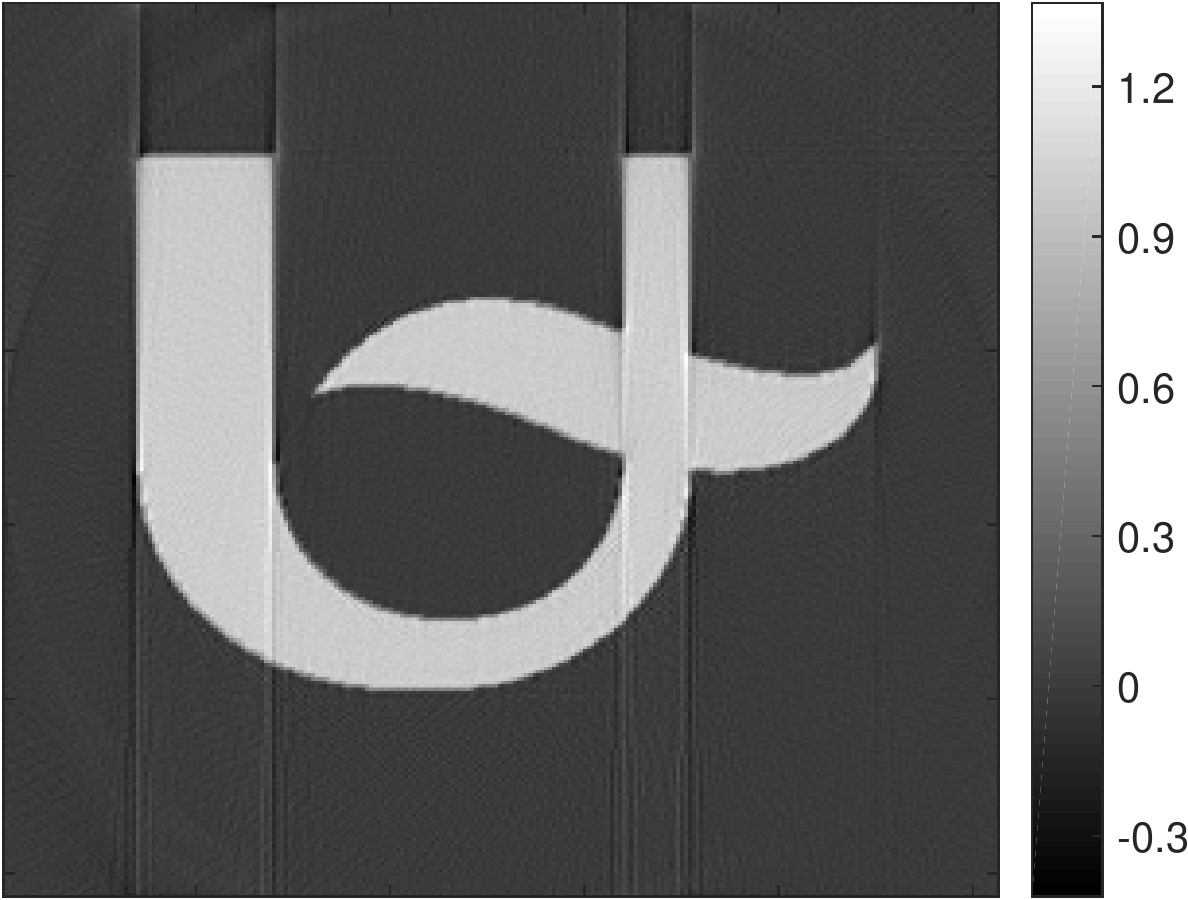}    
        \caption{}
  \end{subfigure} \\
  \begin{subfigure}[b]{0.33\textwidth}
        \includegraphics[scale=0.366, clip]{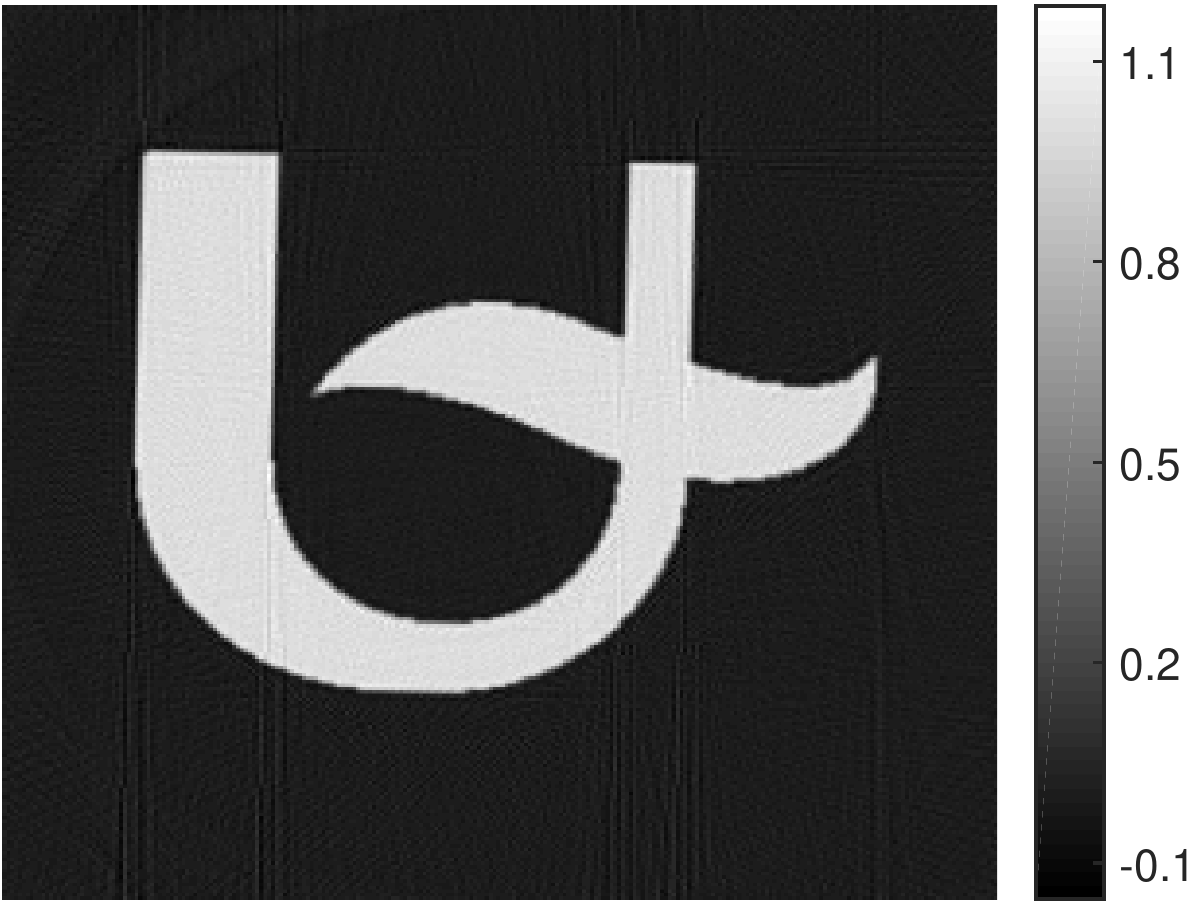}    
        \caption{}       
  \end{subfigure}
    \begin{subfigure}[b]{0.33\textwidth}
        \includegraphics[scale=0.366, clip]{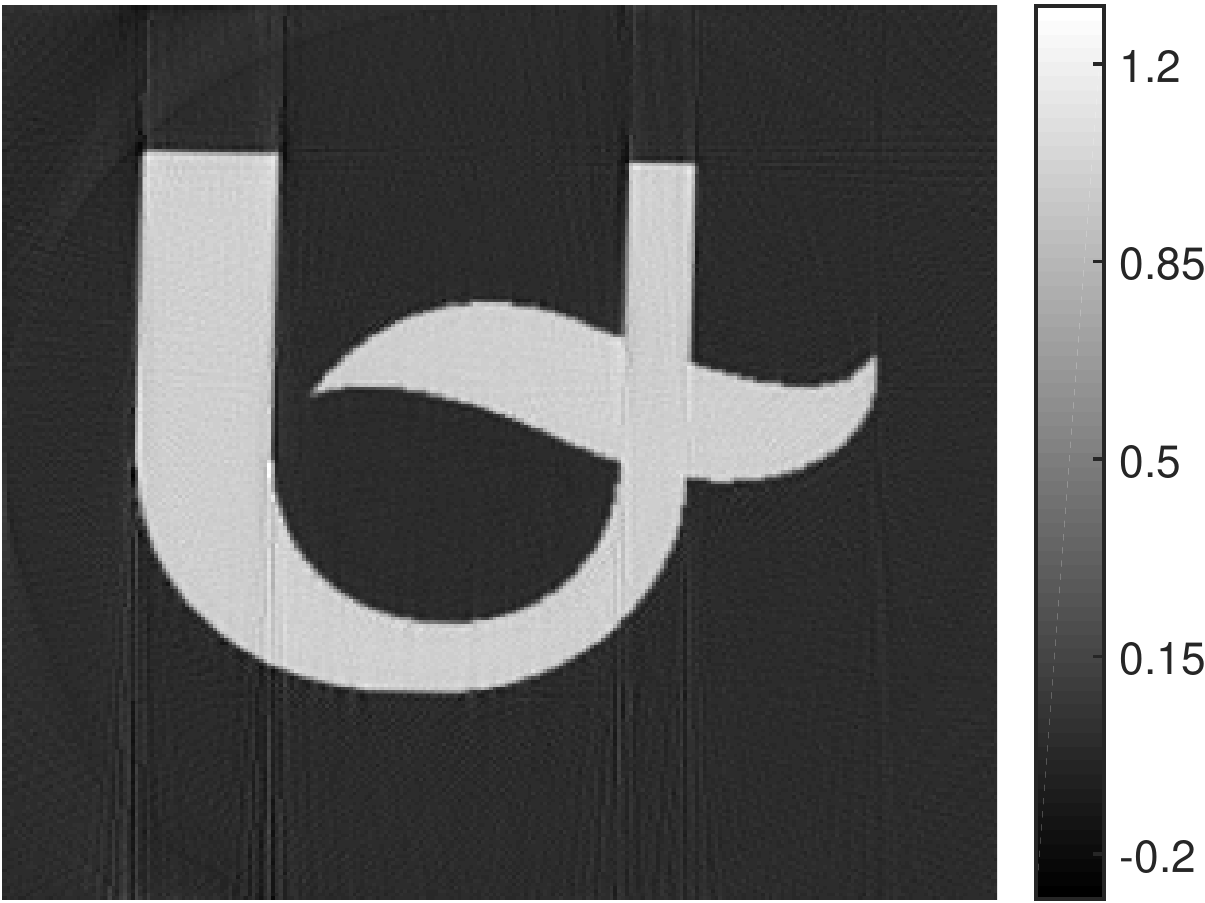}    
        \caption{}       
  \end{subfigure}
  \begin{subfigure}[b]{0.33\textwidth}
        \includegraphics[scale=0.366, clip]{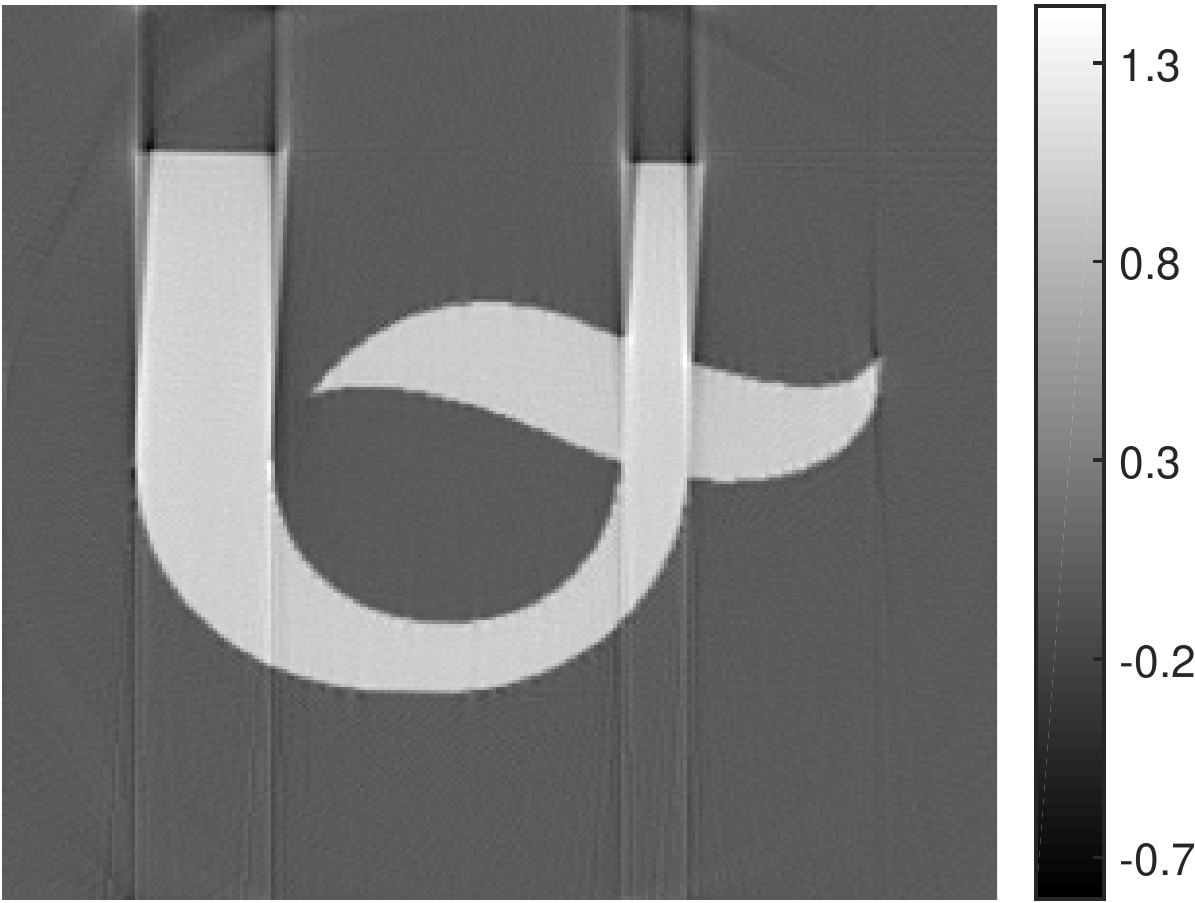}    
        \caption{}       
  \end{subfigure} \\
  \begin{subfigure}[b]{0.33\textwidth}
        \includegraphics[scale=0.366, clip]{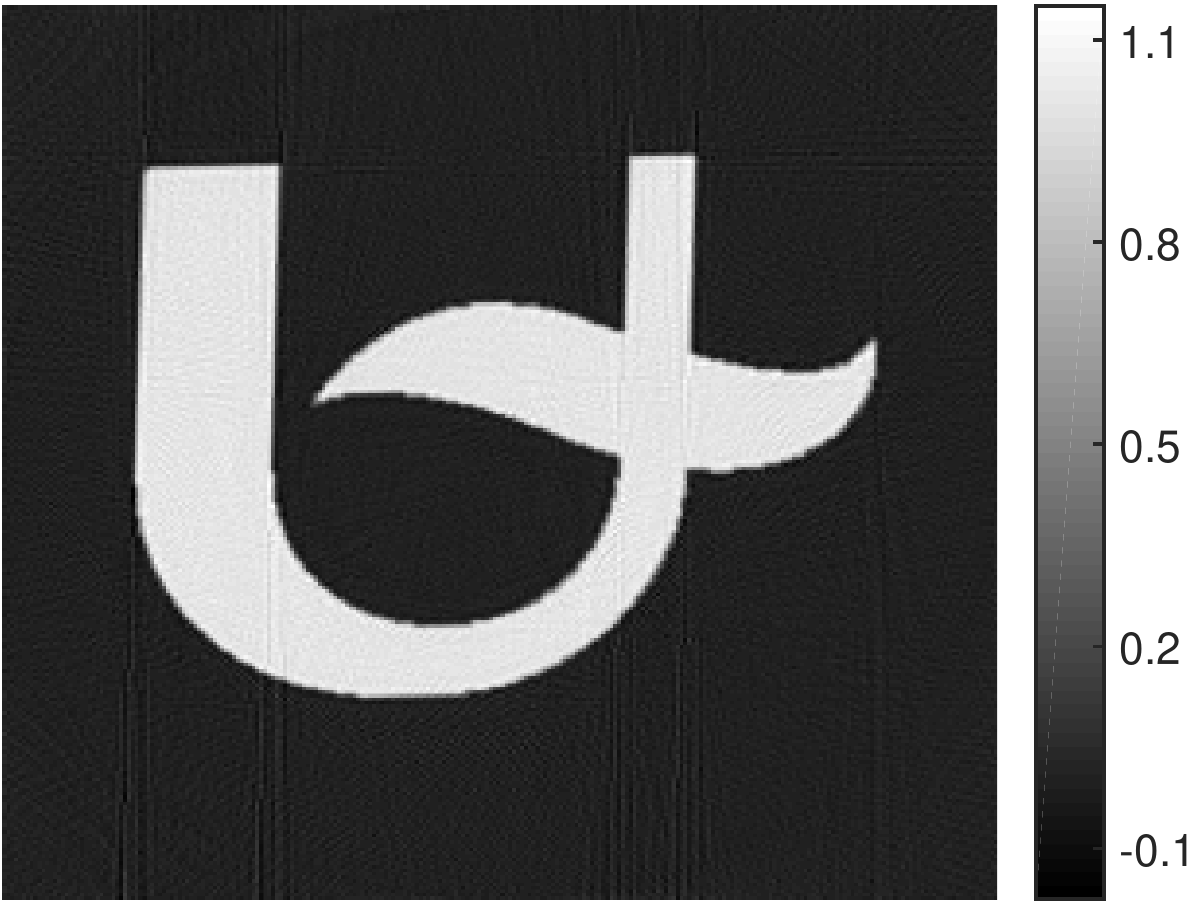}    
        \caption{}       
  \end{subfigure}
  \begin{subfigure}[b]{0.33\textwidth}
        \includegraphics[scale=0.366, clip]{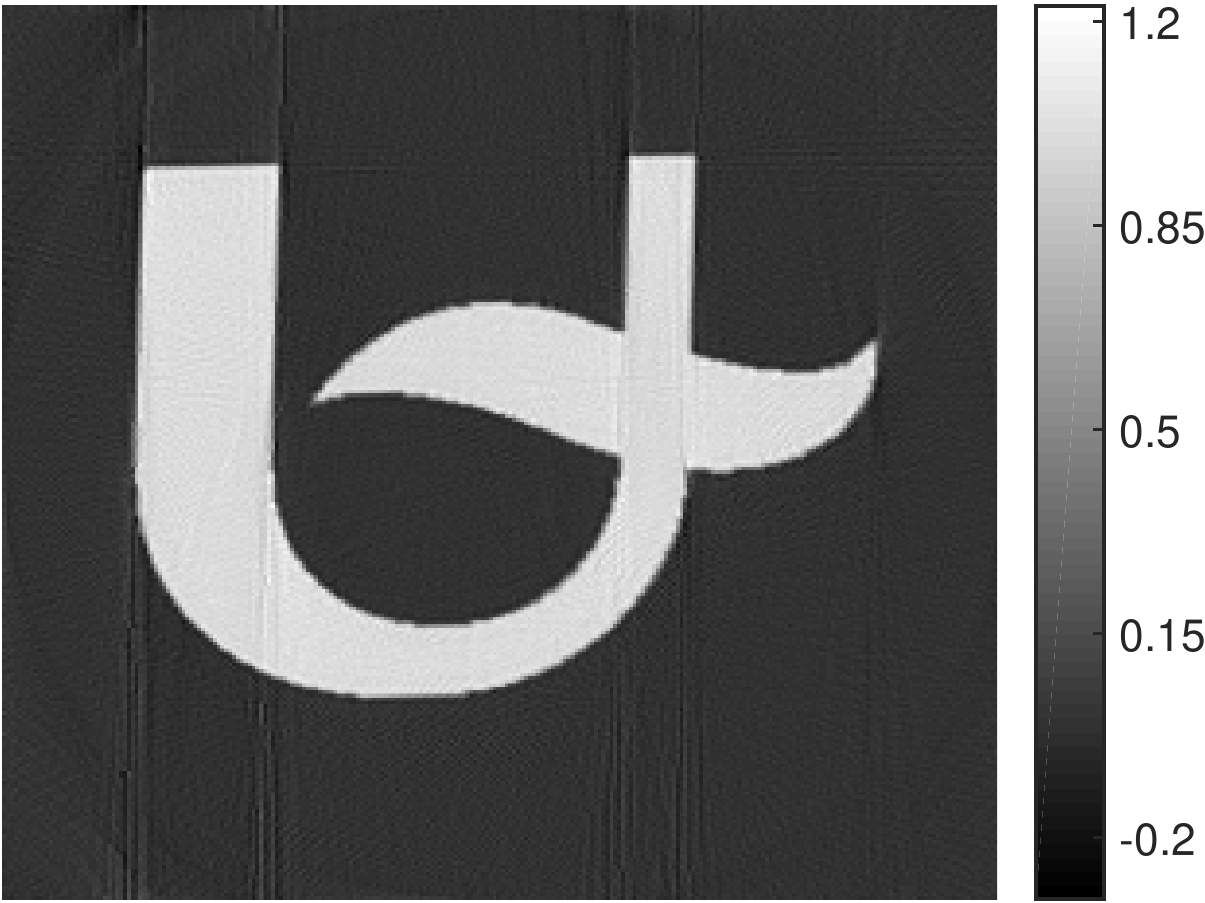}   
        \caption{}       
  \end{subfigure}
  \begin{subfigure}[b]{0.33\textwidth}
        \includegraphics[scale=0.366, clip]{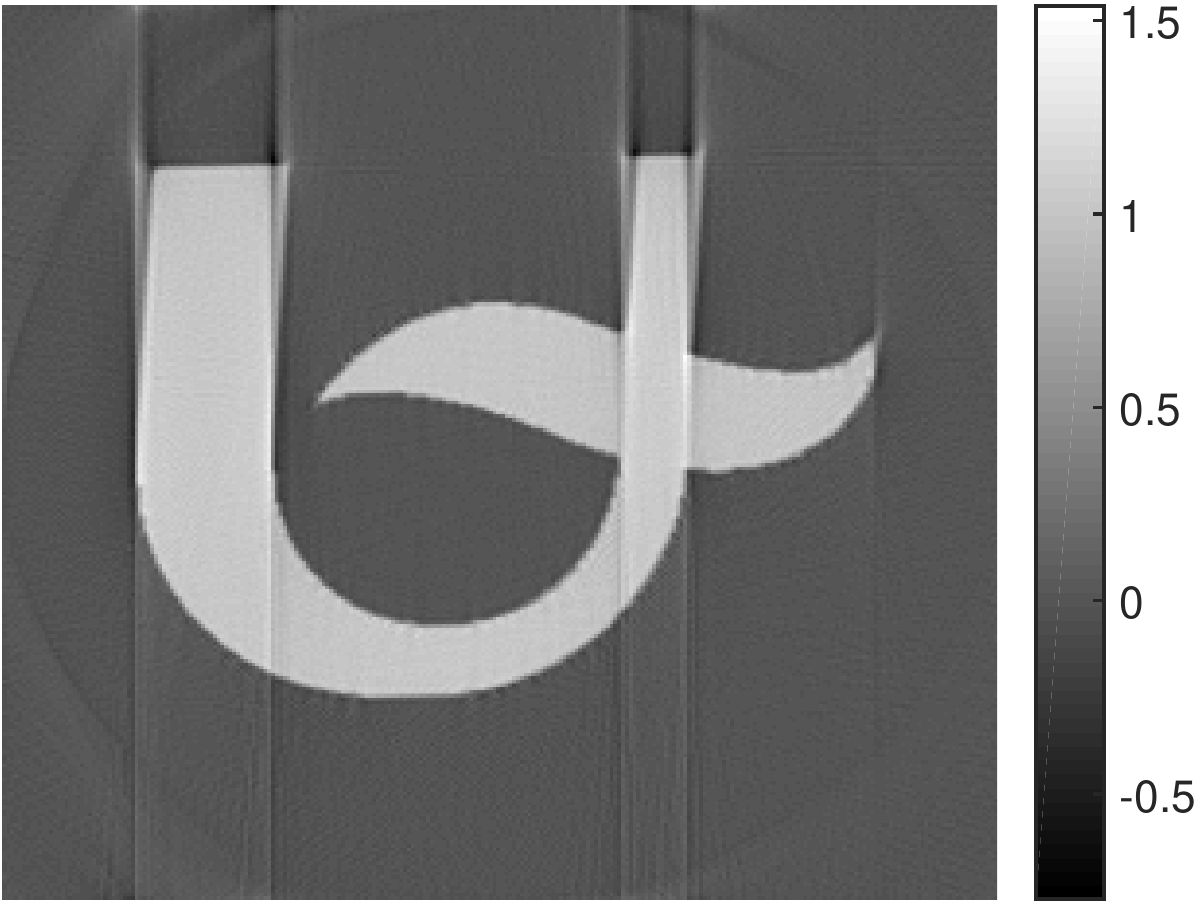}    
        \caption{}       
  \end{subfigure}
\caption{Reconstructions for the 3 different examples from figure \ref{fig:examples}.  (a)-(d)-(g) Reconstruction $f_{t^{l/2}}^{\text{corr}}$ for each example. (b)-(e)-(h) Reconstruction $\hat{f}_{t^{l/2}}^{\text{corr}}$ for each example (c)-(f)-(i) Reconstruction $f^{\text{rec}}_{t^{l/2}}$ without correcting for motion. We can see that the reconstruction with the exact motion works the best (first column).}\label{fig:correcting}
\end{figure}

In this section we validate the algorithm \ref{algo:correcting} in section \ref{sec:cor_images} by applying it on the examples from figure \ref{fig:examples}. For this we assume we have access to the exact motion $\mathbf{v}$ or the approximation $\hat{ \mathbf{v}}$ made in the previous section \ref{sec:motion_est_num}. We assume we have the exact scan data, in the next section we add noise to our data. As mentioned before, we make our reconstructions at the middle of the first scan, i.e. at time $\Delta t/2$. For this we only need the sinogram of the first scan. 
 In table \ref{table:rec_ex} we compare the error with the exact object $f_{\Delta t/2}$ of the following reconstructions: a) the error we make when we perform a reconstruction $f^{\text{ex}}_{\Delta t/2}$ of object $f_{\Delta t/2}$ assuming the object is stationary in this state during the data acquisition,  
 b) reconstructed figure $f_{\Delta t/2}^{\text{corr}}$ when correcting the image for the exact motion $\mathbf{v}$, c) reconstructed figure $\hat{f}_{\Delta t/2}^{\text{corr}}$ when correcting the image for the estimated motion $\hat{ \mathbf{v}}$ and d) a reconstruction $f^\text{rec}_{\Delta t/2}$ where we do not correct for the motion. It is logical that it is not possible to do better than the reconstruction when there is no motion present so if the reconstruction when correcting for motion has a similar error then we can conclude our method works. We can conclude from this table that the error the algorithm make (columns 2) is similar to the error we make when there is no motion (column 1), so the algorithm to correct for motion works. If we compare the last column with all the other columns we can see that correcting the system for the motion has a significant effect on the quality of the reconstruction. This can also be seen in figure \ref{fig:correcting}.
 
\begin{table}[H]
\begin{center}
\begin{tabular}{ccccc}
\hline
& $\left \| f^{\text{ex}}_{\Delta t/2} - f_{\Delta t/2} \right \|_2$  & $\left \| f_{\Delta t/2}^{\text{corr}} - f_{\Delta t/2} \right \|_2$ &   $\left \| \hat{f}_{\Delta t/2}^{\text{corr}} - f_{\Delta t/2}\right \|_2$  & $\left \| f_{\Delta t/2}^{\text{rec}} - f_{\Delta t/2} \right \|_2$\\
\hline
Motion 1: Shift & $1.3589$ &  $2.7775$ &  $4.0105$  & $5.4246$ \\
Motion 2: Rotation & $2.6729$ &  $2.0863$  & $3.2873$ & $12.8958$ \\
Motion 3& $2.5758$ &  $2.0819$ &  $2.9457$ &  $13.0636$  \\
\hline 
\end{tabular}
\caption{The error with respect to the exact solution $f_{\Delta t/2}$. First column is the error of the reconstruction if object $f$ is stationary in the state at time $\Delta t/2$. The second column is the error when we correct for the exact motion. The third column is the error when we correct for the estimated motion. In the last column we put the error when we do not correct for the motion. We see that the error in the first two columns is similar which means our method works. We can see that when using the estimated motion, we still have a far more precise solution than when we do not correct at all.} \label{table:rec_ex}
\end{center}
\end{table}

\subsection{Correcting CT-images with noise}
We have  seen that our algorithm works given exact scan data. The question arises how it performs with added noise on the data. We apply the same noise as in section~\ref{sec:motion_est_num}. We first check the quality of the reconstruction when we correct for the exact motion using the noisy data and then we test how the method performs when the motion is calculated using the noisy data and the reconstruction is corrected for this estimated motion.

\begin{figure}[H]
   \begin{subfigure}[b]{0.33\textwidth}
        \includegraphics[scale=0.366, clip]{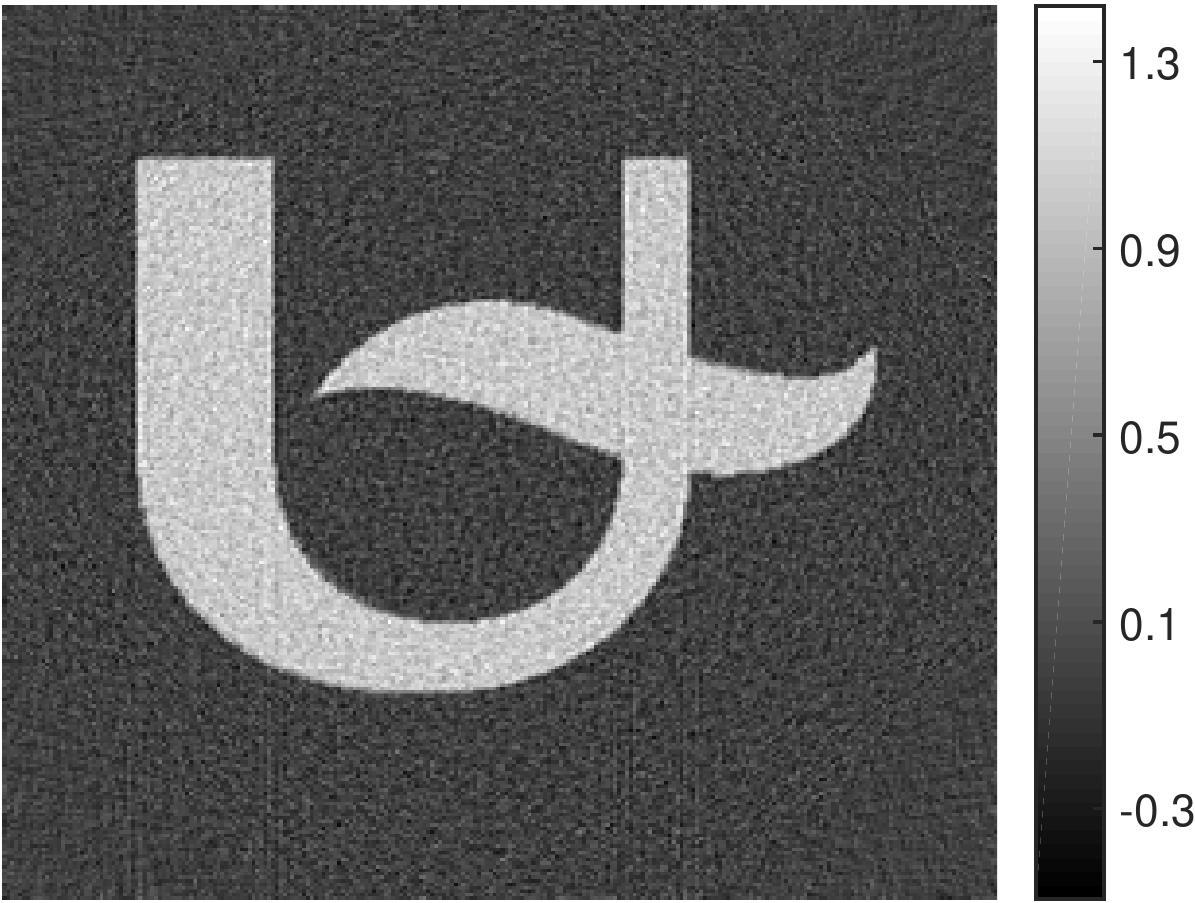}    
        \caption{}       
  \end{subfigure} 
  \begin{subfigure}[b]{0.33\textwidth}
        \includegraphics[scale=0.366, clip]{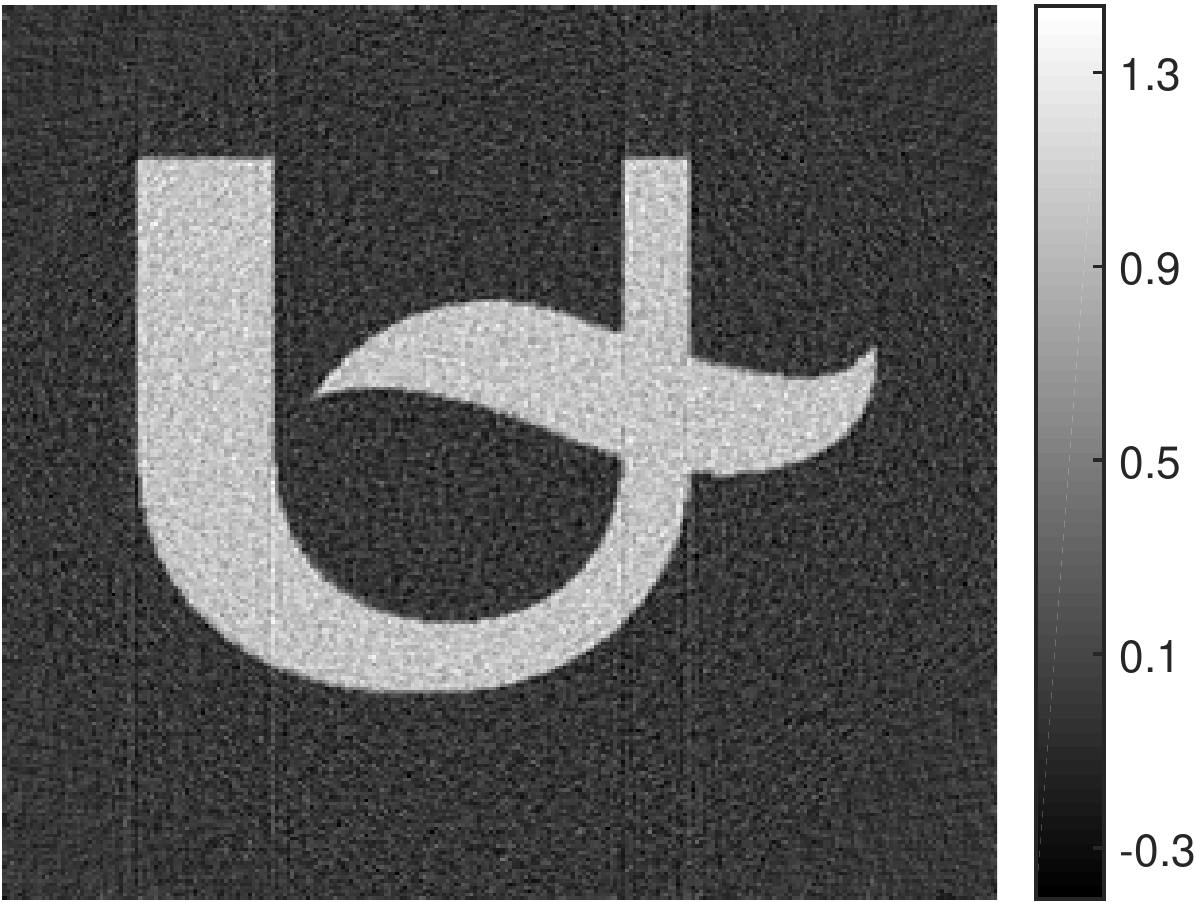}    
        \caption{}
  \end{subfigure}
  \begin{subfigure}[b]{0.33\textwidth}
        \includegraphics[scale=0.366, clip]{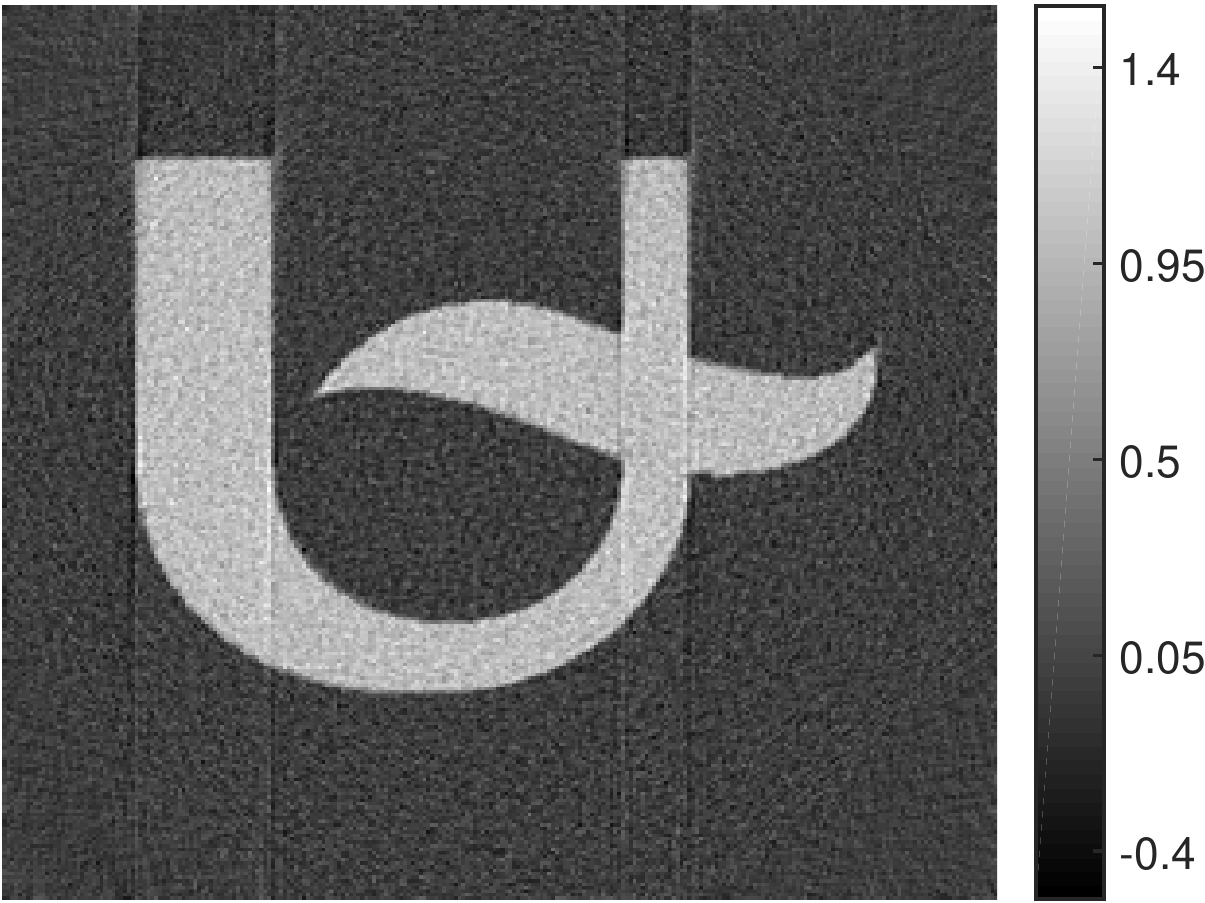}    
        \caption{}
  \end{subfigure} \\
  \begin{subfigure}[b]{0.33\textwidth}
        \includegraphics[scale=0.366, clip]{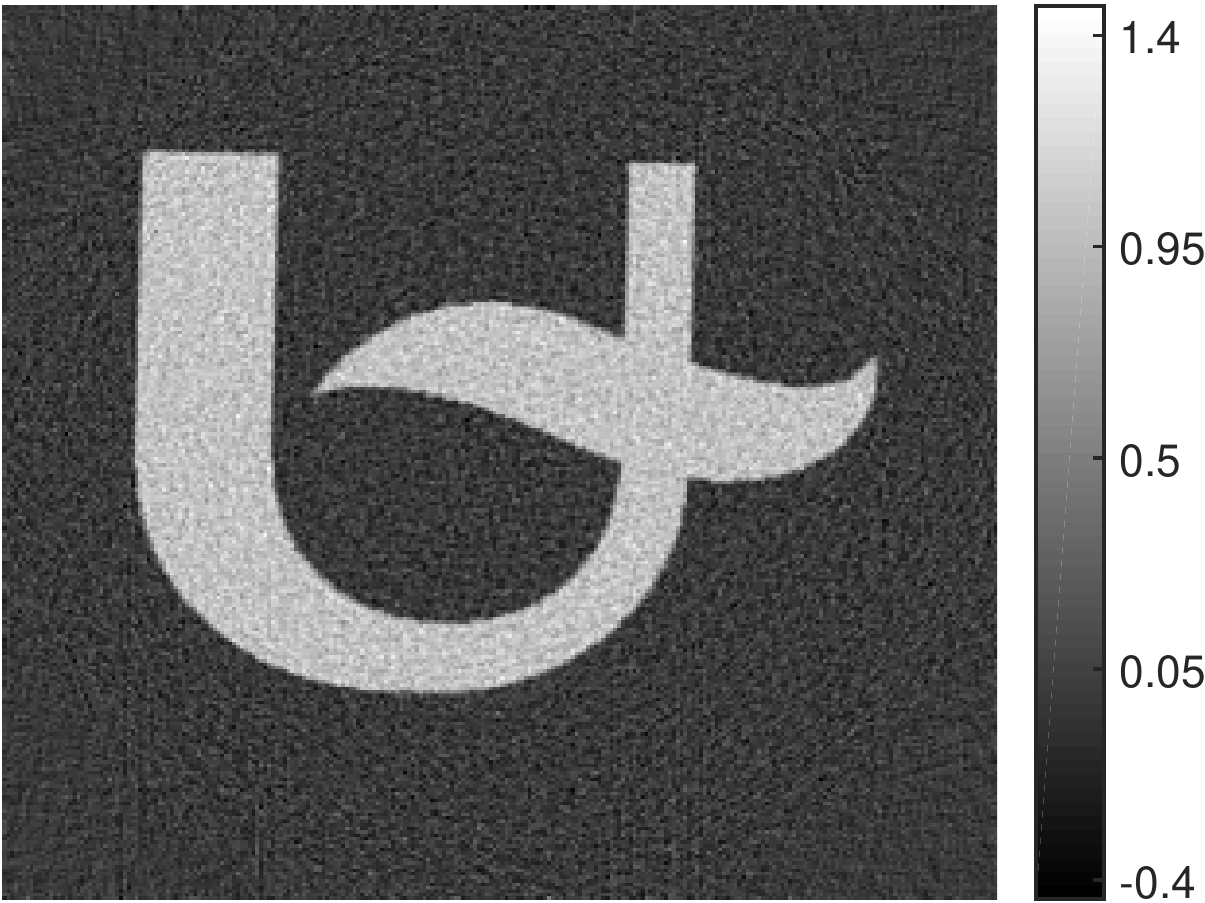}    
        \caption{}       
  \end{subfigure}
    \begin{subfigure}[b]{0.33\textwidth}
        \includegraphics[scale=0.366, clip]{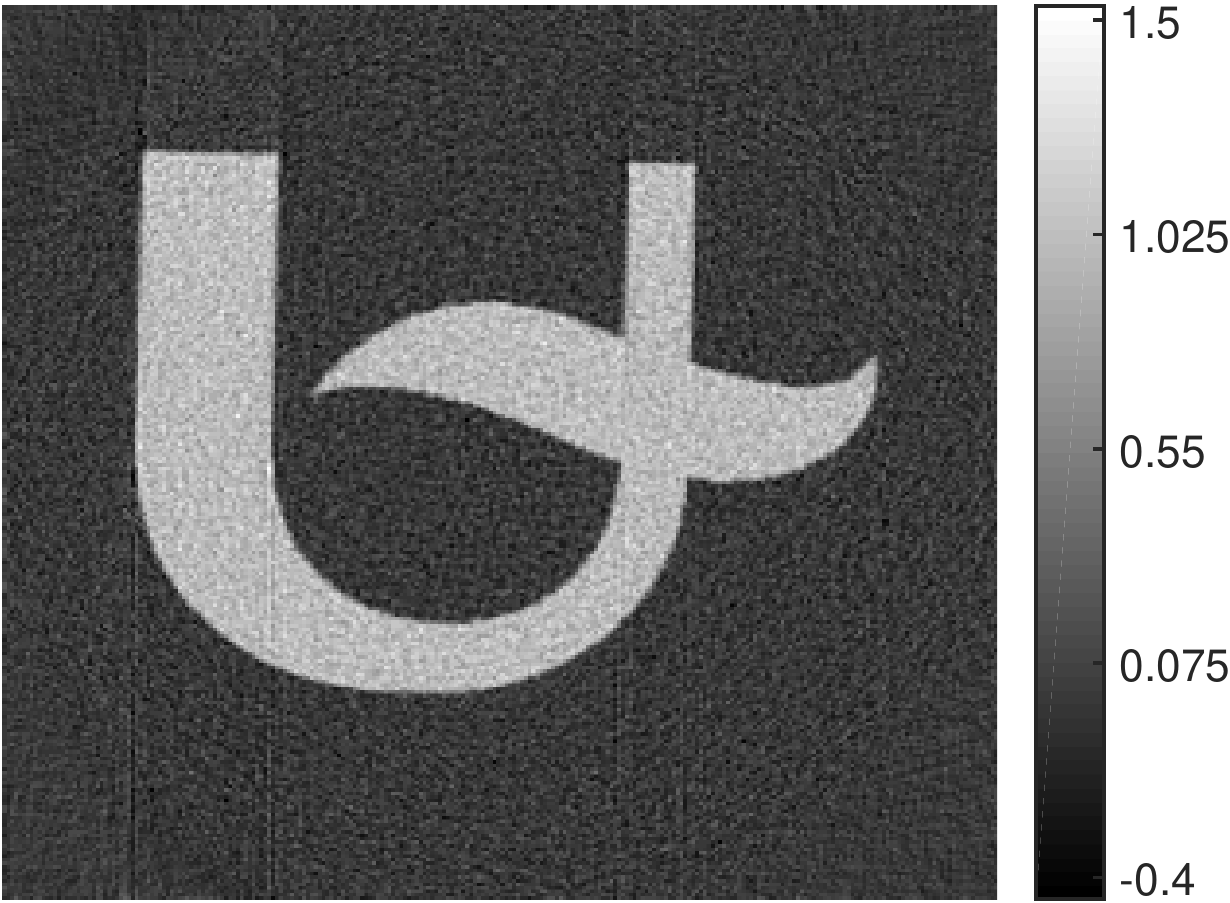}    
        \caption{}       
  \end{subfigure}
  \begin{subfigure}[b]{0.33\textwidth}
        \includegraphics[scale=0.366, clip]{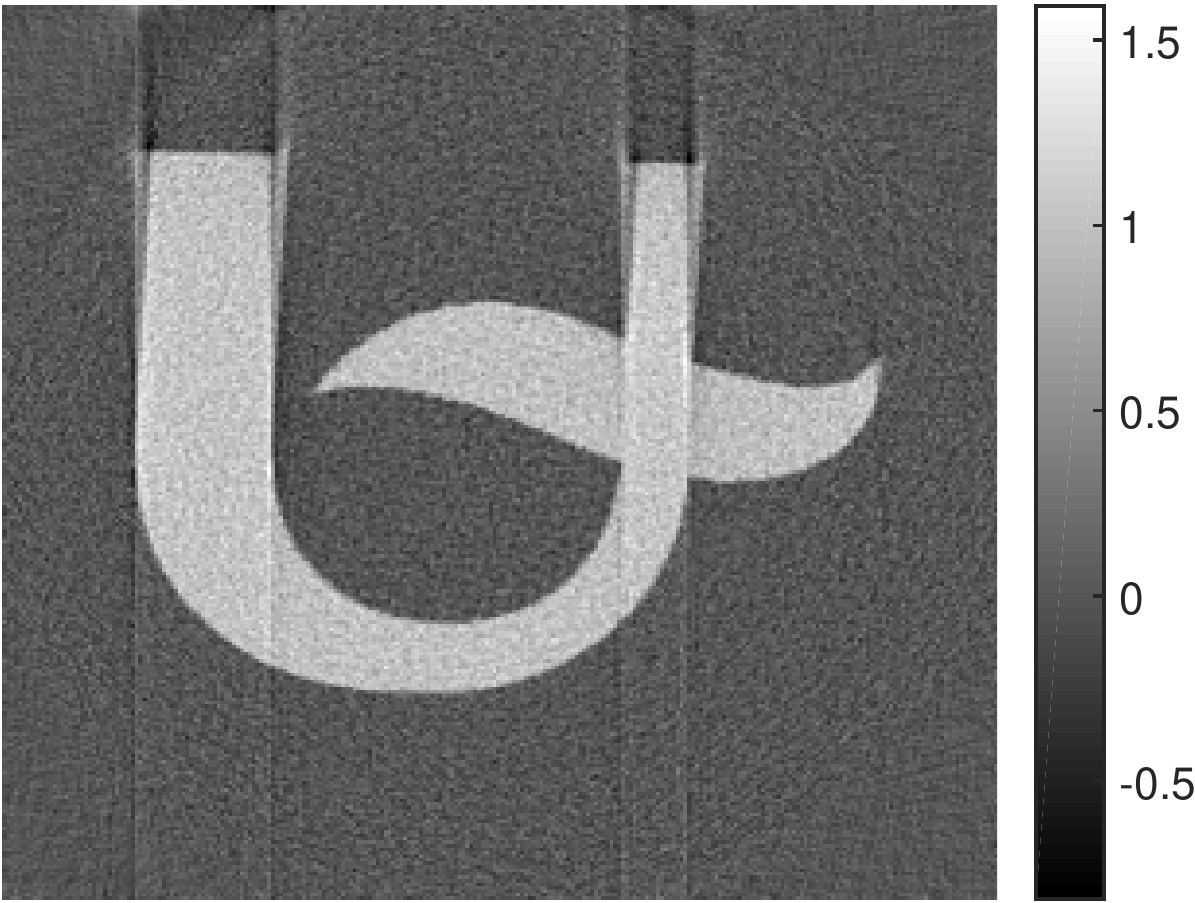}    
        \caption{}       
  \end{subfigure} \\
  \begin{subfigure}[b]{0.33\textwidth}
        \includegraphics[scale=0.366, clip]{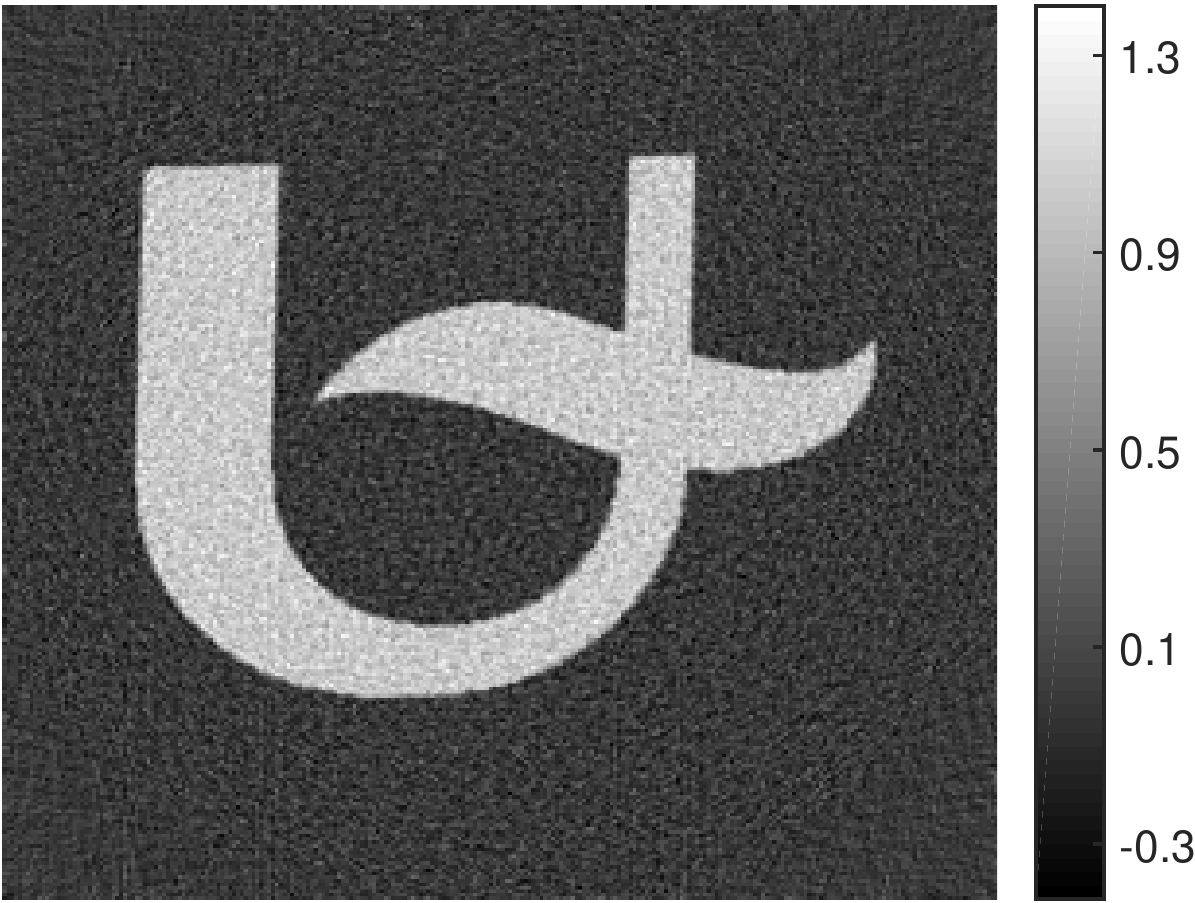}    
        \caption{}       
  \end{subfigure}
  \begin{subfigure}[b]{0.33\textwidth}
        \includegraphics[scale=0.366, clip]{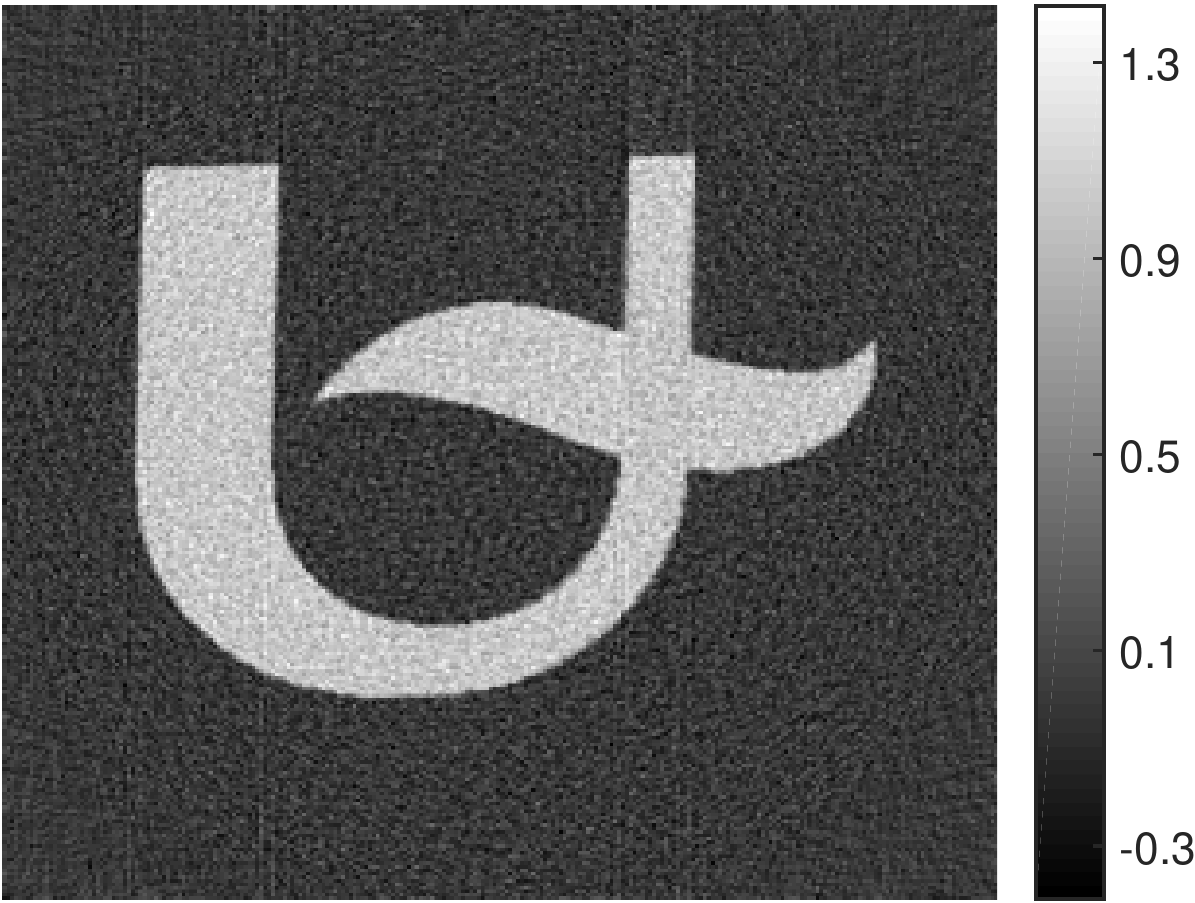}   
        \caption{}       
  \end{subfigure}
  \begin{subfigure}[b]{0.33\textwidth}
        \includegraphics[scale=0.366, clip]{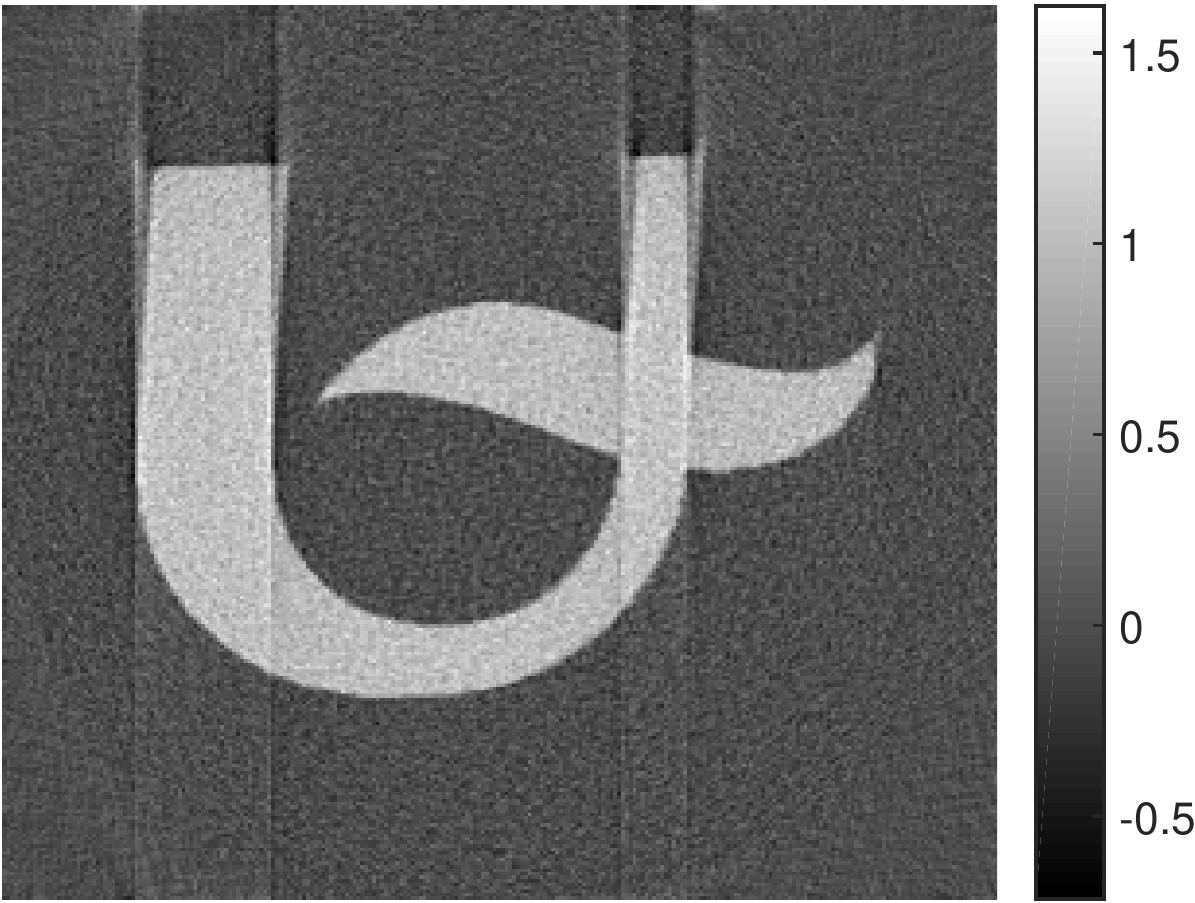}    
        \caption{}       
  \end{subfigure}

\caption{Reconstructions for the 3 different examples from figure \ref{fig:examples}. Each row represents a different example. (a)-(d)-(g) Reconstruction $f_{t^{l/2}}^{\text{corr}}$ for each example. (b)-(e)-(h) Reconstruction $\hat{f}_{t^{l/2}}^{\text{corr}}$ for each column (c)-(f)-(i) Reconstruction $f^{\text{rec}}_{t^{l/2}}$ without correcting for motion. We can see that the reconstruction with the exact motion works the best (first column).}\label{fig:correcting_noise}
\end{figure}

The results of the reconstruction can be found in the next table. We use the same abbreviations as in table \ref{table:rec_ex}. We see that adding normal distributed error does not affect much the quality of the reconstruction. The error with the exact motion (see second column in table~\ref{table:rec_ex_noise}) is also in this case comparable with the error we make when the  object is stationary (see first column in table~\ref{table:rec_ex_noise}). The same conclusions can be drawn from the images of the reconstructions (see figure~\ref{fig:correcting_noise}). 
\begin{table}[H]
\begin{center}
\begin{tabular}{ccccc}
\hline
& $\left \| f^{\text{ex}}_{\Delta t/2} - f_{\Delta t/2} \right \|_2$  & $\left \| f_{\Delta t/2}^{\text{corr}} - f_{\Delta t/2} \right \|_2$ &   $\left \| \hat{f}_{\Delta t/2}^{\text{corr}} - f_{\Delta t/2}\right \|_2$  & $\left \| f_{\Delta t/2}^{\text{rec}} - f_{\Delta t/2} \right \|_2$\\
\hline
Motion 1: Shift & $3.3296$ & $4.4059$ & $5.0106$ & $6.1845$ \\
Motion 2: Rotation & $3.5557$ & $3.4496$ & $4.2850$ & $13.386$ \\
Motion 3& $3.4983$ & $3.4416$ & $3.6837$ & $13.506$   \\
\hline 
\end{tabular} 
\caption{The error with respect to the exact solution $f_{\Delta t/2}$ when noise is added to the images. The first column is the error of the reconstruction if object $f$ is stationary in the state at time $\Delta t/2$. The second column is the error when we correct for the exact motion. The third column is the error when we correct for the estimated motion. The last column shows the error in absence of motion correction. We see that the error in the first two columns is similar which means our method works. We can see that when using the estimated motion, we get an error of similar magnitude} \label{table:rec_ex_noise}
\end{center}
\end{table}

We see that adding normal distributed noise has no big impact on the quality of the restrictions. This can be seen  in the table as well as in the above figure.
\section{Conclusion} \label{sec:conclusion}
We have shown in this paper that the motion can be determined starting from CT-scan data and that we can use this to correct CT-scan images. In fact, we show that techniques from imaging such as optical flow an be used for the dynamic CT-problem. Furthermore, the correction for the motion can be done very efficiently because all x-rays can be calculated in parallel. Although the results look promising, there is still work to refine the proposed method. First of all we have only tested the presented algorithms with simulated data so it would be very interesting to look how it performs with real data. Secondly, we calculate the motion in every pixel although in certain areas we can see that there is no motion present. Certainly when we look at extending our methods to the 3D case, it is not possible anymore to calculate the motion in every single pixel. We already did some experiments how to split up sinogram data into parts where there is motion and into parts where there is no motion present. In the areas where we know that the object stay stationary, we obviously do not need to calculate the motion here. Moreover, we only need to reconstruct this data once. The question arises how to adapt the methods if we only need to perform everything on a small area. Also the automatic determination of the regularisation parameter for the estimation of the motion is something that needs to be done. If this would work, this could significantly improve our methods.




\bibliographystyle{elsarticle-harv}
\bibliography{references}

\begin{thebibliography}{25}
\expandafter\ifx\csname natexlab\endcsname\relax\def\natexlab#1{#1}\fi
\providecommand{\url}[1]{\texttt{#1}}
\providecommand{\href}[2]{#2}
\providecommand{\path}[1]{#1}
\providecommand{\DOIprefix}{doi:}
\providecommand{\ArXivprefix}{arXiv:}
\providecommand{\URLprefix}{URL: }
\providecommand{\Pubmedprefix}{pmid:}
\providecommand{\doi}[1]{\href{http://dx.doi.org/#1}{\path{#1}}}
\providecommand{\Pubmed}[1]{\href{pmid:#1}{\path{#1}}}
\providecommand{\bibinfo}[2]{#2}
\ifx\xfnm\relax \def\xfnm[#1]{\unskip,\space#1}\fi
\bibitem[{Anandan(1989)}]{anandan1989computational}
\bibinfo{author}{Anandan, P.}, \bibinfo{year}{1989}.
\newblock \bibinfo{title}{A computational framework and an algorithm for the
  measurement of visual motion}.
\newblock \bibinfo{journal}{International Journal of Computer Vision}
  \bibinfo{volume}{2}, \bibinfo{pages}{283--310}.
\newblock \URLprefix \url{https://doi.org/10.1007/BF00158167},
  \DOIprefix\doi{10.1007/BF00158167}.
\bibitem[{Andersen and Kak(1984)}]{andersen1984simultaneous}
\bibinfo{author}{Andersen, A.H.}, \bibinfo{author}{Kak, A.C.},
  \bibinfo{year}{1984}.
\newblock \bibinfo{title}{Simultaneous algebraic reconstruction technique
  (sart): a superior implementation of the art algorithm}.
\newblock \bibinfo{journal}{Ultrasonic imaging} \bibinfo{volume}{6},
  \bibinfo{pages}{81--94}.
\bibitem[{Baker et~al.(2010)Baker, Bennett, Kang and Szeliski}]{rol_shut}
\bibinfo{author}{Baker, S.}, \bibinfo{author}{Bennett, E.},
  \bibinfo{author}{Kang, S.B.}, \bibinfo{author}{Szeliski, R.},
  \bibinfo{year}{2010}.
\newblock \bibinfo{title}{Removing rolling shutter wobble}, in:
  \bibinfo{booktitle}{Computer Vision and Pattern Recognition (CVPR), 2010 IEEE
  Conference on}, \bibinfo{organization}{IEEE}. pp.
  \bibinfo{pages}{2392--2399}.
\bibitem[{Bardow et~al.(2016)Bardow, Davison and
  Leutenegger}]{bardow2016simultaneous}
\bibinfo{author}{Bardow, P.}, \bibinfo{author}{Davison, A.J.},
  \bibinfo{author}{Leutenegger, S.}, \bibinfo{year}{2016}.
\newblock \bibinfo{title}{Simultaneous optical flow and intensity estimation
  from an event camera}, in: \bibinfo{booktitle}{Proceedings of the IEEE
  Conference on Computer Vision and Pattern Recognition}, pp.
  \bibinfo{pages}{884--892}.
\bibitem[{Basnayake et~al.(2013)Basnayake, Luttman and Bollt}]{HS_TV}
\bibinfo{author}{Basnayake, R.}, \bibinfo{author}{Luttman, A.},
  \bibinfo{author}{Bollt, E.}, \bibinfo{year}{2013}.
\newblock \bibinfo{title}{A lagged diffusivity method for computing total
  variation regularized fluid flow}.
\newblock \bibinfo{journal}{Contemp. Math} \bibinfo{volume}{586},
  \bibinfo{pages}{57--64}.
\bibitem[{Batenburg and Sijbers(2011)}]{batenburg2011dart}
\bibinfo{author}{Batenburg, K.J.}, \bibinfo{author}{Sijbers, J.},
  \bibinfo{year}{2011}.
\newblock \bibinfo{title}{Dart: a practical reconstruction algorithm for
  discrete tomography}.
\newblock \bibinfo{journal}{IEEE Transactions on Image Processing}
  \bibinfo{volume}{20}, \bibinfo{pages}{2542--2553}.
\bibitem[{Bruhn et~al.(2005)Bruhn, Weickert and Schn{\"o}rr}]{Bruhn2005}
\bibinfo{author}{Bruhn, A.}, \bibinfo{author}{Weickert, J.},
  \bibinfo{author}{Schn{\"o}rr, C.}, \bibinfo{year}{2005}.
\newblock \bibinfo{title}{Lucas/kanade meets horn/schunck: Combining local and
  global optic flow methods}.
\newblock \bibinfo{journal}{International Journal of Computer Vision}
  \bibinfo{volume}{61}, \bibinfo{pages}{211--231}.
\bibitem[{Gregor and Benson(2008)}]{gregor2008computational}
\bibinfo{author}{Gregor, J.}, \bibinfo{author}{Benson, T.},
  \bibinfo{year}{2008}.
\newblock \bibinfo{title}{Computational analysis and improvement of sirt}.
\newblock \bibinfo{journal}{IEEE Transactions on Medical Imaging}
  \bibinfo{volume}{27}, \bibinfo{pages}{918--924}.
\bibitem[{Hahn(2014)}]{hahn2014efficient}
\bibinfo{author}{Hahn, B.}, \bibinfo{year}{2014}.
\newblock \bibinfo{title}{Efficient algorithms for linear dynamic inverse
  problems with known motion}.
\newblock \bibinfo{journal}{Inverse Problems} \bibinfo{volume}{30},
  \bibinfo{pages}{035008}.
\bibitem[{Horn and Schunck(1981)}]{HS}
\bibinfo{author}{Horn, B.K.}, \bibinfo{author}{Schunck, B.G.},
  \bibinfo{year}{1981}.
\newblock \bibinfo{title}{Determining optical flow}, in:
  \bibinfo{booktitle}{1981 Technical symposium east},
  \bibinfo{organization}{International Society for Optics and Photonics}. pp.
  \bibinfo{pages}{319--331}.
\bibitem[{Joseph(1982)}]{joseph1982improved}
\bibinfo{author}{Joseph, P.M.}, \bibinfo{year}{1982}.
\newblock \bibinfo{title}{An improved algorithm for reprojecting rays through
  pixel images}.
\newblock \bibinfo{journal}{IEEE transactions on medical imaging}
  \bibinfo{volume}{1}, \bibinfo{pages}{192--196}.
\bibitem[{Li et~al.(2005)Li, Schreibmann, Yang and Xing}]{li2005motion}
\bibinfo{author}{Li, T.}, \bibinfo{author}{Schreibmann, E.},
  \bibinfo{author}{Yang, Y.}, \bibinfo{author}{Xing, L.}, \bibinfo{year}{2005}.
\newblock \bibinfo{title}{Motion correction for improved target localization
  with on-board cone-beam computed tomography}.
\newblock \bibinfo{journal}{Physics in medicine and biology}
  \bibinfo{volume}{51}, \bibinfo{pages}{253}.
\bibitem[{Louis(1996)}]{sota_4}
\bibinfo{author}{Louis, A.K.}, \bibinfo{year}{1996}.
\newblock \bibinfo{title}{Approximate inverse for linear and some nonlinear
  problems}.
\newblock \bibinfo{journal}{Inverse Problems} \bibinfo{volume}{12},
  \bibinfo{pages}{175}.
\bibitem[{Lu et~al.(2006)Lu, Parikh, Hubenschmidt, Bradley and
  Low}]{lu2006comparison}
\bibinfo{author}{Lu, W.}, \bibinfo{author}{Parikh, P.J.},
  \bibinfo{author}{Hubenschmidt, J.P.}, \bibinfo{author}{Bradley, J.D.},
  \bibinfo{author}{Low, D.A.}, \bibinfo{year}{2006}.
\newblock \bibinfo{title}{A comparison between amplitude sorting and
  phase-angle sorting using external respiratory measurement for 4d ct}.
\newblock \bibinfo{journal}{Medical physics} \bibinfo{volume}{33},
  \bibinfo{pages}{2964--2974}.
\bibitem[{Lucas et~al.(1981)Lucas, Kanade et~al.}]{LK}
\bibinfo{author}{Lucas, B.D.}, \bibinfo{author}{Kanade, T.}, et~al.,
  \bibinfo{year}{1981}.
\newblock \bibinfo{title}{An iterative image registration technique with an
  application to stereo vision.}, in: \bibinfo{booktitle}{IJCAI}, pp.
  \bibinfo{pages}{674--679}.
\bibitem[{Mang and Biros(2015)}]{mang2015inexact}
\bibinfo{author}{Mang, A.}, \bibinfo{author}{Biros, G.}, \bibinfo{year}{2015}.
\newblock \bibinfo{title}{An inexact newton--krylov algorithm for constrained
  diffeomorphic image registration}.
\newblock \bibinfo{journal}{SIAM Journal on Imaging Sciences}
  \bibinfo{volume}{8}, \bibinfo{pages}{1030--1069}.
\bibitem[{Mooser et~al.(2013)Mooser, Forsberg, Hack, Sz{\'e}kely and
  Sennhauser}]{mooser2013estimation}
\bibinfo{author}{Mooser, R.}, \bibinfo{author}{Forsberg, F.},
  \bibinfo{author}{Hack, E.}, \bibinfo{author}{Sz{\'e}kely, G.},
  \bibinfo{author}{Sennhauser, U.}, \bibinfo{year}{2013}.
\newblock \bibinfo{title}{Estimation of affine transformations directly from
  tomographic projections in two and three dimensions}.
\newblock \bibinfo{journal}{Machine vision and applications}
  \bibinfo{volume}{24}, \bibinfo{pages}{419--434}.
\bibitem[{Natterer(2001)}]{math_ct}
\bibinfo{author}{Natterer, F.}, \bibinfo{year}{2001}.
\newblock \bibinfo{title}{The mathematics of computerized tomography}.
\newblock Classics in applied mathematics 32, \bibinfo{publisher}{Society for
  Industrial and Applied Mathematics}.
\bibitem[{Palenstijn et~al.(2011)Palenstijn, Batenburg and
  Sijbers}]{palenstijn2011performance}
\bibinfo{author}{Palenstijn, W.}, \bibinfo{author}{Batenburg, K.},
  \bibinfo{author}{Sijbers, J.}, \bibinfo{year}{2011}.
\newblock \bibinfo{title}{Performance improvements for iterative electron
  tomography reconstruction using graphics processing units (gpus)}.
\newblock \bibinfo{journal}{Journal of structural biology}
  \bibinfo{volume}{176}, \bibinfo{pages}{250--253}.
\bibitem[{Pan et~al.(2009)Pan, Sidky and Vannier}]{pan2009commercial}
\bibinfo{author}{Pan, X.}, \bibinfo{author}{Sidky, E.Y.},
  \bibinfo{author}{Vannier, M.}, \bibinfo{year}{2009}.
\newblock \bibinfo{title}{Why do commercial ct scanners still employ
  traditional, filtered back-projection for image reconstruction?}
\newblock \bibinfo{journal}{Inverse problems} \bibinfo{volume}{25},
  \bibinfo{pages}{123009}.
\bibitem[{Simoncini and Szyld(2007)}]{simoncini2007recent}
\bibinfo{author}{Simoncini, V.}, \bibinfo{author}{Szyld, D.B.},
  \bibinfo{year}{2007}.
\newblock \bibinfo{title}{Recent computational developments in krylov subspace
  methods for linear systems}.
\newblock \bibinfo{journal}{Numerical Linear Algebra with Applications}
  \bibinfo{volume}{14}, \bibinfo{pages}{1--59}.
\bibitem[{Tarnec et~al.(2014)Tarnec, Destrempes, Cloutier and Garcia}]{HS_conv}
\bibinfo{author}{Tarnec, L.L.}, \bibinfo{author}{Destrempes, F.},
  \bibinfo{author}{Cloutier, G.}, \bibinfo{author}{Garcia, D.},
  \bibinfo{year}{2014}.
\newblock \bibinfo{title}{A proof of convergence of the horn-schunck optical
  flow algorithm in arbitrary dimension}.
\newblock \bibinfo{journal}{SIAM Journal on Imaging Sciences}
  \bibinfo{volume}{7}, \bibinfo{pages}{277--293}.
\bibitem[{Van~Eyndhoven et~al.(2014)Van~Eyndhoven, Batenburg and
  Sijbers}]{van2014region}
\bibinfo{author}{Van~Eyndhoven, G.}, \bibinfo{author}{Batenburg, K.J.},
  \bibinfo{author}{Sijbers, J.}, \bibinfo{year}{2014}.
\newblock \bibinfo{title}{Region-based iterative reconstruction of structurally
  changing objects in ct}.
\newblock \bibinfo{journal}{IEEE Transactions on Image Processing}
  \bibinfo{volume}{23}, \bibinfo{pages}{909--919}.
\bibitem[{Van~Eyndhoven et~al.(2012)Van~Eyndhoven, Sijbers and
  Batenburg}]{van2012combined}
\bibinfo{author}{Van~Eyndhoven, G.}, \bibinfo{author}{Sijbers, J.},
  \bibinfo{author}{Batenburg, J.}, \bibinfo{year}{2012}.
\newblock \bibinfo{title}{Combined motion estimation and reconstruction in
  tomography}, in: \bibinfo{booktitle}{European Conference on Computer Vision},
  \bibinfo{organization}{Springer}. pp. \bibinfo{pages}{12--21}.
\bibitem[{Webb(1990)}]{webb1990watching}
\bibinfo{author}{Webb, S.}, \bibinfo{year}{1990}.
\newblock \bibinfo{title}{From the watching of shadows: the origins of
  radiological tomography}.
\newblock \bibinfo{publisher}{CRC Press}.

\end{thebibliography}




%
\end{document}